\documentclass[12pt]{article}

\usepackage{layout}
\usepackage[textwidth=460pt, textheight=630pt,voffset=15pt]{geometry}

\usepackage{graphicx}%
\usepackage{multirow}%
\usepackage{amsmath,amssymb,amsfonts}%
\usepackage{amsthm}%
\usepackage{mathrsfs}%
\usepackage{mathtools}
\usepackage{stmaryrd} % for \llbrackets
\usepackage{accents} % for \llbrackets
\usepackage{tensor}

\usepackage[title]{appendix}%
\usepackage{xcolor}%
\usepackage{textcomp}%
\usepackage{manyfoot}%
\usepackage{booktabs}%
\usepackage{algorithm}%
\usepackage{algorithmicx}%
\usepackage{algpseudocode}%
\usepackage{listings}%
\usepackage{etoolbox}
\usepackage{placeins} % to force placement of float figures.
\usepackage{afterpage}

\usepackage[pdftex,colorlinks=true, pdfstartview=FitV, linkcolor= Mlink, citecolor=Mlink, urlcolor= Mlink, hyperindex=true, hyperfigures=true]{hyperref}
\usepackage{authblk}
 \usepackage[numbers,sort&compress]{natbib}
\IfFileExists{aastexv6.sty}{\input{aastexv6.sty}}{\input{/Users/quentinvigneron/Documents/Travail/Research/tex_/aastexv6.sty}}% For ADS abbreviations
\input{Short_cut_NR_Topo.sty}

\title{\bf Non-relativistic regime and topology: topological term in the Einstein equation}
\author[1]{{Quentin} {Vigneron}\footnote{\href{mailto:quentin.vigneron@umk.pl}{quentin.vigneron@umk.pl}
}}
\affil[1]{{Institute of Astronomy, Faculty of Physics, Astronomy and Informatics}, {Nicolaus Copernicus University}, {{Grudziadzka 5}, {Toru\'n}, {87-100}, {Poland}}}

\begin{document}
%\layout

\maketitle

\begin{abstract}{We study the non-relativistic (NR) limit of relativistic spacetimes in relation with the topology of the Universe. We first show that the NR limit of the Einstein equation is only possible in Euclidean topologies, i.e. for which the covering space is~$\mE^3$. 
We interpret this result as an inconsistency of general relativity in non-Euclidean topologies and propose a modification of that theory which allows for the limit to be performed in any topology. For this, a second reference non-dynamical connection is introduced in addition to the physical spacetime connection. The choice of reference connection is related to the covering space of the spacetime topology. Instead of featuring only the physical spacetime Ricci tensor, the modified Einstein equation features the difference between the physical and the reference Ricci tensors. This theory should be considered instead of general relativity if one wants to study a universe with a non-Euclidean topology and admitting a non-relativistic limit.}
\end{abstract}

%\keywords{keyword1, Keyword2, Keyword3, Keyword4}



\newpage
\section{Introduction and summary}
%\subsection{Motivations}

%Expansion and topology are two characteristics of the Universe general relativity has made possible to study from the early XX$^{\rm th}$ century. Only later we managed to include the description of expansion in Newton's theory of gravitation, which describes the non-relativistic regime, first for homogeneous expanding models \citep{1955_Heckmann_et_al, 1956_Heckmann_et_al}, then for a general (inhomogeneous) solution \citep{1997_Buchert_et_al, 2021_Vigneron}. However, the study of different topologies has remained until recently only described by general relativity. This is because the spatial Ricci curvature tensor in Newton's theory is exactly zero, imposing the topology to be Euclidean (i.e. when the covering space $\tilde\Sigma$ is the Euclidean space $\mE^3$), and thus preventing the study of universe models with a spatial non-Euclidean topology, i.e. $\tilde\Sigma \not= \mE^3$ (see section~\ref{sec::def_topo} for more details on the definition of Euclidean and non-Euclidean topologies). In other words, while we have a relativistic theory defined for any topology, the non-relativistic theory we have is only defined for Euclidean topologies. 
 
\begin{figure}[t]
	\centering
	\includegraphics[width=440pt]{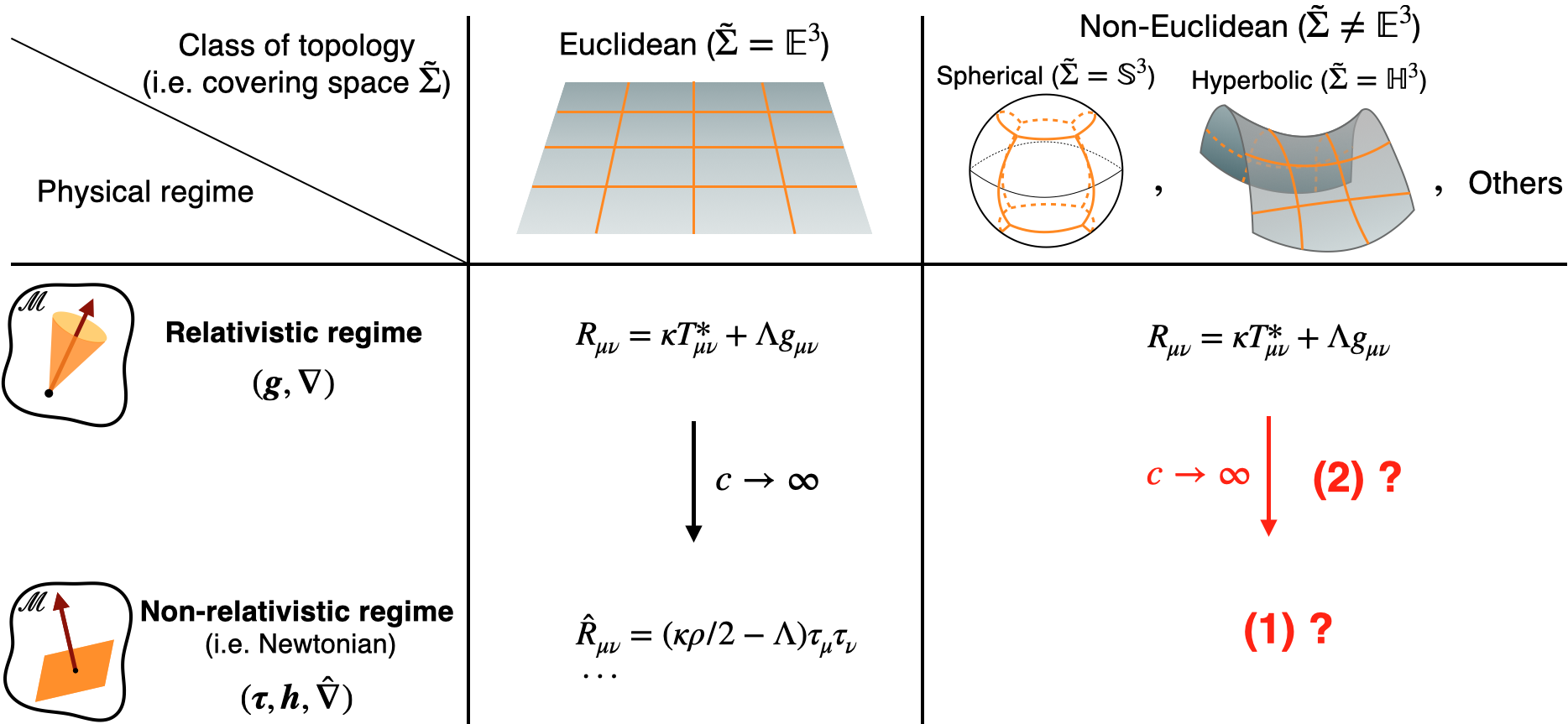}
	\caption{In this scheme, the second and third columns represent a Universe with, respectively, a Euclidean spatial topology and a non-Euclidean one. The second and third lines represent the relativistic and non-relativistic regimes. The former is defined from local Lorentzian invariance, and is mathematically described by Lorentzian structures solution of the Einstein equation, which is defined in any topology. The latter is defined from local Galilean invariance and is mathematically described via Galilean structures (Section~\ref{sec::Gal_struct}) solution of the Newton-Cartan equation (i.e. Newton's theory), \textit{only} in the Euclidean case. ``(1)~What non-relativistic equation should we consider in non-Euclidean topologies?'', and ``(2)~Is the non-relativistic limit of the Einstein's equation possible in non-Euclidean topologies?'' were the two questions that initiated the work of this paper.}
	\label{fig::sch0}
\end{figure}

\begin{figure}[b]
	\centering
	\includegraphics[width=15cm]{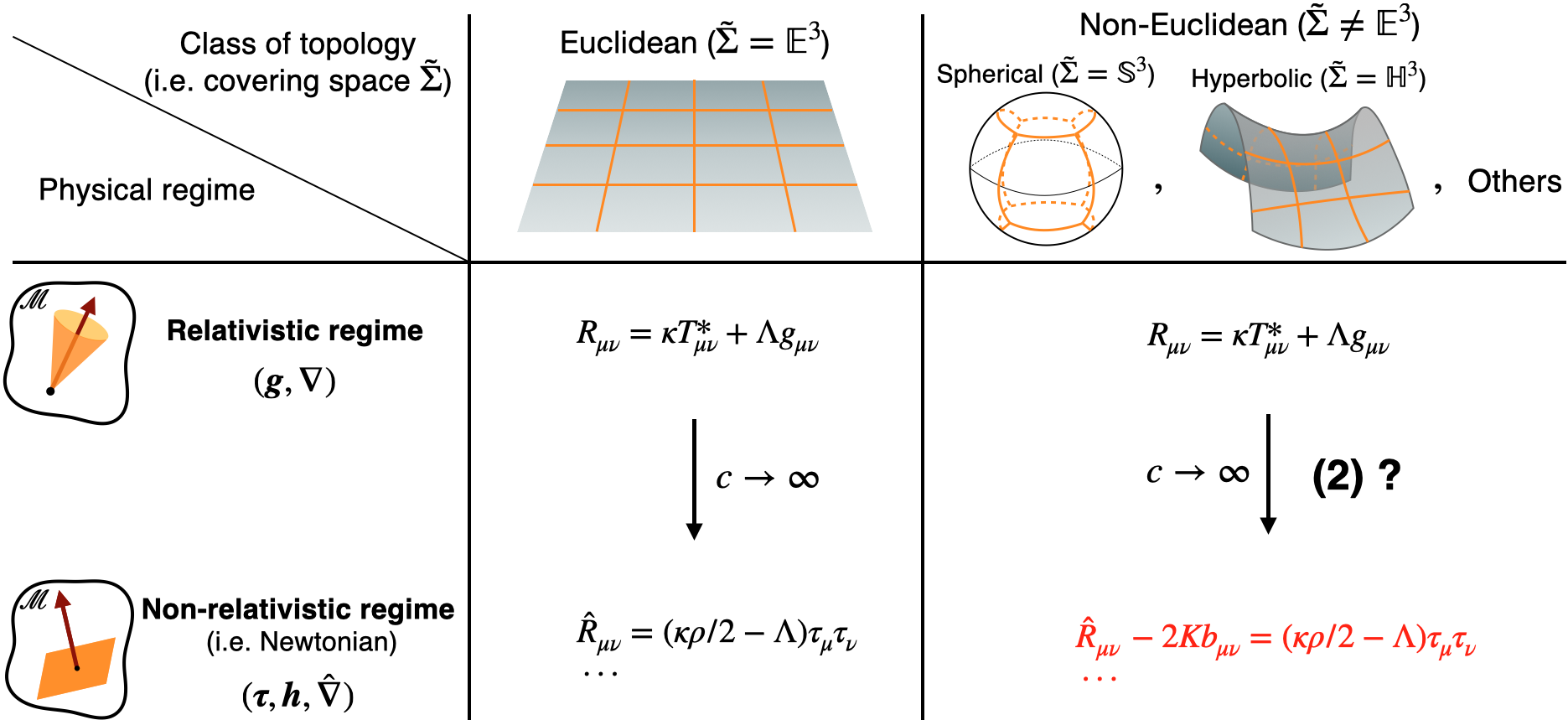}
	\caption{The answer to the first question was found in \citep{1976_Kunzle, 2022_Vigneron_b}: the non-relativistic theory on non-Euclidean topologies is given by a modification of the Newton-Cartan equation, in which a spatial curvature term is added.}
	\label{fig::sch1}
\end{figure}

\begin{figure}[t]
	\centering
	\includegraphics[width=15cm]{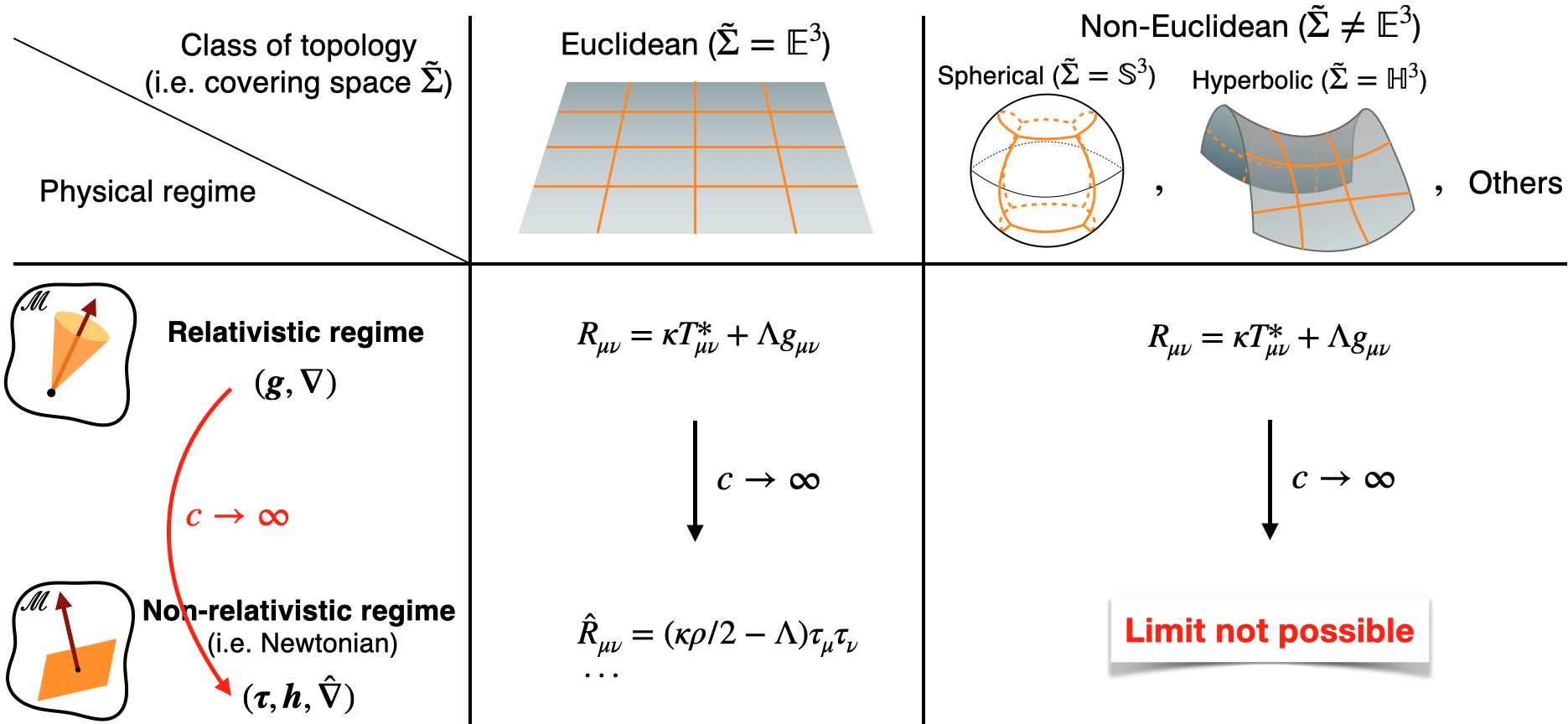}
	\caption{Two new results appear on this third scheme: i) the NR limit fundamentally corresponds to a limit of structures, from a Lorentzian structure to a Galilean structure (Sections~\ref{sec::gen_the_limit} and \ref{sec::the_limit}), i.e. it is a limit between Lorentzian and Galilean invariance; ii) the NR limit of the Einstein equation is only possible in 4-manifolds with a Euclidean spatial topology (Section~\ref{sec::limit_Einstein}).}
	\label{fig::sch2}
\end{figure}

\begin{figure}[!hb]
	\centering
	\includegraphics[width=15cm]{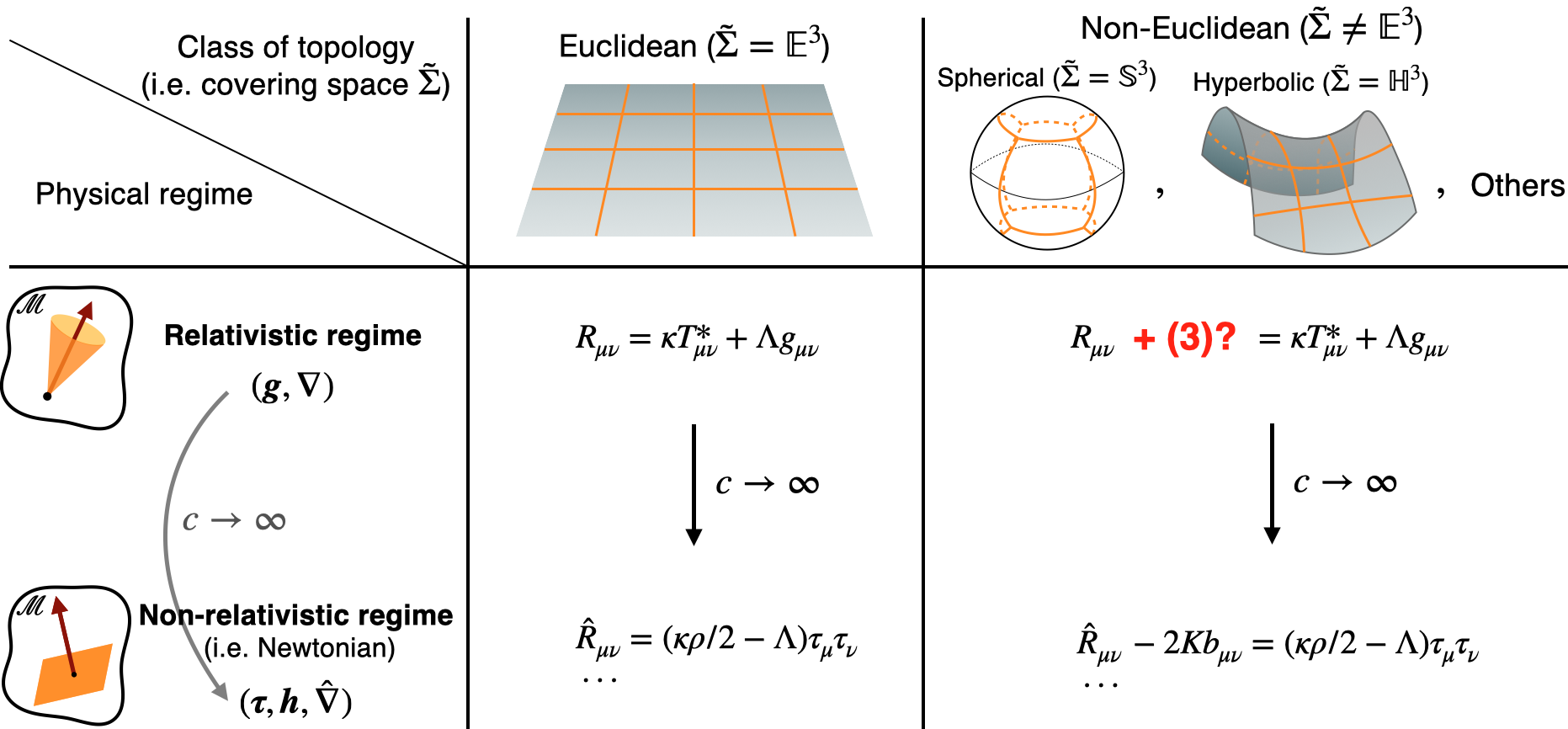}
	\caption{If we require the NR limit to exist in any topologies, %i.e. we require compatibility between the NR and the relativistic regime, 
	then the relativistic equation to consider should feature an additional term with respect to the Einstein equation. ``(3)~What is that term?'' is the third question we are interested in in this paper. Several arguments for why this question is fundamental are drawn in Section~\ref{sec::Question}.}
	\label{fig::sch3}
\end{figure}

\begin{figure}[t]
	\centering
	\includegraphics[width=15cm]{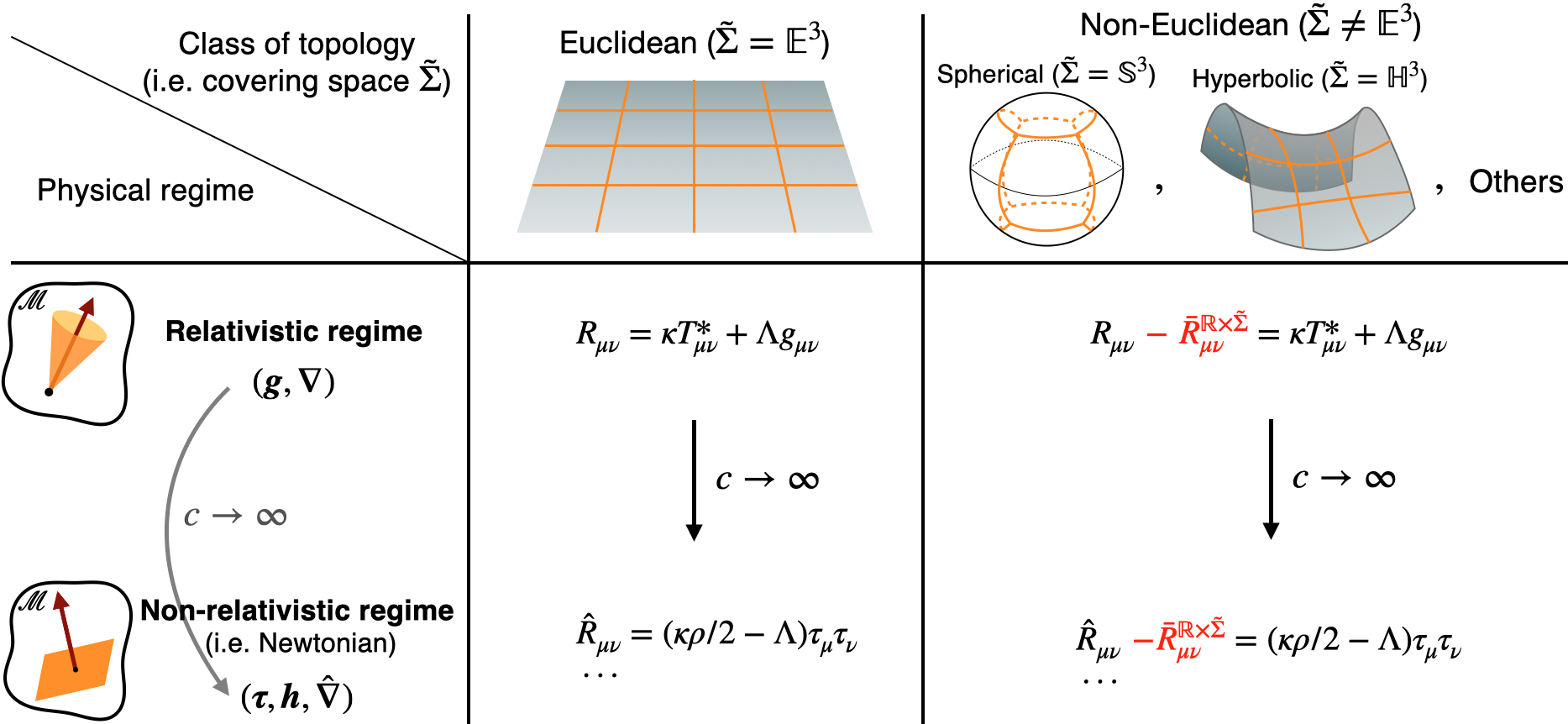}
	\caption{By adding in the Einstein equation a non-dynamical, reference, spacetime curvature $\bar R_{\mu\nu}^{\mathbb R\times\tilde\Sigma}$ related to the covering space ${\mathbb R\times\tilde\Sigma}$ of the spacetime topology, we allow for the NR limit to be performed in any topologies. In the follow up paper of this study \citep{2023_Vigneron_et_al_b}, we show that this new relativistic equation implies that the expansion becomes blind to the spatial curvature (i.e. $\Omega_{\not= K} = 1, \ \forall \Omega_K$).}
	\label{fig::sch4}
\end{figure}

The goal of this paper is to study the link between the non-relativistic (NR) regime, the relativistic regime and the topology of the Universe. In particular, a new modification of the Einstein equation in relation with the topology will be proposed. The logic of this study can be summarised into three main questions which are presented below, along with the main results of the paper. To give a visual understanding if these results, we have also condensed them into a sequence of figures at the end of this introduction (Figure~\ref{fig::sch0}, ~\ref{fig::sch1},~\ref{fig::sch2},~\ref{fig::sch3}, and~\ref{fig::sch4}). Each new implementation between one figure to the next one is highlighted in red.\saut

\textit{Non-relativistic theory and topology}: In cosmology, we usually use two types of theories: Lorentz invariant theories, i.e. relativistic theories, such as general relativity in the current paradigm; and a Galilean invariant theory, i.e. a non-relativistic theory, which is Newton's theory. While general relativity is well defined for any topology we can physically consider for the Universe, Newton's theory can only be defined on a spatial Euclidean topology, i.e. when the covering space $\tilde\Sigma$ is the Euclidean space $\mE^3$ (see section~\ref{sec::def_topo} for more details on the definition of Euclidean and non-Euclidean topologies). In other words, while we have a relativistic theory defined for any topology, the non-relativistic theory we have is only defined for Euclidean topologies. This leads to a first question (Figure~\ref{fig::sch0}):
 \begin{itemize}
	\item[(1)] What equations, i.e. theory, should we consider to describe the non-relativistic regime of gravitation in non-Euclidean topologies?
\end{itemize}
This question was physically motivated in \citep{2022_Vigneron} and properly formalised in \citep{1976_Kunzle, 2021_Vigneron, 2022_Vigneron} using a mathematical object called \textit{Galilean structures} (see section~\ref{sec::Gal_struct}), which enables us to write Galilean invariant equations in the same language as general relativity, i.e. differential geometry. Using this object, we can formulate Newton's equations as 4-dimensional equations on a 4-manifold: these are the so-called Newton-Cartan equations~\citep[e.g.][]{1972_Kunzle}.

The answer to question (1) was given by \citet{1976_Kunzle} and fully studied in \citet{2022_Vigneron_b, 2023_Vigneron_et_al_a} (see also the heuristic approach of \citep{2009_Roukema_et_al, 2020_Barrow}): the non-relativistic theory in non-Euclidean topologies is given by a modified Newton-Cartan equation (Figure~\ref{fig::sch1}). This theory could be called ``non-relativistic-non-Euclidean theory'', and in this sense Newton's theory would be called the ``non-relativistic-Euclidean theory''. In~\citep{2022_Vigneron_b}, we used the term \textit{non-Euclidean Newtonian theory} (hereafter NEN theory) which we keep for the remaining of this paper. This theory has to be understood as \textit{an extension} (to include non-Euclidean topologies, e.g. spherical) and \textit{not a modification} of Newtonian gravitation, since Newton's second law is still valid and Galilean invariance is kept, contrary to MOND theory for instance. In other words, Newton's theory and the NEN theory are essentially the same NR theory of gravitation, but defined on a different topology. We showed in \citep{2022_Vigneron_b}, that this theory is the only physical Galilean invariant theory which can be defined on non-Euclidean topologies (spherical or hyperbolic in the case of \citep{2022_Vigneron_b}) and is second order. The theory is presented in Appendix~\ref{sec::NEN}. It has the same equations as Newton's theory, but with the presence of a non-zero spatial curvature in the spatial derivatives.\saut

\textit{Non-relativistic limit and the Einstein equation}: While we generally do not consider the non-relativistic regime to be a fundamental feature of the Universe, it is still used to study gravitation at the classical and quantum levels \citep[e.g.][]{1971_Iwasaki, 2022_Hartong_et_al}. For this reason, it is generally assumed that relativistic gravitation should include in a certain limit NR gravitation. Our current understanding of the link between these two regimes is that Einstein's equation is compatible with NR gravitation in a Euclidean topology, i.e. compatible with Newton's theory. We therefore raise a second question (Figure~\ref{fig::sch0}):
 \begin{itemize}
	\item[(2)] Is the Einstein equation compatible with the non-relativistic regime for any topology?
\end{itemize}
This question is somehow equivalent as asking if Einstein's equation is compatible with the NEN theory. Checking this compatibility is the first goal of this paper.% This question is schematised in Figure~\ref{fig::sch1}. %It is of growing importance for cosmology, for which both NR and relativistic calculations are used to study structure formation, due to a rising debate on the value of the global spatial curvature (hence on the type of topology of our Universe) that should be inferred from cosmological data \citep[e.g.][]{2020_Di-Valentino_et_al}.\saut

After properly defining the non-relativistic limit in Section~\ref{NR_limit}, we will show in Section~\ref{sec::Limit_EE} that the answer to that question is ``No'': the NR limit of the Einstein equation only exists if the topology is Euclidean (Figure~\ref{fig::sch2}).\saut

\textit{Topological term in the Einstein equation}: We argue in Section~\ref{sec::Question} that this result can be interpreted as an inconsistency of general relativity for non-Euclidean spatial topologies, and that full mathematical compatibility between the relativistic and the NR regimes of gravitation for any topology should be required. Consequently, Einstein's equation needs to be modified, which leads to a third question (Figure~\ref{fig::sch3}):
 \begin{itemize}
	\item[(3)] What relativistic equation admitting a non-relativistic limit in any topology should we consider?
\end{itemize}
The second goal of this paper is to propose what is to our opinion the right relativistic equation. It is given by a slight variation of the bi-connection theory of \citet{1980_Rosen}, where a term related to topology is added to the Einstein equation (Figure~\ref{fig::sch4}). 
This term depends on a reference curvature $\bar R_{\mu\nu}$ defined from a reference connection $\bar\nabla$ (and not necessarily a reference metric) that is chosen as function of the spacetime topology. Therefore, this theory is a bi-connection theory (the usual physical Lorentzian metric of general relativity is still present), which only differs from general relativity in non-Euclidean topologies, for which $\bar R_{\mu\nu} \not=0$, and should be considered instead of general relativity if one wants to study a universe with a non-Euclidean topology and admitting a NR limit. 

We describe in more details that theory in Section~\ref{sec::Part_II}, and show in Section~\ref{sec::Galilean_limit} that its NR limit is well defined for non-Euclidean topologies, the resulting theory in the limit being, as expected, the NEN theory derived in \citep{2022_Vigneron_b}. As a consistency check, we also derive the non-relativistic dictionary of the bi-connection theory in Section~\ref{sec::consist_dico}.  We conclude in Section~\ref{sec::ccl_limit_NEN}.

\newpage

%\vfill

\clearpage

%\afterpage{\FloatBarrier}
%\clearpage

\section{Definitions}
\label{sec:def}

\subsection{What we mean by Euclidean and non-Euclidean topologies}
\label{sec::def_topo}

In this paper, we will always consider 4-manifolds $\CM$ of the form $\mathbb{R}\times\Sigma$, with $\Sigma$ a \textit{closed} 3-manifold. When talking about (non)-Euclidean topology, this will refer to the topology of $\Sigma$.\saut

The classification of 3-dimensional topological spaces (Thurston's classification \citep{1982_Thurston, 2006_Morgan_et_al}) says that any topology can be decomposed into a connected sum over eight irreducible classes of topology. Each class, or `prime topology', is characterised by its covering space $\tilde\Sigma$. A topology within a class is characterised by a discrete isometry group~$\Gamma$, such that $\Sigma = \tilde\Sigma/\Gamma$. In particular, a topology is said to be \textit{Euclidean} if its covering space $\tilde\Sigma$ is the Euclidean space $\mE^3$, and it is said to be \textit{non-Euclidean} if the covering space is different from $\mE^3$ (example: spherical if $\tilde\Sigma = \mS^3$, hyperbolic if $\tilde\Sigma = \mH^3$). More precision on these notions are given \citet[e.g.][]{1995_La_Lu}.\saut

An important property which is of particular interest for this paper is:
\begin{align}
	\T\CR = \T 0 \ \Rightarrow \ \textrm{Euclidean topology}, \label{eq::property}
\end{align}
or equivalently,
\begin{align}
	\textrm{non-Euclidean topology} \ \Rightarrow \ \T\CR \not= \T0, \nonumber
\end{align}
where $\T\CR$ is the Ricci curvature tensor of any Riemannian metric defined on $\Sigma$. In other words, Euclidean does not imply that the Ricci curvature tensor must be zero, but can be zero. It is not a geometrical notion but a topological notion, which is slightly more general as shown by~\eqref{eq::property}.\saut

We stress that this notion of Euclidean, spherical and hyperbolic where curvature is not necessarily constant is in fact, implicitly, the notion used in Cosmology. Indeed, the constant curvature $K$ is only valid for the background spatial metric. The real curvature, even for $K=0$, i.e. the `Euclidean' or `flat' case, is non zero due to the presence of inhomogeneities. Furthermore, considering closed spatial sections is not a restriction with respect to cosmology. Indeed, as will be presented in Section~\ref{sec::local_global}, because spatial boundary conditions are needed to solve either Einstein's or Newton's equations, a closed spatial manifold is (implicitly) always assumed in cosmology. Therefore for this reason, the type of manifolds and topologies considered in the Standard Model of Cosmology are exactly the ones described by the Thurston classification and used in the present paper. In particular, the Standard Model describes either a Euclidean, spherical or hyperbolic topology in the same sense given above (see \citep[e.g.][]{2020_Galloway_et_al} for cosmological models not considering prime topologies).
%We use the word `topology' instead of `geometry' to make clear , i.e while the covering space of $\Sigma$ is $\mE^3$ this does not mean that we should consider a flat metric on $\Sigma$.
%In other words, the manifolds, and their topology, considered in this paper are the one usually considered in cosmology. The aim of this section was the avoid any ambiguities
\saut

We also stress that the term ``non-Euclidean'' in the name ``non-Euclidean Newtonian theory'' refers to topology in the same sense. This reflects the fact that the purpose of this theory is not to take into account curvature effects which could come from relativistic corrections, but effects of the topology on the non-relativistic regime.\saut

\subsection{Notations}

In this paper, we will use three different spacetime structures on the same manifold: a Lorentzian structure $(\T g, \T\nabla)$, a reference connection $\T{\bar\nabla}$, and a Galilean structure $(\hat{\T h}, \T \tau, \hat{\T\nabla})$. Because similar quantities will be defined in each case, we will use the following notations:
\begin{itemize}
	\item The quantities related to the Lorentzian structure will not have specific notation, except in Section~\ref{sec::Galilean_limit} for the NR limit.
	\item The quantities related to the reference structure will be denoted with a bar. We have $\bar{\nabla}_\alpha$, $\bar\Gamma^\gamma_{\alpha\beta}$, $\bar R_{\alpha\beta}$, $\bar \CR_{\alpha\beta}$, $\bar h_{ij}$, $\bar D_i$.
	\item The quantities related to the Galilean structure will be denoted with a hat. We have $\hat\nabla_\mu$, $\hat\Gamma^\gamma_{\alpha\beta}$, $\hat R_{\alpha\beta}$, $\hat \CR^{\alpha\beta}$, $\hat h^{\alpha\beta}$, $\hat D_i$.
\end{itemize}

We denote $\kappa \coloneqq 8\pi G$ where $G$ is the gravitational constant.\saut

Throughout this paper, we denote indices running from 0 to 3 by Greek letters and indices running from 1 to 3 by Roman letters.

\section{The Non-relativistic limit of Lorentzian spacetimes}
\label{NR_limit}

We present in this section the NR limit used in this paper. As this limit is defined from a Lorentzian structure which is not necessarily solution of the Einstein equation, for now we will not apply the limit to this equation. This will be done in section~\ref{sec::Limit_EE}.

\subsection{What limit?}
\label{sec::Limit}

%As for now, the NR limit is usually defined as a weak field limit around the Minkowski metric (see e.g. chapter 4 in \citep{2019_Piotr}), or with the standard Post-Newtonian limit in which the components of the spacetime Lorentzian metric are expressed as a Taylor expanse of the speed of light for which each order depends on the Newtonian gravitational potential (e.g. chapter 39 in \citep{1973_MTW}). The problem with these limits is that they generally assume spatial flatness and no expansion. This is something we do not want as we necessarily need a non-zero spatial curvature in non-Euclidean topologies. 
Two limits are usually considered in general relativity. First, the Post-Minkowskian limit, where the Lorentzian metric $\T g$ is a perturbation of the Minkowski metric $\T \eta$ as $g_{\alpha\beta} = \eta_{\alpha\beta} + f_{\alpha\beta}$ where $|f_{\alpha\beta}| \ll 1$ and $|\partial_\gamma f_{\alpha\beta}| \ll 1$. The Einstein equation is then linearised at first order in $f_{\alpha\beta}$ and its derivatives. This approach has a main drawback for the purpose of this paper: because the perturbation is defined with respect to the Minkowski metric, the spatial topology is necessarily Euclidean. One way to allow for a non-Euclidean topology would be to replace $\T \eta$ by a FLRW metric, which is the approach used in cosmology \citep{1980_Bardeen}. However, this would imply by definition the spatial expansion to be necessarily given by the Friedmann equations. This is not appropriate as the expansion law should not be an assumption but a result coming from the NR equations at the limit (as shown in \citep{1997_Buchert_et_al, 2021_Vigneron} for Newton's theory). Second, the standard Post-Newtonian limit, which also has drawbacks making it not suited for the purpose of this paper. In this limit, the components of the spacetime Lorentzian metric are expressed as a Taylor series of the speed of light for which each order depends on the Newtonian gravitational potential (e.g. chapter 39 in \citep{1973_MTW}). The main drawback is the assumption of a flat spatial background metric (with or without expansion) incompatible with non-Euclidean topologies. \saut%This limit remains also essentially heuristic as, to our knowledge, there is no fundamental justification for the Taylor series being solely dependant on the Newtonian potential.\saut

Therefore, we need a general definition for what we call a NR limit regardless of the topology considered, along with a precise mathematical framework to perform this limit. In \citep{2022_Vigneron_b}, we argued that the best way of defining a NR theory regardless of the topology is to consider Galilean invariance as a fundamental principle and to use a mathematical object called \textit{Galilean structures} (presented in section~\ref{sec::Gal_struct}). These objects allow for a formulation of Newton's theory close to general relativity (this is the Newton-Cartan formulation \citep{1972_Kunzle, 2022_Hartong_et_al}). In this formalism the physical equation relates the energy content of a (spacetime) 4-manifold to its Galilean (spacetime) curvature, similarly to general relativity with Lorentzian structures. This formulation of Newton's theory also allows for a better conceptual description of expansion \citep{2022_Vigneron}. Therefore, by considering Lorentzian and Galilean invariances as the key properties characterising a NR limit, we give the following definition:\saut

\begin{definition}[Non-relativistic limit]
{A non-relativistic limit, regardless of the topology, starts from a locally Lorentz invariant theory and tends in the limit $c\rightarrow\infty$ to a locally Galilean invariant theory. The former theory is described by a Lorentzian structure and the latter by a Galilean structure.}\saut
\end{definition}

Such a limit was first developed by \citet{1976_Kunzle}, and further studied by \citep[e.g.][]{1990_Dautcourt_a, 1992_Rendall, 2011_Tichy_et_al, 2022_Hartong_et_al}. It does not have the problems listed above for the usual limits used in general relativity: there are no assumptions on the curvature or topology of the spatial sections; there is no assumed background expansion; non-linearities for the fluid dynamics are present at the limit. For these reasons, this approach is usual considered to be the fundamental way of defining the NR limit.\saut

The status of the 4-manifold on which is performed the limit is detailed in Section~\ref{sec::gen_the_limit} after having introduced Galilean structures in Section~\ref{sec::Gal_struct}. The limit is explicitly given in Section~\ref{sec::the_limit}, with additional properties and discussions in Sections~\ref{sec::interp_lambda}, \ref{sec::Limit_timelike} and \ref{sec::limit_R_munu}.

\subsection{Galilean structures}
\label{sec::Gal_struct}

A \textit{Galilean structure} defined on a differentiable 4-manifold $\CM$ is a set $(\T\tau, \hat{\T h}, \hat{\T\nabla})$, where $\T\tau$ is an exact 1-form, $\hat{\T h}$ is a symmetric positive semi-definite (2,0)-tensor of rank 3, i.e. its signature is (0,+,+,+), with $\hat h^{\alpha\mu}\tau_\mu = 0$, and $\hat{\T\nabla}$ is a connection compatible with $\T\tau$ and $\hat{\T h}$, called a \textit{Galilean connection}:
\begin{equation}
	\hat \nabla_\alpha \tau_\beta = 0 \quad ; \quad \hat \nabla_\gamma \hat h^{\alpha\beta} = 0. \label{eq::NC_def_structure}
\end{equation}
An important feature to have in mind regarding Galilean structures is the fact that because there is no spacetime metric in $(\T\tau, \hat{\T h}, \hat{\T\nabla})$, i.e. rank 4 semi-positive symmetric tensor, then {there is no duality between forms and vectors, and therefore no means of raising or lowering indices from $(\T\tau, \hat{\T h}, \hat{\T\nabla})$ alone}.\saut

In this framework, a vector $\T u$ is called a \textit{$\tau$-timelike vector} if $u^\mu\tau_\mu \not= 0$ or a \textit{unit} $\tau$-timelike vector if $u^\mu\tau_\mu = 1$, and a (n,0)-tensor $\T T$ is called \textit{spatial} if $\tau_\mu {T^{... \overset{\overset{\alpha}{\downarrow}}{\mu} ...}} = 0$ for all $\alpha \in \llbracket1,n\rrbracket$. The exact 1-form $\T \tau$ defines a foliation $\folGR$ in $\CM$, where $\Sigma_t$ are spatial hypersurfaces in $\CM$ defined as the level surfaces of the scalar field $t$, with $\T \tau = \T\dd t$. The tensor $\T h$ naturally defines a (spatial) Riemannian metric on $\folGR$. Finaly, the coefficients $\hat\Gamma^\sigma_{\alpha\beta}$ of the Galilean connection define a (spacetime) Riemann tensor ${\hat R^{\sigma}}_{\alpha\beta\gamma}$ on $\CM$ with the usual formula
\begin{equation}
	{\hat R^{\sigma}}_{\alpha\beta\gamma} \coloneqq 2 \, \partial_{[\beta}\hat\Gamma^\sigma_{\gamma]\alpha} + 2 \,  \hat\Gamma^\sigma_{\mu[\beta} \hat\Gamma^\mu_{\gamma]\alpha}. \label{eq::Gal_Riem}
\end{equation}
Unlike for Lorentzian structures, from the knowledge of $\T\tau$ and $\hat{\T h}$, the Galilean connection~$\hat{\T\nabla}$ is not unique and is defined up to a unit timelike vector $\T B$ (or equivalently, up to a spatial vector) and a two form $\kappa_{\alpha\beta}$. Its coefficients $\hat\Gamma_{\alpha\beta}^\gamma$ take the form
\begin{equation}
	\hat\Gamma_{\alpha\beta}^\gamma = \GB_{\alpha\beta}^\gamma + 2\tau_{(\alpha}\kappa_{\beta)\mu} \hat h^{\mu\gamma}, \label{eq::NC_connection}
\end{equation} 
where
\begin{equation}
	\GB_{\alpha\beta}^\gamma \coloneqq \hat h^{\gamma\mu}\left(\partial_{(\alpha} \bb{B}_{\beta)\mu} - \frac{1}{2} \partial_\mu \bbB_{\alpha\beta}\right) + B^\gamma \partial_{(\alpha}\tau_{\beta)}, \label{eq::NC_Gamma_B}
\end{equation}
and where $\tensor[^{\T B}]{\T b}{}$ is the projector orthonormal to the unit $\tau$-timelike vector $\T B$, defined with
\begin{equation}
	\bbB_{\alpha\mu} B^\mu \coloneqq 0 \quad ; \quad \bbB_{\alpha\mu}\hat h^{\mu\beta} \coloneqq {\delta_\alpha}^\beta - \tau_\alpha B^\beta. \label{eq::NC_def_bb}
\end{equation}
This type of structures is invariant under local Galilean transformations \citep{1972_Kunzle}. $\T \tau$ and $\hat{\T h}$ define a \textit{preferred foliation}, i.e. a preferred space direction, in $\CM$, hence corresponding to the Newtonian picture of spacetime.\saut

Finally, the spatial Riemanian tensor $\hat\CR^{\alpha\beta\mu\nu}$ induced by $\hat h^{\mu\nu}$ on the $\tau$-foliation is given by (see Section~3. of \citep{1986_Malament_a})
\begin{align}
	\hat\CR^{\alpha\beta\mu\nu} = {\hat R^{\alpha}}_{\gamma\sigma\upsilon} \hat h^{\beta\gamma} \hat h^{\sigma\mu} \hat h^{\upsilon\nu}, \label{eq::induced_Riem}
\end{align}
and the induced spatial Ricci tensor $\hat\CR^{\alpha\beta}$ is given by
\begin{align}
	\hat\CR^{\alpha\beta} = {\hat R}_{\mu\nu} \hat h^{\alpha\mu} \hat h^{\beta\nu}. \label{eq::induced_Ricc}
\end{align}

%\begin{remark}{Interestingly, $(\T\tau, \hat{\T h}, \hat{\T\nabla})$ alone does not define a preferred time direction which would be given by a vector field. The reason for this is that such a vector would be given by the dual to $\T \tau$ obtained via a raise of indices, which is not possible with $(\T\tau, \hat{\T h}, \hat{\T\nabla})$ as mentioned above. A preferred time direction will only appear once we consider a physical equation (like the Newton-Cartan equation), and will be given by a vector $\T G$ called Galilean vector (Section~\ref{sec::Gal_existence}).}\end{remark}

\subsection{Status of the 4-manifold within the limit}
\label{sec::gen_the_limit}

As presented in section~\ref{sec::Limit}, the principle behind a NR limit is to start from a Lorentzian structure ($\accentset{\lambda}{\T g}, \accentset{\lambda}{\Gamma}^\gamma_{\alpha\beta}$) and end-up with a Galilean structure $(\T\tau, \hat{\T h}, {\hat\Gamma}^\gamma_{\alpha\beta})$ for $\lambda \coloneqq 1/c^2 \rightarrow 0$, with $c$ the speed of light. We explain in Section~\ref{sec::interp_lambda} why we parametrise as function of $\lambda \coloneqq 1/c^2$.\saut

Because both structures must be defined on a manifold, then the most general way of defining a NR limit is to assume that the manifold on which the Lorentzian structure is defined is also parametrised by $\lambda$:
\begin{equation}
	\accl{\CM} {\rm \ equipped \ with\ } (\accentset{\lambda}{\T g}, \accentset{\lambda}{\Gamma}^\gamma_{\alpha\beta}) \quad \overset{\lambda \rightarrow 0}{\longrightarrow} \quad \accentset{(0)}{\CM} {\rm \ equipped \ with\ }(\T\tau, \hat{\T h}, {\hat\Gamma}^\gamma_{\alpha\beta}). \label{eq::chien}
\end{equation}
We argue in what follows why we should actually consider $\accl{\CM}$ to be the same for any $\lambda$.\saut

An interpretation we can make of the mathematical objects used in general relativity is the following: the 4-manifold $\CM$ is the object describing our Universe and the Lorentzian structure $(\T g, \T \nabla)$ defined of $\CM$ is the object describing physical measurements in the Universe. The metric defines measurement of distances; the connection defines how we compare physical properties between two points in the Universe. That measurement is constrained by a physical equation: the Einstein equation. Then, a NR limit corresponds to approximating this `relativistic measurement', for which, in particular, their exists an absolute velocity, by a `non-relativistic measurement', for which there are a preferred space $\T{\hat h}$ and time $\T \tau$ directions with respect to which distances should be measured.\saut

In this view, a NR limit should only be a change of definition of physical measurement, and not a change of Universe, i.e. the `manifold Universe' $\CM$ should not change under the procedure $c\rightarrow\infty$. Therefore, it seems more physical, or natural, to require a NR limit to be defined as follows:
\begin{equation}
	\CM {\rm \ equipped \ with\ } (\accentset{\lambda}{\T g}, \accentset{\lambda}{\Gamma}^\gamma_{\alpha\beta}) \quad \overset{\lambda \rightarrow 0}{\longrightarrow} \quad \CM {\rm \ equipped \ with\ }(\T\tau, \hat{\T h}, {\hat\Gamma}^\gamma_{\alpha\beta}).\label{eq::Limit_M=M}
\end{equation}
It is only \textit{a limit of structures}, i.e. the manifold on which is defined the Lorentzian and the Galilean structures does not change and is the same for both. \textit{Therefore, the topology of $\CM$, and especially the topology of $\Sigma$ (since we consider $\CM = \mathbb{R}\times\Sigma$), is unchanged when taking the NR limit.} This is a very important property that will be considered when applying the limit on the Einstein equation.\saut

{The approach~\eqref{eq::chien} is similar to the `active point of view' and the approach~\eqref{eq::Limit_M=M} to the `passive point of view' used to describe perturbation theory in cosmology \citep{2008_Malik_et_al}. Therefore, we expect~\eqref{eq::chien} and~\eqref{eq::Limit_M=M} to be equivalent and the above property (i.e. topology being fixed when performing the limit) to hold even in the first approach. In other words, approach~\eqref{eq::chien} might require $\accentset{{\lambda_1}}{\CM}$ to be diffeomorphic to $\accentset{{\lambda_2}}{\CM}$, which would imply their topology to be the same, and therefore, the topology to be still unchanged during the limiting procedure.}

\subsection{The limit}
\label{sec::the_limit}

A NR limit satisfying property~\eqref{eq::Limit_M=M} was developed by \citet{1976_Kunzle} and generalised by \citet{1992_Rendall}. The principle is to consider a 4-manifold $\CM$ and a 1-parameter family of Lorentzian structures $\{\overset{\lambda}{\T g}, \overset{\lambda}{\T \nabla}\}_{\lambda>0}$ on $\CM$ that depends smoothly on $\lambda$, such that
\begin{align}
	&\accentset{\lambda}{g}^{\alpha\beta} = \hat h^{\alpha\beta} + \lambda \, \accentset{(1)}{g}^{\,\alpha\beta} +  \lambda^2 \, \accentset{(2)}{g}^{\,\alpha\beta} + \mathcal{O}(\lambda^3), \label{eq::Limit_g^ab}\\
	&\accentset{\lambda}{g}_{\alpha\beta} = -\frac{1}{\lambda} \tau_\alpha \tau_\beta + \accentset{(0)}{g}_{\alpha\beta} +  \lambda \, \accentset{(1)}{g}_{\alpha\beta}+ \mathcal{O}(\lambda^2), \label{eq::Limit_cov_g_lambda}
\end{align}
where $\hat{\T h}$ is a positive semi-definite (2,0)-tensor of rank 3 on $\CM$, and $\T\tau$ a 1-form. An important property is that the $n$-th order of the metric contravariant components are not the inverse of the $n$-th order of the covariant components, i.e. $\accentset{(n)}{g}^{\,\alpha\mu} \, \accentset{(n)}{g}_{\mu\beta} \not= \delta^\alpha_\beta$. Instead we have by definition $\accentset{\lambda}{g}^{\alpha\mu} \accentset{\lambda}{g}_{\mu\beta} = \delta^\alpha_\beta$, which, with \eqref{eq::Limit_g^ab} and \eqref{eq::Limit_cov_g_lambda}, leads to
\begin{align}
	& -\tau_\mu \hat h^{\mu\alpha} = 0, \label{eq::loup}\\
	&-\accentset{(1)}{g}^{\,\alpha\mu} \,\tau_\mu\tau_\beta + \hat h^{\alpha\mu} \accentset{(0)}{g}_{\mu\beta} = \delta^\alpha_\beta, \label{eq::Starship} \\
	&-\accentset{(2)}{g}^{\,\alpha\mu} \,\tau_\mu\tau_\beta + \accentset{(1)}{g}^{\,\alpha\mu} \,\accentset{(0)}{g}_{\mu\beta} + \hat h^{\alpha\mu} \,\accentset{(1)}{g}_{\mu\beta} = 0, \label{eq::soupe}
\end{align}
for the first three orders. Equation~\eqref{eq::Starship} leads to
\begin{equation}
	\hat h^{\alpha\mu}\,\accentset{(0)}{g}_{\mu\beta} = \delta^\alpha_\beta - \tau_\beta B^\alpha,
\end{equation}
where we denote $B^\alpha \coloneqq -\tau_\mu \accentset{(1)}{g}^{\,\mu\alpha}$, with $B^\mu\tau_\mu = 1$. Then we can write $\accentset{(1)}{g}^{\,\alpha\beta}$ and $\accentset{(0)}{g}_{\alpha\beta}$ as
\begin{align}
	\accentset{(1)}{g}^{\,\alpha\beta}	&= -B^\alpha B^\beta + k^{\alpha\beta}, \label{eq::g^1_limit}\\
	\accentset{(0)}{g}_{\alpha\beta}		&= \bb{B}_{\alpha\beta} - 2\phi\tau_\alpha\tau_\beta, \label{eq::g_0_limit}
\end{align}
where $k^{\mu\nu}$ is spatial (i.e. $k^{\mu\alpha}\tau_\mu = 0$), $\phi$ is an arbitrary scalar, and $\bb{B}_{\alpha\beta}$ is defined in equation~\eqref{eq::NC_def_bb} and corresponds to the projector orthogonal to $\T B$ with respect to the tensors $\T \tau$ and $\hat{\T h}$. Then, computing the Levi-Civita connection of $\accentset{\lambda}{\T g}$, we find
\begin{align}
	\accentset{\lambda}{\Gamma}^\gamma_{\alpha\beta} &= 
			\frac{1}{\lambda} \hat h^{\gamma\mu}\left(\tau_\alpha \partial_{[\mu}\tau_{\beta]} + \tau_\beta \partial_{[\mu}\tau_{\alpha]}\right) 
			+ \left(2\phi \hat h^{\gamma\mu} + \accentset{(1)}{g}^{\,\gamma\mu}\right)\left(\tau_\alpha \partial_{[\mu}\tau_{\beta]} + \tau_\beta \partial_{[\mu}\tau_{\alpha]}\right) \\
		&\qquad+ \hat h^{\gamma\mu}\left(\partial_{(\alpha} \bb{B}_{\beta)\mu} - \frac{1}{2} \partial_\mu \bb{B}_{\alpha\beta}\right) + B^\gamma \partial_{(\alpha}\tau_{\beta)} 
			+ \tau_\alpha\tau_\beta \hat h^{\gamma\mu}\partial_\mu \phi + \mathcal{O}(\lambda). \nonumber
\end{align}
If the 1-form $\T\tau$ is exact, which we will suppose from now\footnote{It is possible to consider $\T\tau$ not exact, and even $\T d \T\tau \not=0$, but this requires torsion which we do not consider in this paper (see e.g. \cite{2020_Hansen_et_al}).}, the connection $\accentset{\lambda}{\Gamma}^\gamma_{\alpha\beta}$ has a regular limit for $\lambda \rightarrow 0$ which corresponds to a Galilean connection $\hat\Gamma^\gamma_{\alpha\beta}$ compatible with $\T\tau$ and $\hat{\T h}$:
\begin{align}
	\accentset{\lambda}{\Gamma}^\gamma_{\alpha\beta} &= \hat\Gamma^\gamma_{\alpha\beta} + \mathcal{O}(\lambda) \nonumber \\
	&= \tensor[^{\T B}]{\Gamma}{}^\gamma_{\alpha\beta} + \tau_\alpha\tau_\beta \hat h^{\gamma\mu}\partial_\mu \phi + \mathcal{O}(\lambda), \label{eq::Limit_connection_coeff}
\end{align}
where $ \tensor[^{\T B}]{\Gamma}{}^\gamma_{\alpha\beta}$ is defined in equation~\eqref{eq::NC_Gamma_B}. This Galilean connection is not the most general possible because the 2-form $\kappa_{\alpha\beta}$ [defined in equation~\eqref{eq::NC_connection}] is necessarily exact as we have $\kappa_{\alpha\beta} = \tau_{[\alpha} \partial_{\beta]} \phi$.\saut

In summary, from the Taylor series~\eqref{eq::Limit_g^ab} and \eqref{eq::Limit_cov_g_lambda} with $\T\tau = \T\dd t$, property~\eqref{eq::Limit_M=M} is fulfilled. To our knowledge, there is no proof that~\eqref{eq::Limit_g^ab} and \eqref{eq::Limit_cov_g_lambda} is the only way to define a limit fulfilling that property. However, the leading orders of these Taylor series are motivated by the Minkowski metric, as shown in the next section. For this reason, we think it is unlikely that another form than~\eqref{eq::Limit_g^ab} and \eqref{eq::Limit_cov_g_lambda} is possible. The same conclusion is drawn in \citep{1992_Rendall}, where the author analyses the mathematical properties of this limit starting from a restricted set of axioms in the case of a flat spatial {metric~$\hat h^{\alpha\beta}$}.\footnote{\citep{1992_Rendall} considers the more general case where odd powers of $1/c$ are present in the Taylor series. However, they are only physical, i.e. cannot be set to zero with a gauge transformation, from the order $c^{-1}$ in $g_{\mu\nu}$ and the order $c^{-5/2}$ for $g^{\mu\nu}$. Therefore, this does not change the main results of the present paper as it relies on the leading orders.}\saut

\begin{remark}
{$\accl{g}_{\mu\nu}$ is singular for $\lambda \rightarrow 0$. It is possible to define a non-singular limit of the metric by changing the property $\accentset{\lambda}{g}^{\alpha\mu} \accentset{\lambda}{g}_{\mu\beta} = \delta^\alpha_\beta$ into $\accentset{\lambda}{g}^{\alpha\mu} \accentset{\lambda}{g}_{\mu\beta} = \lambda \delta^\alpha_\beta$ as in \citep{2019_Ehlers}. The subsequent calculation remains however the same. So we will stick to the definition~\eqref{eq::Limit_cov_g_lambda} which is most often used.}
\end{remark}

\subsection{Local or global limit?}
\label{sec::local_global}

As we will see in Section~\ref{sec::limit_Einstein}, this limit applied on the Einstein equation will imply a topological constraint. This will be possible only if we consider the limit to hold globally on $\CM$. But, other than for pure investigative purposes, why one would want to consider this hypothesis? Indeed, instead of considering $\T\tau$ to be exact, implying the limit to be defined globally, $\T\tau$ could be assumed only closed, and therefore the limit would hold only locally. From this, no topological constraints could be drawn.

%A criticism that can be addressed to the above framed statement would be to say that instead of considering the limit to hold globally, i.e. with $\T\tau$ being exact, it could be considered only locally, i.e. $\T\tau$ being only closed. A consequence is that the $\tau$-foliation would not necessarily be a $\Sigma$ manifold, and therefore the zero curvature tensor result at the limit would not anymore give a constraint on spatial topology. 
This argument, while in general correct, does not apply from the moment we are solving the equations (which is generally what we want to do in physics). The reason is that in general relativity as well as in Newton's theory, elliptic equations need to be solved: the Gauss-Codazzi equations for the former theory and the Poisson equation for the latter. These equations being elliptic, they require boundary conditions which can be of two forms: either an infinite spatial manifold with suitable integrability (in other words an isolated system), or a compact manifold (closed for a physically relevant situation).\footnote{The only case when boundary conditions are not necessary is when exact homogeneity is assumed. This is however of little physical relevance, even for cosmology, as inhomogeneities (at least as perturbations) are always needed to make physical measurements and predictions. Therefore, boundary conditions are required.} This need for boundary conditions is present everywhere in general relativity (see \citep{2009_Dyer_et_al} for a discussion on this topic): e.g.
\begin{itemize}
	\item the Scalar-Vector-Tensor decompositions of symmetric 3-dimensional tensors used in standard perturbations is uniquely defined only with the above conditions \citep{1973_York},
	\item the use of the Fourier transform, or more generally harmonic decomposition, to study the linear regime in cosmology requires the same conditions (in fact for these reasons, the $\Lambda$CDM model is mathematically a model of a closed Universe),
	\item most definitions of mass require asymptotical flatness,
\end{itemize}
as well as in Newton's theory: e.g.
\begin{itemize}
	\item boundary conditions are required to obtain the Poisson equation (and therefore Newton's law) from the Newtonian limit of general relativity \citep[e.g.][]{1990_Dautcourt_a, 2021_Vigneron}. %For that reason, saying that Newton's theory does not hold globally is actually erroneous.
	\item expansion, usually introduced with a Hubble flow \citep[e.g.][]{1955_Heckmann_et_al, 1980_Peebles}, is in fact fundamentally a non-local field appearing when compact boundary conditions are considered (instead of an isolated system) in the Newton-Cartan equation \citep{2021_Vigneron}. 
\end{itemize}

These boundary conditions require the limit to hold everywhere on the domain they define. The boundary conditions considered in the present paper are the one of a closed manifold (the only physically relevant condition in cosmology, i.e. fulfilling the Copernican principle). Therefore, the limit should be defined everywhere on the manifold.
{These arguments also hold at the level of the theory (general relativity or Newton's theory): strictly speaking we never, and we cannot, use these theories only locally, as any resolution (in general) of their constitutive equations require boundary conditions.}

In conclusion, from the moment we are solving the equations, we need to consider general relativity, its NR limit and the NR theory at the limit to hold globally.\saut

\begin{remark}`Holding globally' does not mean that the theory obtained in the limit, e.g. Newton's theory, should give accurate physical results at all scales. It is a mathematical statement and requirement regarding the Taylor series~\eqref{eq::Limit_g^ab} and~\eqref{eq::Limit_cov_g_lambda} defining the limit, and the solutions of the equations. We expect on very large scales (illustrated by small wave numbers in the matter power spectrum) and around compact objects, the leading order to be not accurate anymore and post-non-relativistic (i.e. post-Newtonian) orders should be taken into account. {But, in any cases, these relativistic orders need the leading order to exist to be defined, i.e. need the full Taylor series to be mathematically well-defined.}
\end{remark}

\subsection{Interpretation of $\lambda$}
\label{sec::interp_lambda}

The Minkowski metric, in coordinates where $x^0$ has the dimension of a time, can be written as $\eta_{\alpha\beta} = \text{diag}(-c^2, 1, 1, 1)$, where $c$ is the speed of light. Its inverse is $\eta^{\alpha\beta} = \text{diag}(-c^{-2}, 1, 1, 1)$. Then assuming $c\rightarrow \infty$, these two matrices become
\begin{align*}
	\eta_{\alpha\beta} &\overset{c\rightarrow\infty}{\sim} \text{diag}(-c^2, 0, 0, 0), \\
	\eta^{\alpha\beta} &\overset{c\rightarrow\infty}{\sim}\text{diag}(0, 1, 1, 1).
\end{align*}
These leading orders for $c\rightarrow \infty$ have the same form as the leading orders of $\accentset{\lambda}{\T g}$ for $\lambda\rightarrow 0$: in coordinates adapted to the foliation given by $\T\tau$ (i.e. $\tau_\alpha = \delta^0_\alpha$ and $h^{\alpha\beta} = \delta^\alpha_a \delta^\beta_b h^{ab}$), we have $\accentset{(0)}{g}^{\,\alpha\beta} = \delta^\alpha_a \delta^\beta_bh^{ab}$ and $\accentset{(-1)}{g}_{\alpha\beta} = \text{diag}(-1/\lambda, 0, 0, 0)$. This shows that the limit $\lambda \rightarrow 0$ can be seen as a limit where $c\rightarrow\infty$ with $\lambda = c^{-2}$.\saut

\begin{figure}[t]
	\centering
	\includegraphics[width=7cm]{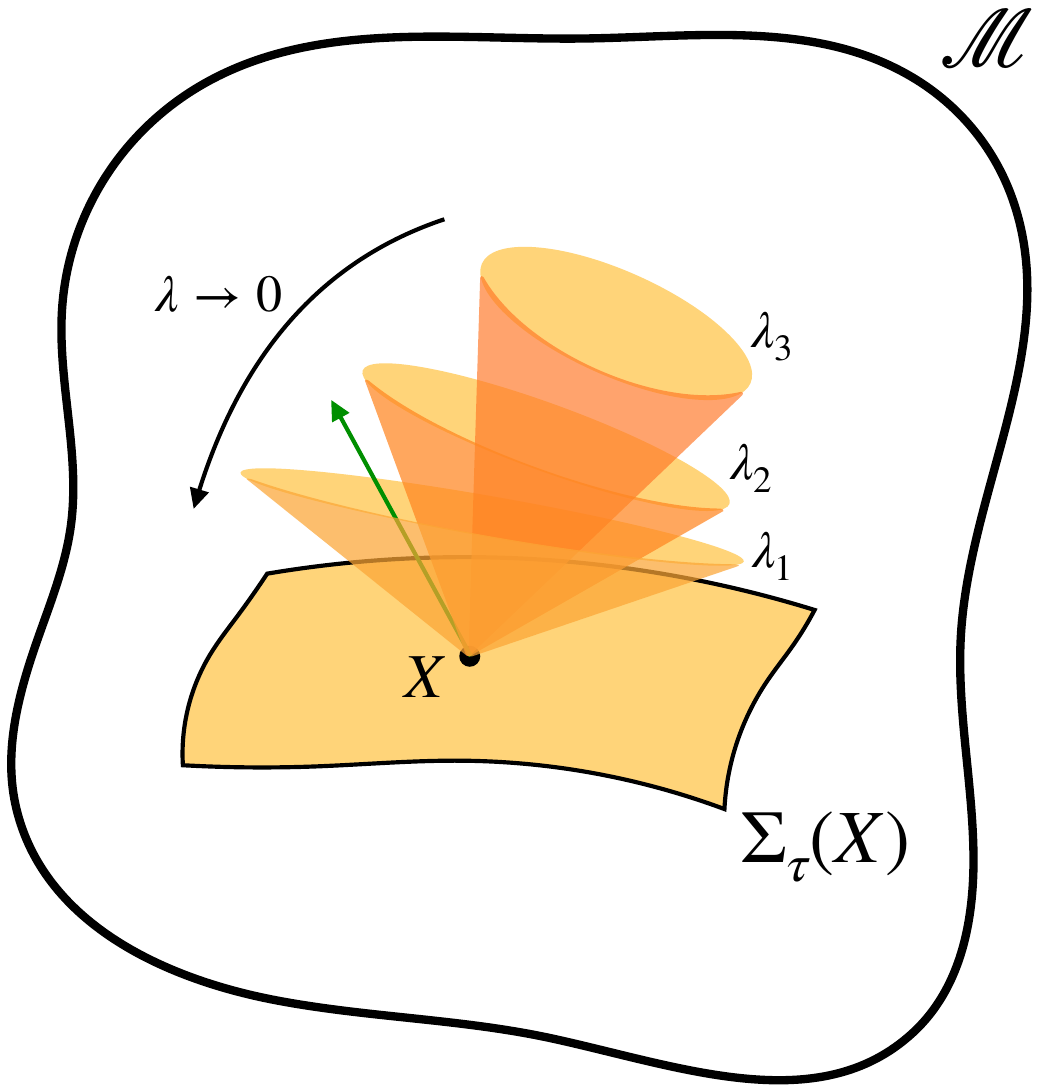}
	\caption{Scheme of three light-cones of the family $\{\overset{\lambda}{\T g}\}_{\lambda>0}$ of Lorentzian metrics at a point $X$ in $\CM$. The smaller $\lambda$ is, the more open the light-cone is: $\lambda_1 < \lambda_2 < \lambda_3$. The slice $\Sigma_\tau(X)$ corresponds to the only hypersurface member of the foliation defined by $\T\tau$ and passing through $X$. The limit of the family of light-cones is this hypersurface. The green vector is an example of a vector which is g-timelike for the Lorentzian metric $\overset{\lambda_1}{\T g}$, but spacelike for $\overset{\lambda_2}{\T g}$ and  $\overset{\lambda_3}{\T g}$, hence showing that no vector can be timelike for every $\overset{\lambda}{\T g}$ (see section~\ref{sec::Limit_timelike}).}
	\label{fig::Limit_Cone}
\end{figure}

Therefore, the speed of light related to a Lorentzian metric $\accentset{\lambda}{\T g}$ of the family $\{\overset{\lambda}{\T g}\}_{\lambda>0}$ depends on $\lambda$ and is given by
\begin{equation}
	c_\lambda = \lambda^{-1/2}.
\end{equation}
This means that the family $\{\overset{\lambda}{\T g}\}_{\lambda>0}$ of Lorentzian metrics defines a family of light-cones at each point of $\CM$. The light-cone related to a metric $\accentset{\lambda_1}{\T g}$ is more open than the one related to a metric $\accentset{\lambda_2}{\T g}$ if $\lambda_1 < \lambda_2$. This is represented in Figure~\ref{fig::Limit_Cone}.\saut

If one wants to set the speed of light to be 1, which corresponds to choosing a coordinate system such that $x^0 = ct$, this is only possible for one Lorentzian metric $\accentset{\lambda}{\T g}$. For all the other metrics, the speed of light in this coordinate system will differ from~$1$. This property is really important as it tells us that when we will consider equations (and their NR limit) which should feature the speed of light, we cannot take $c=1$, and are obliged to take $c = \lambda^{-1/2}$. This will be the case for the norm of timelike vectors (section~\ref{sec::Limit_timelike}) and for the Einstein equation (section~\ref{sec::Limit_EE}).

\subsection{Limit of $g$-timelike vectors}
\label{sec::Limit_timelike}

Because the notion of timelike vectors is defined in both structures, but is not equivalent, to distinguish between the two, we will call g-timelike vectors the ones related to a Lorentzian structure, and we already call $\tau$-timelike vectors the ones related to the Galilean structure.\saut

Let us consider a vector $\T u$ which is ``unit'' g-timelike\footnote{Because the speed of light related to $\accentset{\lambda}{\T g}$ cannot be set to 1 for all $\lambda$, then  ``unit'' means that the norm of $\T u$ is $c_\lambda$.} for every member of the family $\{\overset{\lambda}{\T g}\}_{\lambda>0}$, i.e. $u^\mu u^\nu\accentset{\lambda}{g}_{\mu\nu} = -1/\lambda$. If we assume that this vector does not depend on~$\lambda$, then for a sufficiently small $\lambda$ we have $(u^\mu\tau_\mu)^2 = 1$. As $\T u$ is \textit{any} unit g-timelike vector, this cannot be possible. So, $\T u$ needs to depend on $\lambda$. We write it $\accentset{\lambda}{\T u}$ (more precisely we define a family $\{\accentset{\lambda}{\T u}\}_{\lambda>0}$ of vectors). Because it is unit for all~$\accentset{\lambda}{\T g}$, then we have
\begin{equation}
	\accentset{\lambda}{u}^\alpha = \accentset{(0)}{u}^{\,\alpha} + \lambda\accentset{(1)}{u}^{\,\alpha} + \mathcal{O}(\lambda^2), \label{eq::u_limit_1}
\end{equation}
with $\accentset{(0)}{u}^{\,\mu}\,\tau_\mu = 1$, and $\accentset{(1)}{u}^{\,\mu}\,\tau_\mu = \frac{1}{2}\accentset{(0)}{u}^{\,\mu}\,\accentset{(0)}{u}^{\,\nu}\,\accentset{(0)}{g}_{\mu\nu}$. In addition, we can compute the covariant components of $\accentset{\lambda}{\T u}$, defined as $\accentset{\lambda}{u}_\alpha \coloneqq \accentset{\lambda}{u}^\mu\accentset{\lambda}{g}_{\mu\alpha}$, which become
\begin{equation}
	\accentset{\lambda}{u}_\alpha = -\frac{1}{\lambda}\tau_\alpha + \accentset{(0)}{u}_\alpha +\mathcal{O}(\lambda),
\end{equation}
where $\accentset{(0)}{u}_\alpha = \accentset{(0)}{u}^{\,\mu}\, \accentset{(0)}{g}_{\mu\alpha} - \frac{1}{2}\accentset{(0)}{u}^{\,\mu}\,\accentset{(0)}{u}^{\,\nu}\,\accentset{(0)}{g}_{\mu\nu}\tau_\alpha$. Because $\accentset{(0)}{u}^{\,\mu}\,\tau_\mu = 1$, this implies that \textit{a unit g-timelike vector $\accl{\T u}$ for all members of $\{\overset{\lambda}{\T g}\}_{\lambda>0}$ corresponds to a unit $\tau$-timelike vector at the limit $\lambda\rightarrow 0$}. However, the reverse is not possible: a $\tau$-timelike vector can never correspond to a g-timelike vector for all $\lambda$.

\subsection{Limit of the Lorentzian Riemann tensor}
\label{sec::limit_R_munu}

For the purpose of taking the limit of the Einstein equation, we present in this section the limit of the Riemann tensor related to a Lorentzian metric.\saut

The family $\{\overset{\lambda}{\T g}\}_{\lambda>0}$ defines a family of Riemann tensors $\{{\accentset{\lambda}{R}^\sigma}_{\gamma\alpha\beta}\}_{\lambda>0}$ on $\CM$. Using the expression of ${\accentset{\lambda}{R}^\sigma}_{\gamma\alpha\beta}$ in terms of the coefficients of the connection,
\begin{equation}
	{\accentset{\lambda}{R}^\sigma}_{\gamma\alpha\beta} = 2\partial_{[\alpha}\accentset{\lambda}{\Gamma}^\sigma_{\beta]\gamma} + 2\accentset{\lambda}{\Gamma}^\sigma_{\mu[\alpha}\accentset{\lambda}{\Gamma}^\mu_{\beta]\gamma},
\end{equation}
we easily see that the expansion of ${\accentset{\lambda}{R}^\sigma}_{\gamma\alpha\beta}$ in series of $\lambda$ takes the form
\begin{equation*}
	{\accentset{\lambda}{R}^\sigma}_{\gamma\alpha\beta} = {\hat{R}^\sigma}_{\gamma\alpha\beta} + \mathcal{O}(\lambda),
\end{equation*}
where ${\hat R^\sigma}_{\gamma\alpha\beta}$ is the Riemann tensor associated with the Galilean connection coefficients $\hat\Gamma^\gamma_{\alpha\beta}$ given by~\eqref{eq::Limit_connection_coeff}. Consequently, the leading order of the Ricci tensor coefficients $\overset{\lambda}{R}_{\alpha\beta}$ corresponds to the Ricci tensor of the Galilean connection:
\begin{equation*}
	\accentset{\lambda}{R}_{\alpha\beta} = {\hat R}_{\alpha\beta} + \mathcal{O}(\lambda).
\end{equation*}
Because Galilean structures do not feature non-degenerate metrics, raising and lowering indices is not possible. Therefore the Riemann and Ricci tensors of Galilean connections are only defined as, respectively, a (1,3)-tensor and a (0,2)-tensor. Consequently, identities (for instance Bianchi identities) where the coefficients ${\accentset{\lambda}{R}^\sigma}_{\gamma\alpha\beta}$ are raised or lowered by $\overset{\lambda}{\T g}$ have no equivalent in Galilean structures. However, the limit of these identities gives additional constraints on the (Galilean) Riemann tensor ${\hat R^\sigma}_{\gamma\alpha\beta}$. From the leading order of the interchanged symmetry $\tensor{\accentset{\lambda}{R}}{^\alpha_\gamma^\beta_\sigma} = \tensor{\accentset{\lambda}{R}}{^\beta_\sigma^\alpha_\gamma}$, we obtain
\begin{equation}
	\hat h^{\mu\beta}\tensor{\hat R}{^\alpha_\gamma_\mu_\sigma} - \hat h^{\mu\alpha}\tensor{\hat R}{^\beta_\sigma_\mu_\gamma} = 0, \label{eq::Limit_Bianchi_1}
\end{equation}
and from the contracted second Bianchi identity $\accentset{\lambda}{g}^{\mu\nu}\accentset{\lambda}{\nabla}_{\mu}\accentset{\lambda}{R}_{\nu\alpha} - \frac{1}{2}\accentset{\lambda}{\nabla}_\alpha \accentset{\lambda}{R}_{\mu\nu}\accentset{\lambda}{g}^{\mu\nu} = 0$, we obtain
\begin{equation}
	\hat h^{\mu\nu}\hat{\nabla}_{\mu}{\hat R}_{\nu\alpha} - \hat{\nabla}_\alpha\frac{\hat{R}_{\mu\nu}\hat h^{\mu\nu}}{2} = 0, \label{eq::Limit_Bianchi_2}
\end{equation}
at leading order. Both of these equations are geometrical constraints on the Galilean structures. In particular, equation~\eqref{eq::Limit_Bianchi_1}, called the Trautman condition, is essential to have an irrotational gravitational field \citep[see e.g][]{2019_Ehlers, 2021_Vigneron}, as expected for a NR theory.\saut

Apart for the leading orders of the Bianchi identities, no other constraints on the Galilean structure appears from the limit. In particular, the spatial Ricci curvature $\hat\CR^{\alpha\beta}$ related to $\hat{\T h}$ is totally free, and therefore also the spatial topology. Contraints on this tensor can only appear once we assume the Lorentzian structures to be solutions of the Einstein equation.\saut

Additional formulas related to the NR limit are presented in Appendix~\ref{app::formulas}.

\newpage
\section{Non-relativistic limit of the Einstein equation}
\label{sec::Limit_EE}

In this section, we give the answer to the question (2) raised in the introduction, namely: {\it Is the Einstein equation compatible with the non-relativistic regime for any topology?}

\subsection{Limit of the energy-momentum tensor}
\label{sec::limit_T_munu}

The energy-momentum tensor $\accl{T}^{\alpha\beta}$ of a general fluid of 4-velocity $\accl{u}^\alpha$ can be written as
\begin{equation}
	\accl{T}^{\alpha\beta} = \accl{\epsilon} \, \accl{u}^\alpha\accl{u}^\beta + 2\accl{q}^{(\alpha}\accl{u}^{\beta)} +\accl{p} \, \tensor[^u]{\accl{b}}{^\alpha^\beta} + \accl{\pi}^{\alpha\beta}, \label{eq::Limit_N_T}
\end{equation}
where $\tensor[^u]{\accl{b}}{_\alpha_\beta} \coloneqq \accl{g}_{\alpha\beta} + \lambda \accl{u}_\alpha\accl{u}_\beta$ is the projector orthogonal to $\accl{\T u}$. The expansion series of $\tensor[^u]{\accl{\T b}}{}$ is detailed in~\ref{app::Limit_Formulas}. As $\accl{\T T}$ has the dimension of an energy and $\accl{\T u}$ the dimension of a velocity, then $\accl{\epsilon}$ has the dimension of a mass density. Therefore in the case of a dust fluid, $\accl{\epsilon}$ is a zeroth order term given by the rest mass density of the fluid: $\accl{\epsilon} = \rho$ for all~$\lambda$. For a more general \textit{matter fluid}, where the fluid elements have internal energy, $\accl{\epsilon}$ will have higher order terms \citep{1976_Kunzle}. Using equations~\eqref{eq::u_limit_1} and~\eqref{eq::app_b_3} on the conservation law $\accl{\nabla}_\mu\accl{T}^{\mu\alpha} = 0$, we obtain the Newtonian mass conservation law, i.e. $\hat{\nabla}_\mu\left(\rho\accentset{0}{u}^\mu\right) = 0$, \textit{only if}~\citep[see][]{1976_Kunzle}
\begin{align}
	\accl{p} = \mathcal{O}(1) \quad ; \quad \accl{q}^\alpha = \mathcal{O}(\lambda) \quad ; \quad \accl{\pi}^{\alpha\beta} = \mathcal{O}(1). \label{eq::fluid_regular}
\end{align}
That is, the pressure $\accl{p}$, the heat flux $\accl{q}^\alpha$ and the anisotropic stress $\accl{\pi}^{\alpha\beta}$ are all regular for $\lambda \rightarrow 0$. Therefore, once we consider that there is no non-relativistic process for which mass is not conserved, i.e. once we require to have $\hat{\nabla}_\mu\left(\rho\accentset{0}{u}^\mu\right) = 0$ in a non-relativistic theory\footnote{See \citet{2016_Stichel} for an example of $
\hat{\nabla}_\mu\left(\rho\accentset{0}{u}^\mu\right) \not= 0$.}, then the conditions~\eqref{eq::fluid_regular} must hold. In that case, $\accl{T}^{\alpha\beta}$ has a regular limit, and we have
\begin{align}
	\accl{T}_{\alpha\beta} &= \frac{1}{\lambda^2}\left[\rho \, \tau_\alpha\tau_\beta\right] + \mathcal{O}(\lambda^{-1}), \label{eq::T_munu} \\
	\accl{T}_{\alpha\beta} - \frac{1}{2} \accl{T}^{\mu\nu}\accl{g}_{\mu\nu} \accl{g}_{\alpha\beta} &= \frac{1}{\lambda^2}\left[\frac{\rho}{2}\tau_\alpha\tau_\beta\right] + \mathcal{O}(\lambda^{-1}), \label{eq::T_munu_star}\\
	\accl{T}^{\alpha\beta} - \frac{1}{2} \accl{T}^{\mu\nu}\accl{g}_{\mu\nu} \accl{g}^{\alpha\beta} &= \frac{1}{\lambda}\left[\frac{\rho}{2}\hat h^{\alpha\beta}\right] + \mathcal{O}(1). \label{eq::T^munu_star}
\end{align}
This regularity of $\accl{T}^{\alpha\beta}$ for $\lambda \rightarrow 0$ is generally assumed in papers using the NR limit in order to obtain the Newton-Cartan equation from Einstein's equation \citep[e.g.][]{1976_Kunzle, 1990_Dautcourt_a}. The present section aimed at giving some physical insights supporting this hypothesis, without requiring \textit{a priori} the Newton-Cartan equation to hold.

\subsection{Limit of the Einstein equation}
\label{sec::limit_Einstein}

The Einstein equation features two constants: the cosmological constant $\Lambda$ and the gravitational constant $G$. The energy-momentum tensor $\accl{\T T}$ having the dimension of an energy, it must appear as ``$\kappa \lambda^2 \accl{\T T}\,$'' in the equation. However, from that equation alone the cosmological constant does not have, \textit{a priori}, a preferred dimension. For instance, if we consider $\Lambda$ to have the dimension of a curvature, then Einstein's equation will feature the term $\Lambda\accl{g}_{\alpha\beta}$. Instead, if we consider that $\Lambda$ has the dimension of a time$^{-2}$, there will be $\lambda\Lambda\accl{g}_{\alpha\beta}$.

The limit of the Einstein equation needs to be performed on the twice covariant version, i.e. as a (0,2)-tensor equation, featuring the Ricci tensor. Considering the (2,0) or the (1,1) versions, or the formulation featuring the Einstein tensor, give either no or redundant information at the limit. Therefore we have two possibilities:
\begin{align}
	\accl{R}_{\alpha\beta} = \kappa\lambda^2\left(\accl{T}_{\alpha\beta} - \frac{1}{2}\accl{T}\accl{g}_{\alpha\beta}\right) + \Lambda\accl{g}_{\alpha\beta}, \label{eq::Limit_E_eq_2}
\end{align}
or
\begin{align}
	\qquad\accl{R}_{\alpha\beta} = \kappa\lambda^2\left(\accl{T}_{\alpha\beta} - \frac{1}{2}\accl{T}\accl{g}_{\alpha\beta}\right) + \lambda\Lambda\accl{g}_{\alpha\beta}. \label{eq::Limit_E_eq_2_bis}
\end{align}
For the choice ``$\Lambda\accl{g}_{\alpha\beta}$'', the leading order of~\eqref{eq::Limit_E_eq_2} is $\frac{1}{\lambda} \Lambda \tau_\alpha\tau_\beta + \mathcal{O}(1) = 0$, implying $\Lambda = 0$. Therefore, only the choice ``$\lambda\Lambda\accl{g}_{\alpha\beta}$'' allows for a non-zero cosmological constant. The leading order of~\eqref{eq::Limit_E_eq_2_bis} is
\begin{align}
	{\hat R}_{\alpha\beta} = \left(\frac{\kappa}{2}\rho - \Lambda\right)\tau_\alpha\tau_\beta. \label{eq::Newton-Cartan}
\end{align}
This equation along with the geometrical constraint~\eqref{eq::Limit_Bianchi_1} and the conservation law ${\hat\nabla}_\mu\left(\rho\accentset{0}{u}^\mu\right) = 0$ corresponds to the Newton-Cartan system.

This system has been derived from the Einstein equation without any assumptions on the spatial topology or on the spatial Ricci curvature tensor related to $\hat h^{\mu\nu}$ and given by ${\hat\CR}^{\alpha\beta} = \hat h^{\mu\alpha}\hat h^{\nu\beta}\hat R_{\mu\nu}$. Still, equation~\eqref{eq::Newton-Cartan} necessarily implies ${\hat\CR}^{\alpha\beta} = 0$, which can be seen by projecting~\eqref{eq::Newton-Cartan} twice along $\hat{\T h}$\footnote{This result can also be derived directly from the 1+3-Einstein system of equations related to the fluid velocity $\overset{\lambda}{\T u}$ \citep[see chapter~4 in][]{2021_Vigneron_PhD}.}. As presented in Section~\ref{sec::local_global}, this result holds globally.

Therefore, this implies that on the $\tau$-foliation, for which we recall the hypersurfaces are $\Sigma$, it is possible to define a metric (here $\hat h^{ij}$) whose Ricci tensor is zero. From property~\eqref{eq::property} this means that the topology of each hypersurface member of this foliation is a Euclidean topology, i.e. that $\Sigma$ has an Euclidean topology. We recall that, as explained in section~\ref{sec::gen_the_limit}, the NR limit does not change the topology of $\CM$. Therefore, we have the following result:\saut

\begin{theorem}
A solution of the Einstein equation on a 4-manifold $\CM = \mathbb{R}\times\Sigma$ with $\Sigma$ closed, and which has a non-relativistic limit everywhere, as defined in Section~\ref{sec::the_limit}, requires the topology of the 3-manifold $\Sigma$ to be Euclidean. The limit is the cosmological Newtonian theory.
\end{theorem}\saut

%We recall that the reasons why we consider the limit to hold everywhere are presented Section~\ref{sec::local_global}. %Another justification for this hypothesis is to consider the Copernican principle.
The only assumption on the energy-momentum tensor for this result to hold is that it describes a matter fluid whose mass is conserved at the limit. In any case, the result holds in vacuum. The result is schematised in Figure~\ref{fig::sch2}.

The fact that the NR limit of the Einstein equation leads to the Newton-Cartan equation and imposes the Galilean structure to be spatially flat is not a new result of course. It was first derived in \citep{1976_Kunzle}, and later emphasised by \citep[][]{1986_Malament_a, 1986_Malament_b}. 
However, Ref. \citep{1976_Kunzle} imposes the limit of the energy-momentum tensor so that it leads to the Newton-Cartan equation, while we showed that this equation is always obtained (for a matter energy-momentum tensor or in vacuum), even with the presence of the cosmological constant. Refs. \citep[][]{1986_Malament_a, 1986_Malament_b} also only focus on discussing the spatial curvature,  while the (fundamental) link to topology and global properties is not made. The present paper aims at emphasising the importance of this link. In particular, the discussions made in Sections~\ref{sec::local_global} and~\ref{sec::Question} lead to the interpretation that this zero spatial curvature result should be considered more as an inconsistency within GR, than as a natural (logical) result that we should accept.

\subsection{Missing term in the Einstein equation}
\label{sec::Question}

The previous result leads to the third question raised in the introduction: \textbf{What relativistic equation should we consider in a 4-manifold with a non-Euclidean spatial topology if we want compatibility with the NR regime?} In other words, what term $\Top_{\alpha\beta}$ should we add in Einstein's equation, i.e. considering
\begin{equation}
	{G}_{\alpha\beta} = \kappa\lambda^2 {T}_{\alpha\beta} - \lambda \Lambda {g}_{\alpha\beta} + {{\Top}}_{\alpha\beta}, \label{eq::EE_Top}
\end{equation}
such that the NR limit exists in any topology (Figure~\ref{fig::sch3}). That question always stands on a purely mathematical point of view. In the next two sections, we motivate it with physical arguments, and propose a possible origin for the term $\Top_{\mu\nu}$. An answer to the question is given in Section~\ref{sec::Part_II}, and represented in Figure~\ref{fig::sch4}.

\subsubsection{Motivations for the modification}

When one constructs a new relativistic theory, or even historically when Einstein proposed his equation, it is generally required that this theory should be compatible with Newton's theory. The reason given is that Newton's law has been tested to be successful in (e.g.) the solar system. However, as shown in \citep{2023_Vigneron_et_al_a}, in a NR theory (Newton's theory or the NEN theory for spherical or hyperbolic topologies), the gravitational potential~$\Phi$ \textit{close} to a mass point is always given by Newton's law, i.e. $\Phi = -GM/r + \bigO{r}$, and correcting terms appear as function of the topology chosen ($\mathbb{R}^3$, $\mathbb{T}^3$, $\mathbb{S}^3$, ...), i.e. as function of the boundary conditions chosen. Those terms are always negligible on scales small compared to the finite size of the Universe (i.e. size of the boundary conditions), which are the scales on which the law has been tested. This means that from observations, we cannot distinguish between different non-relativistic theories. In other words, observations only tell us that gravitation is well described by a NR calculation on (e.g.) solar system scales, but the precise NR theory to consider depends on the type of topology of the Universe, which is not known. Therefore, requiring a relativistic theory to be only compatible with Newton's theory is a restriction with respect to what we learn from observations.

This suggests that the right constraint on any relativistic theory should be to require the compatibility with the NR regime in general (Newton's theory, spherical NEN theory, hyperbolic NEN theory, ...), the right NR theory depending on the topology of the Universe.
%, which is \textit{a priori} not known when constructing the relativistic theory, and can only be known \textit{a posteriori}, when comparing the relativistic theory with observations. 
If we consider this more general constraint, then, because Einstein's equation only has a NR limit in Euclidean topologies, then this equation needs to be modified to allow the limit to be performed in other topologies. This leads to the question raised in the beginning of Section~\ref{sec::Question}.

Another way of saying it is the following: in the case our Universe is, for instance, spherical, the result of Section~\ref{sec::limit_Einstein} shows that it is strictly mathematically incompatible to use both Einstein's equation and a non-relativistic calculation. Saying that we consider the calculation only locally does not save the game, since global conditions are always necessary (see Section~\ref{sec::local_global}). In other words, Einstein's equation cannot describe the NR regime of a universe with a non-Euclidean topology, and this at any scales. Of course this is not a problem if the topology of our Universe is Euclidean. But if it is not, which is not forbidden by observations and even suggested \citep[e.g.][]{2020_Di-Valentino_et_al, 2021_Handley}, then, because our Universe is filled with (mostly) NR matter (we could say the Universe is in a NR regime), we should modify the Einstein equation such that it has a NR limit regardless of the topology.

%You might or might not be motivated be the above arguments, still the question raised above always stands on a purely mathematical point of view.

\subsubsection{A possible origin: topological lagrangian multiplier}
\label{eq::Lag_mult}

As explained in Section~\ref{sec::gen_the_limit}, the NR limit does not change the topology of the 4-manifold $\CM$. If we consider Einstein's equation as resulting from a variational principle, the same arises. The action $S$ is defined on a 4-manifold $\CM$ and among all the equations possible on $\CM$ for the physical variable (i.e. $\T g$), we consider that the physical one is the one which extremalises the action, and is obtained with $\delta S = 0$. Therefore the variational principle is performed on the manifold $\CM$ which is \textit{a priori} defined, i.e. the variation of $S$ does not change $\CM$. This means that, when varying the action, the topology should already be fixed, implying that we are forbidden to consider paths for $\T g$ (when varying the action) for which the Riemann tensor is incompatible with the topology.

In classical mechanics, when constraints are applied on a variational principle, an additional term with a Lagrange multiplier must be added in the action. The same might be needed for the topology in general relativity where we would have an action of the form:
\begin{align}
	S = \int \sqrt{g} \left(R_{\mu\nu} g^{\mu\nu} + \T{\mathcal{Z}} \cdot \T C + \lambda\kappa\Lie{\rm matter}\right),
\end{align}
where $\T C = \T 0$ would be a tensorial equation constraining the topology of $\CM$ and especially the covering space $\tilde\Sigma$ of $\Sigma$ in the case $\CM = \mathbb{R}\times\Sigma$, and $\T{\mathcal{Z}}$ would be the tensorial Lagrange multiplier associated to this contraint. The idea of the present approach is to say that the additional term $\Top_{\mu\nu}$ in equation~\eqref{eq::EE_Top} would result from the variation of $\sqrt{g} \, \T{\mathcal{Z}} \cdot \T C$, i.e. from the fixation of the topology in the variational principle.

While this idea might be interesting as it would give a justification for equation~\eqref{eq::EE_Top}  outside of considering the NR regime and limit, it is not sure whether or not it can be successfully followed. The reason is that, to our knowledge, there does not exist a local tensorial equation constraining topology in three our four dimensions. Therefore, the solution we will propose in Section~\ref{sec::Part_II} to the question raised in Section~\ref{sec::Question} will not follow the Lagrange multiplier approach.

\section{Topological term in the Einstein equation with a bi-connection theory}
\label{sec::Part_II}

\subsection{Constraints on the missing term}
\label{sec::constraints_Top}

If we require the NR limit to exist in non-Euclidean topologies we need to consider a modified Einstein equation~\eqref{eq::EE_Top} where the additional term $\T{\Top}$ allows for spatial curvature. 
From the limit alone, few constraints can be drawn for this term. For instance, considering the Taylor series~\eqref{eq::T_munu}--\eqref{eq::T^munu_star} for the energy-momentum tensor, along with the relations
\begin{align}
	\accl{{G}}_{\alpha\beta} &= \frac{1}{\lambda}\left[\frac{\hat\CR}{2}\tau_\alpha\tau_\beta\right] + \mathcal{O}(1) \quad ; \quad \accl{{G}}^{\alpha\beta} = \hat\CR^{\alpha\beta} - \frac{\hat\CR}{2}\hat h^{\alpha\beta} + \mathcal{O}(\lambda), \label{eq::T_cond}
\end{align}
this implies $\accentset{(-1)}{{\cal T}}_{\alpha\beta} \propto \tau_\alpha \tau_\beta$ and $\accentset{(0)}{{\cal T}}^{\alpha\beta}$ to be spatial and non-zero.
%Because, has the following properties in the NR limit:
%\begin{align}
%	\accl{\mathcal{T}}_{\alpha\beta} &= \frac{1}{\lambda}\left[\frac{\hat\CR}{2}\tau_\alpha\tau_\beta\right] + \mathcal{O}(1), \label{eq::T_cond_1} \\
%	\accl{\mathcal{T}}^{\alpha\beta} &= \hat\CR^{\alpha\beta} - \frac{\hat\CR}{2}\hat h^{\alpha\beta} + \mathcal{O}(\lambda). \label{eq::T_cond_2}
%\end{align}
%These two conditions allow for $\hat\CR^{\alpha\beta} \not= 0$ in the limit, which is required for non-Euclidean topologies [Property~\eqref{eq::property}]. 
Apart for these few conditions, no other constraint is imposed on $\T{\mathcal{T}}$ by the limit. Therefore, there is substantial freedom on the extrapolation at full order of this term. We found four natural conditions to consider that would constrain this freedom:
\begin{itemize}
	\item[1.] $\T{\mathcal{T}} = 0$ for a Euclidean topology.
	\item[2.] $\T{\mathcal{T}}$ is related to the class of topology of $\CM$, i.e. to the covering space of $\CM$.
	\item[3.] $\T{\mathcal{T}}$ depends on second (or less) derivatives of the spacetime metric.
	\item[4.] The NR limit of the modified Einstein equation should exist in any topology.
\end{itemize}

The reason for {Constraint 1.} is to retrieve the Einstein equation in the Euclidean case. With {Constraint 2.} we reject the possibility of having a coupling term with the fluid. The reason is that we want $\T\Top \not= 0$ even in vacuum. We also require $\T\Top$ to be related only to the class of topology (i.e. the covering space) and not the precise topology. The reason for this comes from the Euclidean case: $\T\Top = 0$ for any Euclidean topology, so we should expect the same behaviour in other classes, with $\T\Top$ being the same within a topological class. {Constraint 3.} is chosen such that we keep a second order theory. We see no reason why the order of a theory should depend on the topology chosen. A consequence is that $\T\Top$ must depend on an additional field other than the metric $\T g$ (due to Lovelock's theorem~\citep{1972_Lovelock}). We can require this field to be related to the class of topology. {Constraint~4.} is the reason for the modification.\saut

Because $\Top_{\mu\nu}$ is related to topology, we call it \textit{the topological term}.

\subsection{The bi-connection theory}
\label{sec::bi-metric}

We found one relativistic equation fulfilling the above natural constraints. It is given by a slightly different theory than the bi-connection theory proposed by Rosen \citep{1980_Rosen}. This might not be the only possibility and another set of constraints could also be proposed. Still we think that the solution we present in this paper is the simplest we can obtain.

\subsubsection{Equations of the theory}
\label{sec::def_bi-metric}

Rosen's bi-connection theory in its 1980 version \citep{1980_Rosen} is defined on a 4-manifold $\CM$ equipped with a physical Lorentzian metric $g_{\alpha\beta}$ (and its inverse $g^{\alpha\beta}$) with its Levi-Civita connection $\T\nabla$, and a non-dynamical reference connection $\T{\bar \nabla}$. This second connection is non-dynamical in the sense that it is the same for any physical metric and energy-momentum tensor. The action is
\begin{align}
	S = \int \sqrt{-g} \left[\left(R_{\mu\nu} - \bar{R}_{\mu\nu}\right) g^{\mu\nu} + \lambda\Lambda + \lambda^2\kappa\, \Lie{\mathrm{matter}}\right] \dd x^4, \label{eq::action1}
\end{align}
where $\sqrt{-g} = \sqrt{-\textrm{det}g_{\alpha\beta}}$, $\lambda = 1/c^2$, and $\bar{R}_{\mu\nu}$ is the Ricci curvature tensor associated with $\T{\bar \nabla}$. The matter Lagrangian $\Lie{\rm matter}$ is unchanged by the introduction of the reference connection.\saut

One of the reasons Rosen introduced this theory was to be able to write a variational principle depending only on first derivatives of the physical metric (something not possible covariantly with Einstein-Hilbert action): by removing boundary terms, we can show that the second order action~\eqref{eq::action1} is equivalent to the covariant first order action
\begin{align}
	S = \int \sqrt{-g} \left(2\, \CC^\mu_{\alpha[\nu}\CC^\nu_{\beta]\mu} g^{\alpha\beta} + \lambda\Lambda + \lambda^2\kappa\, \Lie{\mathrm{matter}}\right) \dd x^4, \label{eq::action2}
\end{align}
where $\CC^\mu_{\alpha\beta} \coloneqq \Gamma^\mu_{\alpha\beta}  - \bar\Gamma^\mu_{\alpha\beta}$, which is a tensor.  Another advantage of having a second reference connection is that the usual non-covariant notion of gravitational energy used in general relativity can be made covariant \citep{1985_Rosen_a}.\saut

When extremalizing the action, the variation of quantities depending only on the reference connection is zero because it is a non-dynamical field. In other words, for all the paths of $\T g$ considered for the variation of $S$, $\bar{\Gamma}^\mu_{\alpha\beta}$ is unchanged. \textit{For this reason, this bi-connection theory is not a bi-gravity theory}. Then, defining the energy momentum tensor as $\sqrt{-g} \, T_{\alpha\beta} \coloneqq -\delta\left(\sqrt{-g} \, \Lie{\rm matter}\right)/\delta g^{\alpha\beta}$, we obtain the following modified Einstein equation
\begin{equation}
	G_{\alpha\beta} = \lambda^2\kappa\, T_{\alpha\beta}- \lambda\Lambda g_{\alpha\beta} + \Top_{\alpha\beta}, \label{eq::ModE1}
\end{equation}
where
\begin{equation}
	\Top_{\alpha\beta} \coloneqq \bar{R}_{\alpha\beta} - \frac{\bar{R}_{\mu\nu}g^{\mu\nu}}{2} g_{\alpha\beta} \label{eq::Top_munu}
\end{equation}
plays the role of an effective (or reference) energy-momentum tensor coming from the presence of the reference spacetime Ricci tensor. The matter energy-momentum tensor $T_{\alpha\beta}$ does not depend on the reference connection, and is the usual one we consider in general relativity for perfect fluid, scalar fields, etc. Equation~\eqref{eq::ModE1} also takes the simplified form
\begin{align}
	R_{\alpha\beta} - \bar{R}_{\alpha\beta} = \lambda^2 \kappa \, \left(T_{\alpha\beta} - \frac{T}{2} g_{\alpha\beta} \right) + \lambda \Lambda g_{\alpha\beta}. \label{eq::ModE2}
\end{align}

From the action~\eqref{eq::action1}, we also obtain the conservation of the matter energy-momentum tensor $\nabla_\mu T^{\mu\alpha} = 0$. This is not an additional hypothesis, as it can be obtained by varying $\int\sqrt{g}\lambda^2\kappa\Lie{\rm matter}\dd^4x$ with respect to infinitesimal changes of coordinates. The same applies for $\int\sqrt{g}\bar R_{\mu\nu}g^{\mu\nu} \dd^4x$ which leads to the conservation of the effective momentum tensor $\nabla_\mu \mathcal{T}^{\mu\alpha} = 0$. This conservation law can be written as
\begin{align}
	g^{\mu\nu}\left(\nabla_\mu \bar R_{\nu\alpha} - \frac{1}{2} \nabla_\alpha \bar R_{\mu\nu}\right) = 0, \label{eq::metrics_cond1}
\end{align}
or in the form
%\begin{align}
%	g^{\mu\nu}\left(\bar\nabla_\mu \bar R_{\nu\alpha} - \frac{1}{2} \bar\nabla_\alpha \bar R_{\mu\nu}\right)  + \bar R_{\alpha\mu} \sqrt{\frac{\bar g}{g}} \bar\nabla_\nu\left(\sqrt{\frac{g}{\bar g}}g^{\mu\nu}\right)= 0. \label{eq::metrics_cond2}
%\end{align}
\begin{align}
	g^{\mu\nu}\left(\bar\nabla_\mu \bar R_{\nu\alpha} - \frac{1}{2} \bar\nabla_\alpha \bar R_{\mu\nu}\right)  - \bar R_{\alpha\mu}\CC^\mu_{\sigma\gamma}g^{\sigma\gamma}= 0. \label{eq::metrics_cond2}
\end{align}
The conservation of $\T{\mathcal{T}}$ is an additional condition with respect to classical general relativity, called \textit{the bi-connection condition}. In the case $\bar{ R}_{\alpha\beta} = 0$, it becomes trivial and we retrieve general relativity. The main interpretation for that additional equation is to constrain the diffeomorphism freedom present when defining $\bar\Gamma^\mu_{\alpha\beta}$ with respect to~$\Gamma^\mu_{\alpha\beta}$ (see Appendix~\ref{app::Gauge} for a detailed explanation).\saut

We see that the introduction of the reference spacetime curvature tensor in the action leads to an additional term $\T\Top$ in Einstein's equation which depends only on second order (or less) derivatives of the physical metric and on an additional reference field which is the reference Ricci tensor $\bar{R}_{\alpha\beta}$. Thus we fulfill Constraint 3. presented in Section~\ref{sec::constraints_Top}. There remains to make a choice for $\bar\Gamma^\mu_{\alpha\beta}$.\saut

%\begin{remark}{Another possibility that we followed for a while was to consider the following equation
%\begin{align}
%	R_{\alpha\beta} - \bar{R}^{\mu\nu}g_{\alpha\mu}g_{\beta\nu} = \lambda^2\kappa \left(T_{\alpha\beta} - \frac{T}{2} g_{\alpha\beta} \right) + \lambda\Lambda g_{\alpha\beta}. \nonumber \label{eq::ModE2bis}
%\end{align}
%instead of equation~\eqref{eq::ModE2}. However, the evolution of the spatial metric at the limit had contradictory relations: one $\propto a^{-2}$ and one $\propto a^2$, making the theory trivial.}\end{remark}\saut

\begin{remark}{The action~\eqref{eq::action1}, respectively action~\eqref{eq::action2}, is not the most general \textit{covariant} second order, respectively first order, action we can build from $\T g$ and $\CC^\mu_{\alpha\beta}$ (contrary to the Einstein-Hilbert action for $\T g$). This choice of action is driven by a physical requirement, namely we want to recover general relativity for $\bar R_{\alpha\beta} = 0$. Thus, for instance, we could not have added $\CC^\alpha_{\mu\nu}  g^{\mu\nu} \CC^\beta_{\gamma\sigma}  g^{\gamma\sigma} g_{\alpha\beta}$.}
\end{remark}

\begin{remark}{We describe this theory as a bi-connection theory, and not a bi-metric theory. Indeed, the presence of a second metric is actually not essential: both the action~\eqref{eq::action2} and the field equations~\eqref{eq::ModE2} and~\eqref{eq::metrics_cond1} depend only on $\bar\Gamma^\mu_{\alpha\beta}$ and not on a reference metric $\bar g_{\mu\nu}$ from which $\bar\Gamma^\mu_{\alpha\beta}$ would derive. In other words, in this theory, there are two ways of differentiating, but only one way of measuring distances.}% Nevertheless, for practical reasons, we will use a reference metric $\bar g_{\mu\nu}$ to describe the choice made for $\bar\Gamma^\mu_{\alpha\beta}$.}
\end{remark}

\subsubsection{Choice of reference connection}
\label{sec::choice_metric}

The choice of reference connection is the only difference between our bi-connection theory and the one of Rosen \citep{1980_Rosen}. As said in the introduction, we always consider the spacetime manifold to be $\CM = \mathbb{R}\times\Sigma$ with $\Sigma$ a closed 3-manifold. %The choice of reference connection such that we will fulfill all the constraint given in Section~\ref{sec::constraints_Top} is to choose $\Gamma^\mu_{\alpha\beta}$ as the Levi-Civita connection of a metric $\bar g_{\mu\nu}$ having the form
We choose $\bar\Gamma^\mu_{\alpha\beta}$ such that there exists a coordinate system $\{x^0, x^i\}$ adapted to a foliation of $\Sigma$ hypersurfaces where
\begin{equation}
	\bar R^\mu{}_{\alpha\nu\beta} = \delta^\mu_a \delta^i_\alpha \delta^b_\nu \delta^j_\beta \,  \bar \CR^a{}_{ibj}(x^k), \label{eq::Riembar_choice}
\end{equation}
implying
\begin{equation}
	\bar R_{\alpha\beta} =  \delta^i_\alpha\delta^j_\beta \, \bar \CR_{ij}(x^k), \label{eq::Riccbar_choice}
\end{equation}
where $\bar \CR^a{}_{ibj}$ and $\bar \CR_{ij}$ are independent of $x^0$. In the case $\Sigma$ is a prime manifold, i.e. belonging to one of the eight Thurston classes of topology, and not a connected sum of such manifolds, $\bar\CR^a{}_{ibj}$ and $\bar \CR_{ij}$ are chosen to be the Riemann and Ricci tensors related to the natural Riemannian metric $\bar{h}_{ij}$ on these topological classes. They are presented in Appendix~\ref{app:Thurston_metric} for each case. In particular, for Euclidean, spherical and hyperbolic topologies we have respectively $\bar \CR^{\mE^3}_{ij} = 0$, $\bar \CR^{\mS^3}_{ij} = 2\bar h_{ij}$ and $\bar \CR^{\mH^3}_{ij} = -2\bar h_{ij}$.\saut

Therefore, with this choice of reference connection we fulfill Constraints~1. and~2. of Section~\ref{sec::constraints_Top} as $\T\Top$ is now directly related to the chosen spacetime topology, and that in the case of a spatial Euclidean topology, we have $\bar R_{\mu\nu} = 0$ implying $\Top_{\mu\nu} = 0$. Equation~\eqref{eq::ModE2} can be interpreted as relating the deviation between the physical spacetime Ricci curvature $R_{\mu\nu}$ and the reference one $\bar R_{\mu\nu}$ to the energy content in $\CM$. %A consequence is that homogeneous vacuum solutions exist for any topology. This is not possible with the Einstein equation, a typical example being the homogeneous solution of a spherical topology, for which the Friedmann equations require the presence of matter.
In other words, contrary to the Einstein equation for which matter directly curves spacetime, with the bi-connection theory, matter only induces a deviation of the physical spacetime curvature from the reference ``topological'' curvature $\bar R_{\mu\nu}$.
This is made more explicit in the follow up paper of this study \citep{2023_Vigneron_et_al_b} in which we will show that expansion does not depend anymore on the curvature (i.e. $\Omega = 1, \ \forall \Omega_K$), i.e. does not depend anymore on the type of topology.\\

%There remains to show that equations~\eqref{eq::ModE2} and \eqref{eq::metrics_cond1} have a NR limit in non-Euclidean topologies.\saut

%\begin{adjustwidth}{1cm}{}
\begin{conjecture}
We conjecture that the above choice for the reference spacetime curvature corresponds to the most symmetric Riemann tensor that can be defined on a manifold with topology $\mathbb{R}\times\Sigma$. In other words, this is the Riemann tensor that admits the most independent symmetry vectors $K^\mu$ with property
\begin{align}
	\Lie{\T K}\bar R^\mu{}_{\alpha\nu\beta} = 0.
\end{align}
\end{conjecture}
\vspace{.2cm}

The vectors $K^\mu$ plays the same role as Killing vectors for metric symmetries. We could call them ``Riemann-Killing vectors''. If the above conjecture is true, then the choice~\eqref{eq::Riembar_choice} corresponds to the ``simplest'' curvature that can be defined given a specific spacetime topology. In this sense, as we do not yet have a fundamental justification for why the choice~\eqref{eq::Riembar_choice} should be made, it follows the strategy of minimal modification.\saut

\begin{remark}{
While not necessary to define the theory, it is possible to write $\bar R^\mu{}_{\alpha\nu\beta}$ as deriving from a reference metric of the form $\bar g_{\alpha\beta} = \pm\delta^0_\alpha\delta^0_\beta + \bar h_{ij}(x^k)\delta^i_\alpha\delta^j_\beta$. The signature of this metric can be either Lorentzian or Riemannian.
}
\end{remark}

\subsubsection{Reference vectors}
\label{sec::prop_gbar}

For any topology, the reference spacetime Ricci tensor is of rank 3 or less. So there exists a vector $\T G$ such that $G^\mu\bar R_{\mu\alpha} = 0$. However, the 1-form $G^\mu g_{\mu\alpha}$ does not necessarily define a foliation in $\CM$, i.e. it is not necessarily exact nor closed, meaning that $G^\mu$ can have vorticity induced by the physical metric. The vector $G^\mu$ is only uniquely defined (up to a factor) for spherical, hyperbolic, $\tilde{SL2(\mathbb{R})}$ and \textit{Nil} topologies. For the other possibilities of prime manifolds, because $\textrm{dim}[\ker(\bar \CR_{ij})] > 0$, then $\textrm{dim}[\ker(\bar R_{\alpha\beta})] > 1$, and  there are at least two independent %\edq{open vectors} 
vectors $\T G_{(1)}$ and $\T G_{(2)}$ such that $G_{(1)}^\mu\bar R_{\mu\alpha} =0$ and $G_{(2)}^\mu\bar R_{\mu\alpha} =0$. Furthermore, any vector $\T G \in \ker(\bar R_{\alpha\beta})$ is a symmetry vector for the reference Ricci tensor:
\begin{align}
	\Lie{\T G} \bar R_{\alpha\beta} = 0. \label{eq::Lie_Rbar}
	%\Lie{\T G} \bar h_{\alpha\beta} = 0. \label{eq::Lie_hbar}
\end{align}

\begin{remark}{There is also a reference exact 1-form $n_\mu$ arising from the definition of $R^\mu{}_{\alpha\nu\beta}$, with property $n_\mu R^\mu{}_{\alpha\nu\beta} = 0$. However, because the field equations~\eqref{eq::ModE2} and~\eqref{eq::metrics_cond1} only depend on the reference Ricci tensor, the physics is blind to the presence of this 1-form.}
\end{remark}

%The reference spacetime Riemann tensor ${{\bar R}^\sigma}_{\alpha\beta\gamma}$ corresponds to the spatial Riemann tensor ${{\bar \CR}^\sigma}_{\alpha\beta\gamma}$ induced by $\bar{\T g}$ on this foliation:
%\begin{equation}
%	{{\bar R}^\sigma}_{\alpha\beta\gamma} = {{\bar \CR}^\sigma}_{\alpha\beta\gamma}.
%\end{equation}
%This spatial reference curvature is the one related to the spatial metric $\bar{h}_{\alpha\beta} = \delta_\alpha^i \delta_\beta^j \bar h_{ij}$ [using the same coordinates as in formula~\eqref{eq::metric_choice}]. Hence in these coordinates we have $\bar R_{\mu\nu} = \delta^i_\mu \delta^j_\nu \bar \CR_{ij}$, where $\bar\CR_{ij}$ is given in Appendix~\ref{app:Thurston_metric} for each type of topology. This is the reason why $\bar R_{\mu\nu} = 0$ in a Euclidean topology, for which $\bar\CR_{ij} = 0$.\saut

\section{Non-relativistic limit of the bi-connection theory}
\label{sec::Galilean_limit}

In this section, we show that the NR limit of the bi-connection theory is well defined for any topology (Section~\ref{sec:Existence_limit}), hence fulfilling Constraint 4. We also show that the NR theory obtained in the limit is the one developed in \citep{2022_Vigneron_b} (Section~\ref{sec:NR_theory_limit}), i.e. the NEN theory, and presented in Appendix~\ref{sec::NEN}. Finally, we derive the dictionary between the gravitational potential in the NR limit and the metric solution of the bi-connection theory (Section~\ref{sec::consist_dico}).

\subsection{Equations at leading order}
\label{sec:Existence_limit}

We consider that the reference curvature is unchanged by the limiting procedure. As explained in Appendix~\ref{app:gauge_Rbar}, this is in general not true, but corresponds to a gauge choice, which does not change physics and the result at the limit. Then using the limit of the physical Riemann tensor and the fluid energy-momentum tensor given in sections~\ref{sec::limit_R_munu} and~\ref{sec::limit_T_munu}, the limit of the interchange symmetry of the physical Lorentzian structures, the modified Einstein equation~\eqref{eq::ModE2} and the twice contracted second Bianchi identity give, respectively,
\begin{align}
	&\hat h^{\mu\beta}{{\hat R}^\alpha}_{\gamma\mu\sigma} - \hat h^{\mu\alpha}{{\hat R}^\beta}_{\sigma\mu\gamma} = 0, \label{eq::Limit_Bianchi_1_bi-metric}\\
	&\hat{R}_{\alpha\beta} - \bar R_{\alpha\beta} = \left(\frac{\kappa}{2} \rho - \Lambda\right) \tau_\alpha \tau_\beta, \label{eq::Limit_Mod_E}\\
	&\hat h^{\mu\nu}\left(\hat{\nabla}_\mu \hat R_{\nu\alpha} - \frac{1}{2} \hat{\nabla}_\alpha \hat R_{\mu\nu}\right) = 0. \label{eq::Limit_Bianchi_2_bi-metric}
\end{align}
The conservation of $T_{\mu\nu}$ and of the topological term $\Top_{\mu\nu}$, i.e. the bi-connection condition~\eqref{eq::metrics_cond1}, give, respectively, at leading order
\begin{align}
	&\hat\nabla_\mu\left(\rho \, u^\mu u^\alpha + \accentset{(0)}{p} \,\hat h^{\mu\alpha} + \accentset{(0)}{\pi}^{\,\mu\alpha}\right)= 0, \label{eq::Limit_nabla_T}\\
	&\hat h^{\mu\nu}\left(\hat{\nabla}_\mu \bar R_{\nu\alpha} - \frac{1}{2} \hat{\nabla}_\alpha \bar R_{\mu\nu}\right) = 0. \label{eq::Limit_bimetric_cond}
\end{align}

The difference with the Newton-Cartan system is the presence of the reference spacetime curvature in equation~\eqref{eq::Limit_Mod_E} and the presence of the bi-connection condition~\eqref{eq::Limit_bimetric_cond}. From the former, we get $\hat R_{\mu\beta} \hat h^{\mu\alpha} = \bar R_{\mu\beta} \hat h^{\mu\alpha}$, which implies that the bi-connection condition~\eqref{eq::Limit_bimetric_cond} is equivalent to equation~\eqref{eq::Limit_Bianchi_2_bi-metric}, and, therefore, is redundant in the limit. There remains to show that these equations are well defined in non-Euclidean topologies (shown in Section~\ref{eq::R_Gal}). For this it is sufficient to show that the spatial curvature in the limit is non-zero and compatible with the chosen topology.\saut

As presented in Section~\ref{sec::Gal_struct}, the spatial curvature $\hat\CR^{\alpha\beta}$ associated to the spatial metric $\hat h^{\alpha\beta}$ of the Galilean structure obtained in the limit is given by $\hat\CR^{\alpha\beta} = \hat h^{\mu\alpha}\hat h^{\nu\beta}\hat R_{\mu\nu}$. In a coordinate system adapted to the $\tau$-foliation, its spatial components $\hat\CR^{ij}$ correspond to the Ricci tensor associated to the spatial components $\hat h^{ij}$ of the spatial metric. The same applies for the spatial components of $\hat\CR^{\mu\nu}\bb{U}_{\mu \alpha}\bb{U}_{\nu \beta}$ for any $\tau$-timelike vector $\T U$:
\begin{align}
	\hat\CR^{\mu\nu}\bb{U}_{\mu i}\bb{U}_{\nu j} = \hat\CR^{kl}\hat h_{ik}\hat h_{jl} \eqqcolon \hat\CR_{ij}, \label{eq::bite_1}
\end{align}
where $\hat h_{ij}$ is the inverse of $\hat h^{ij}$. On the other hand, in any coordinate system $\{\T\partial_t, \T\partial x^i\}$ where $\T\partial_t = \T G$ with $\T G \in \ker(\bar R_{\alpha\beta})$, we have $\bar R_{\alpha\beta} = \bar \CR_{ij}(x^k)\delta^i_\alpha\delta^j_\beta$, where we recall that $\bar \CR_{ij}$ is the Ricci tensor associated to the natural metric on the covering space $\tilde\Sigma$ (presented in Appendix~\ref{app:Thurston_metric} for prime manifolds).\saut

Now, we choose $\T G$ such that $G^\mu\tau_\mu = 1$ and consider a coordinate system $\{\T G, \T\partial x^i\}$ adapted to the $\tau$-foliation\footnote{There always exists a vector $\T G \in \ker(\bar R_{\alpha\beta})$ such that $G^\mu \tau_\mu = 1$. The reason is that $\T\tau$ defines a foliation of $\CM = \mathbb{R}\times\Sigma$ by $\Sigma$ manifolds and that $\bar R_{\alpha\beta}$ always has a zero eigenvalue in the `$\mathbb{R}$-direction'.}. Then, equation~\eqref{eq::Limit_Mod_E} implies
\begin{align}
	\hat\CR^{\mu\nu}\bb{G}_{\alpha\mu}\bb{G}_{\beta\nu} = \bar R_{\alpha\beta}, \label{eq::hemoroides}
\end{align}
which, with relations~\eqref{eq::bite_1} and \eqref{eq::Riccbar_choice}, leads to
\begin{equation}
	\hat\CR_{ij} = \bar \CR_{ij}. \label{eq::R_ij=barR_ij}
\end{equation}
This equation shows that the physical spatial Ricci curvature at the limit is not necessarily zero. It corresponds to a reference Ricci curvature tensor which is well defined in a non-Euclidean topology. In summary, equation~\eqref{eq::R_ij=barR_ij} shows that \textit{the NR limit of the bi-connection theory is well defined in non-Euclidean topologies.}

%and that equations~\eqref{eq::Limit_Mod_E} is equivalent, in the spherical and hyperbolic cases, to equation~\eqref{eq::NEN_eq}. For this an important property to show is the existence of a Galilean vector $\tilde{\T G}$ (defined in Appendix~\ref{app::def_NEN_4D}), i.e. of inertial frames. This is done in Section~\ref{sec::Gal_existence}.

\subsection{Theory obtained in the non-relativistic limit}
\label{sec:NR_theory_limit}

While we showed that the NR limit of the bi-connection theory is well defined for any topology, we did not show that it leads to the NR theory developed in \citep{2022_Vigneron_b}, i.e. the NEN theory as named in the introduction. Equation~\eqref{eq::Limit_Mod_E} is similar to the modified Newton-Cartan equation~\eqref{eq::NEN_eq}, but a key property remains to be shown: there must exist a vector $\tilde G^\mu \in \ker(\bar R_{\alpha\beta})$ which defines inertial frames, i.e. its expansion $\tensor[^{\tilde{G}}]{\Theta}{^\alpha^\beta}$ and vorticity $\tensor[^{\tilde{G}}]{\Omega}{^\alpha^\beta}$ tensors (defined in Appendix~\ref{app::exp_vort}) must have the form:
\begin{align}
	&\tensor[^{\tilde{G}}]{\Theta}{^\alpha^\beta} = H(t)\hat h^{\alpha\beta} + \Xi^{\alpha\beta}, \label{eq::def_G_theta} \\
	&\tensor[^{\tilde{G}}]{\Omega}{^\alpha^\beta} = 0, \label{eq::def_G_omega}
\end{align}
with $\hat D_k \Xi_{ij} = 0$ and $H(t) \coloneqq \dot V_\Sigma/(3V_\Sigma)$ with $V_\Sigma \coloneqq \int_\Sigma\sqrt{\hat h} \dd x^3$ the volume of $\Sigma$. The tensor $\Xi_{ij}$ represents anisotropic expansion, and $H(t)$ represents the global volume expansion.\saut

Requiring the existence of this vector is the main improvement made in \citep{2022_Vigneron_b} compared to \citep{1976_Kunzle} in the development of the NEN theory. As shown in \citep{2022_Vigneron_b}, this requirement is essential for the equations of the theory to be physically coherent.\saut

We show in the next subsections that the limit of the bi-connection theory indeed implies the existence of such a vector, called \textit{Galilean vector}. As a side result we also correct a mistake made in \citep{2022_Vigneron_b} when constructing the NEN theory: we will show that the spatial curvature in that theory is in general homogeneous but anisotropic (even in the spherical and hyperbolic cases), while it was chosen to be necessarily isotropic in~\citep{2022_Vigneron_b}.

\subsubsection{Homogeneous but anisotropic spatial curvature}
\label{eq::R_Gal}
%We show in this section that the reference spacetime Ricci tensor induces a non-zero spatial curvature at the limit directly related to the reference spatial curvature of the reference spatial metric $\bar h_{ij}$.\saut

The equality between the reference curvature and the spatial curvature at the limit [equation~\eqref{eq::R_ij=barR_ij}] does not imply that the spatial metric $\hat h_{ij}$ is equivalent to the Thurston metric $\bar h_{ij}$. In particular, in the spherical or hyperbolic cases, equation~\eqref{eq::R_ij=barR_ij} does not necessarily imply $\hat{\CR}_{ij}$ to be isotropic, i.e. $\hat{\CR}_{ij} - \frac{\hat{\CR}_{cd}\hat h^{cd}}{3} \hat h_{ij} \not=0$. As shown in Appendix~\ref{app:R_ij=barR_ij_sol}, a particular solution is
\begin{align}
	&\hat h_{ij} = A\bar h_{ij} + A_{ij}, \label{eq::mainDA_DH}
\end{align}
for $A_{ij}$ a tensor satisfying $\hat D_i A = 0 \ ; \ \hat D_k A_{ij} = 0 \ ; \ A_{ij}\bar h^{ij} = 0$. It then follows that $\hat D_i = \bar D_i$, where $\hat D_i$ and $\bar D_i$ are respectively the spatial connections related to $\hat h_{ij}$ and $\bar h_{ij}$. In the case of spherical or hyperbolic topologies, i.e. for which $\bar \CR_{ij} = \pm2\bar h_{ij}$, equation~\eqref{eq::mainDA_DH} leads to
\begin{align}
	&\hat\CR_{ij} = \pm2/A \left(\hat h_{ij} - A_{ij}\right), \label{eq::mainSol_R_ij}\\
	&\partial_k\left(\hat\CR_{ij} \hat h^{ij}\right) = 0.
\end{align}
So in general, the physical spatial curvature tensor in the NR limit of the bi-connection theory is \textit{homogeneous but anisotropic}. It is isotropic only if $A_{ij} = 0$. That anisotropy necessarily implies the presence of an anisotropic expansion (Appendix~\ref{app::exp_G}). That property corrects a mistake made in \citep{2022_Vigneron_b}, when constructing the NEN theory, where anisotropic expansion was allowed with isotropic curvature.\saut

We did not manage to prove that~\eqref{eq::mainDA_DH} is the general solution of~\eqref{eq::R_ij=barR_ij}. However, it is a natural choice to take, as it encompasses $\hat\CR_{ij} = \hat\CR\, \hat h_{ij}/3$ for the spherical and hyperbolic topologies. In that case, equation~\eqref{eq::Limit_Mod_E} corresponds to the NEN equation~\eqref{eq::NEN_eq}.\saut

Finally, from property~\eqref{eq::Lie_Rbar} and equation~\eqref{eq::hemoroides}, we have
\begin{equation}
	\Lie{\T G} \left(\hat\CR^{\mu\nu}\bb{G}_{\alpha\mu}\bb{G}_{\beta\nu}\right) = 0, \label{eq::evolv_CR}
\end{equation}
from which equation~\eqref{eq::Limit_Bianchi_2_bi-metric} is redundant. Then, from~\eqref{eq::evolv_CR}, in a coordinate system adapted to the $\tau$-foliation and such that $\T\partial_t = \T G$, we have
\begin{equation}
	\partial_{t|_G} \hat\CR_{ij} = 0. \label{eq::evolv_CR_ij}
\end{equation}
In the isotropic case $\hat\CR_{ij} = \hat\CR(t)\, \hat h_{ij}/3$ (only possible for spherical or hyperbolic topologies), this evolution equation for the Ricci tensor implies $\hat\CR(t) \propto 1/a^2(t)$.

\subsubsection{Existence of a Galilean observer}
\label{sec::Gal_existence}

As explained in Appendix~\ref{sec::NEN}, the presence of vector with properties~\eqref{eq::def_G_theta} and~\eqref{eq::def_G_omega} implies (for isotropic expansion, i.e. $\Xi_{ij} = 0$) the existence of a coordinate system in which $\hat h_{ij}(t,x^k)$ can be separated in space and time dependence \citep{2022_Vigneron_b}: $\hat h_{ij}(t,x^k) = a(t)^2\tilde h_{ij}(x^k)$. In other words, this implies the existence of inertial frames as classically defined in Newton's theory. However, the existence of a Galilean vector is not implied by the definition of Galilean structures alone, but must come from a physical equation. Einstein's equation implies this existence \citep{2021_Vigneron}, there remains to show that this is also the case from the bi-connection equations.\saut

This is shown in Appendix~\ref{app::Gal_met}: any $\tau$-timelike vector $\T G \in \ker(\bar R_{\mu\nu})$ is a Galilean vector: $\T G = \tilde{\T G}$. That vector is only unique if $\textrm{dim}[\ker(\bar \CR_{ij})] = 0$ (equivalently if ${\textrm{dim}[\ker(\bar R_{\mu\nu})] = 1}$). Otherwise, $\tilde{\T G}$ is determined up to a spatial vector $\T w$, i.e. $w^\mu \tau_\mu = 0$, such that $w^i \hat \CR_{ij} = 0$ and $\hat D_k w^i = 0$. In other words, in the direction of zero spatial curvature, inertial frames are defined up to a global translation, as is the case in Newton's theory.\saut

This concludes the proof that the NR limit of the bi-connection theory is well defined for any topology, and that the resulting NR equations correspond (in the spherical or hyperbolic cases) to a corrected version of the NEN theory presented in Appendix~\ref{sec::NEN}. It is corrected in the sense that anisotropic expansion necessarily requires anisotropic spatial curvature, something not originally present in the NEN theory developed in \citep{2022_Vigneron_b}.

%From the solution~\eqref{eq::mainDA_DH}, we show in Appendix~\ref{app::exp_G} that in coordinates adapted to the $\tau$-foliation and such that $\T\partial_t = \T G$, then $\partial_t \hat h_{ij}$ has the required form~\eqref{eq::h_ij_Gal} of the NEN theory. The only difference is that the tensor $\Xi_{ij}$ is not only divergence free, but also gradient free. In Appendix~\ref{app::vort_G}, we show that the vorticity $\TT{\T\Omega}{G}$ of $\T G$ relative to the Galilean structure, defined as $\TT{\Omega}{G}^{\alpha\beta} \coloneqq \hat h^{\mu[\alpha}\hat\nabla_\mu G^{\beta]}$, is zero. Therefore condition~\eqref{eq::Omega_Gal} is fulfilled and $\T G$ corresponds to a Galilean vector, defining an inertial frame in the NEN theory.\saut

%However, as explained in section~\ref{sec::prop_gbar}, this vector is not uniquely defined from $\bar R_{\alpha\beta}$ if $\textrm{dim}[\ker(\bar R_{\alpha\beta})] > 1$, i.e. corresponding to $\mathbb{R}\times\mS^2$, $\mathbb{R}\times\mH^2$ and \textit{Sol} topologies if $\Sigma$ is a prime manifold. In these cases, $\T G$ is defined up to a vector $\T v$, spatial with respect to $\T \tau$, i.e. $v^\mu\tau_\mu = 0$, and with $v^c\hat\CR_{ci} = 0$ and $\hat D_i v^j = 0$ (Appendix~\ref{app::Unique_G}). In classical Newton's theory this is a known property: the theory is invariant under global translations, which corresponds to spatial vectors with zero gradient. We showed that this still holds in non-Euclidean topologies in directions where the Ricci curvature tensor is zero.\saut

\subsection{Dictionary}
\label{sec::consist_dico}

As a complementary result, in this section, we give the dictionary between the (non-Euclidean)-Newtonian theory and the spacetime metric $\accl{g}_{\mu\nu}$ given by the NR limit of either Einstein's equation or the bi-connection theory. The derivation is shown in Appendices~\ref{app::dico} and~\ref{app::Gauge}. We only consider Euclidean, spherical or hyperbolic topologies with isotropic expansion (i.e. $\hat \CR_{ij} = 2K/a^2 \, \hat h_{ij}$ with $K=0$ or $\pm 1$, and $\Xi_{ij} = 0$). We obtain that, as function of the gravitational potential $\Phi$ (solution of the Poisson equation~\ref{eq::Poisson}), the spacetime metric components take the form
\begin{align}
	\accentset{\lambda}{g}^{\alpha\beta} = 
	\left(\begin{array}{cc}
		-\lambda + 2\lambda^2 \left(\Phi + \frac{1}{2}\hat D^k\sigma \hat D_k \sigma  + \partial_{t|_G}\sigma\right) + \bigO{\lambda^3} & -\lambda \hat D^i \sigma + \bigO{\lambda^2}\vspace{.2cm} \\
		-\lambda \hat D^i \sigma + \bigO{\lambda^2} & \begin{array}{c}
												\hat h^{ij} + \lambda \left[2(\Phi - H\sigma)\hat h^{ij} \right.\\
												\left.- 2\hat D^i\hat D^j E - 2\hat D^{(i}F^{j)}\right] + \bigO{\lambda^2}
											\end{array}
	\end{array}\right)\label{eq::dico_g^munu}
\end{align}
and 
\begin{align}
	\accentset{\lambda}{g}_{\alpha\beta} = 
	\left(\begin{array}{cc}
		-\frac{1}{\lambda} - 2 \left(\Phi  + \partial_{t|_G}\sigma\right) + \bigO{\lambda} & - \hat D_i \sigma + \bigO{\lambda} \vspace{.2cm} \\
		- \hat D_i \sigma + \bigO{\lambda} &  \begin{array}{c}
											\hat h_{ij} - \lambda \left[2(\Phi - H\sigma)\hat h_{ij} \right.\\
											\left.\quad - 2\hat D_i\hat D_j E - 2\hat D_{(i}F_{j)} + \hat D_i\sigma \hat D_j \sigma\right] + \bigO{\lambda^2}
										\end{array}
	\end{array}\right), \label{eq::dico_g_munu}
\end{align}
where by definition $\hat D_k F^k = 0$. The variables $E$, $F^i$ and $\sigma$ are gauge fields and can be set to zero without change in the physics, while we can show that $\Phi$ is gauge invariant (Appendix~\ref{app::Gauge}) as expected for the gravitational potential. This dictionary is valid in a coordinate system $\{\T\partial_t, \T\partial x^i\}$ adapted to the $\tau$-foliation and where $\T\partial_t = \tilde{\T G}$. In other words, in an inertial coordinate system (where $\hat h_{ij} = a(t)^2 \bar h_{ij}(x^k)$).\saut

This is exactly the same dictionary between Newton's theory and the spacetime metric given by Einstein's equation. In particular, in the gauge choice $\sigma = 0$, we retrieve the classical formula ``$g_{00} = -\frac{1}{\lambda} - 2\Phi$''. This holds for the gravitational potential $\Phi$ calculated in either a Euclidean topology (with Newton's theory), or in a spherical/hyperbolic topology (with the NEN theory). Furthermore, this also shows that the gravitational slip (i.e. the difference between the potential in $g_{00}$ and in the factor in front of $\hat h_{ij}$) is zero (see Appendix~\ref{app:gauge_NR} for more details). This is an important result as this parameter is usually used as a discriminator between different gravity theories. Therefore, once again, this shows that Newton's theory and the NEN theory are algebraically equivalent, as the equations defining the theories (see the spatial NEN equations in Appendix~\ref{sec::spatial_eq_NEN}) and their dictionaries are equivalent, with the only (implicit) difference being the curvature of the spatial metric and the topology on which they are defined.\saut

\begin{remark}{This dictionary is only a non-relativistic dictionary, and not a `post-non-relativistic' dictionary (we could also use the term `post-(non-Euclidean)-Newtonian'). Post-non-relativistic terms are present at higher orders, in the space components of $\accentset{(2)}{g}^{\,\mu\nu}$ and the time component of $\accentset{(1)}{g}_{\mu\nu}$, which are not considered here.}
\end{remark}
%Then, it seems that we can never choose a gauge such that we are in the Gulstrand-Painlevé coordinates (because in these coordinates, the spatial metric is exactly flat, forbidding \textit{a priori} any first order corrections). The reason might be that, from the standard Schwarzschild coordinates, the coordinate change is $t \rightarrow t + a(r)$, with $a(r) \overset{r \rightarrow \infty}{\longrightarrow} \infty$ (see \href{https://en.wikipedia.org/wiki/Gullstrand–Painlevé_coordinates}{wiki}), and so is not a L2 function, as required by the SVT decomposition, and so as required in the theory.

\newpage
\section{Summary of results and conclusion}
\label{sec::ccl_limit_NEN}

The results and logic followed during this paper are summarised in the evolution of the scheme presented in Figure~\ref{fig::sch0},~\ref{fig::sch1},~\ref{fig::sch2},~\ref{fig::sch3}, and~\ref{fig::sch4}.

The first aim of the paper was to study the existence of a NR limit of the Einstein equation as function of the topology (Question (2) of introduction). We showed that such a limit only exists if the spatial topology is Euclidean, in the sense of the Thurston decomposition, i.e. its universal cover is $\mE^3$. We argued that this result can be interpreted as a signature of an inconsistency of general relativity for non-Euclidean topologies (Section~\ref{sec::Question}), in other words Einstein's equation would be physically wrong if the spatial topology of the Universe would be non-Euclidean.

The second goal was to search for a relativistic equation for which the limit exists in any topology (Question (3) of introduction). We proposed four constraints that such an equation should fulfill (Section~\ref{sec::constraints_Top}) and gave a solution. While this might not be the only answer to question (3) fulfilling the constraints given, we think it is the simplest solution, essentially because no new dynamical field in added to general relativity. The theory is a bi-connection theory defined by the equations
\begin{align}
		&R_{\alpha\beta} - \bar R_{\alpha\beta} = \lambda^2 \kappa \, \left(T_{\alpha\beta} - \frac{T}{2} g_{\alpha\beta} \right) + \lambda \Lambda g_{\alpha\beta}, \label{eq::bi-metric_1} \\
	&g^{\mu\nu}\left(\nabla_\mu \bar R_{\nu\alpha} - \frac{1}{2} \nabla_\alpha \bar R_{\mu\nu}\right) = 0. \label{eq::bi-metric_2}
\end{align}
The Ricci curvature $\bar R_{\alpha\beta}$ is related to a reference non-dynamical connection $\bar\Gamma^\mu_{\alpha\beta}$ chosen as a function of the spacetime topology (which is assumed to be of the form $\CM = \mathbb{R}\times\Sigma$ with $\Sigma$ closed): its Riemann tensor takes the form
\begin{align}
	\bar R^\mu{}_{\alpha\nu\beta} =  \bar \CR^a{}_{ibj}(x^k)\delta^\mu_a \delta^i_\alpha \delta^b_\nu \delta^j_\beta, \label{eq::bi-metric_3}
\end{align}
in a coordinate system adapted to a foliation of $\Sigma$-hypersurfaces, where $\bar \CR^a{}_{ibj}$ is the Riemannian curvature relative to the covering space $\tilde\Sigma$ of $\Sigma$. In this sense, $\bar R_{\alpha\beta}$ is a \textit{topological term} added to the Einstein equation.

%We showed that the NR limit of these equations exists in any topology and leads (in the spherical or hyperbolic cases) to the expect equations: the non-Euclidean Newtonian theory constructed in \citep{2022_Vigneron_b} (see Appendix~\ref{sec::NEN}) with the only difference that the presence of anisotropic expansion implies the spatial curvature to be anisotropic, something not initially present in \citep{2022_Vigneron_b}. As a complementary result, we also derived the dictionary between the non-Euclidean Newtonian theory and the spacetime metric of the bi-metric theory (equation~\eqref{eq::dico_g^munu} and~\eqref{eq::dico_g_munu}). Written as function of the gravitational field $\Phi$, this dictionary is equivalent to the one obtained with Newton's theory and Einstein's equation. In particular, within a certain gauge choice, we have $g_{00} = -c^2 - 2\Phi$, and the gravitational slip is zero. This strengthens the fact that Newton's theory and the NEN theory are essentially the same theory of gravitation, but defined on a different topology.\saut

In conclusion, the above bi-connection theory is compatible with the non-relativistic regime in any topology. Without other candidates and because of its simplicity, this relativistic theory should be used if one wants to study a model universe with a non-Euclidean spatial topology and requiring the NR limit to exist. As for now, the only reason for considering this theory is related to the hypothesis of full compatibility between the relativistic and the NR regimes. It would be interesting to have a more fundamental origin for the presence of the reference spacetime curvature. We expect such a justification to come either from a purely mathematical approach as proposed in Section~\ref{eq::Lag_mult} with a ``topological Lagrange multiplier''; or from physics with quantum gravity theories.

In the follow up paper of this study \citep{2023_Vigneron_et_al_b}, we analyse the consequences for cosmology of the bi-connection theory, in other words the consequences for cosmology of requiring full compatibility between the relativistic and the non-relativistic regimes. In particular, we show a major result: the expansion becomes blind to the spatial curvature (i.e.~{${\Omega = 1, \ \forall \Omega_K}$}), hence giving an elegant solution to the flatness problem.

\section*{Acknowledgements}
This work was supported by the Centre of Excellence in Astrophysics and Astrochemistry of Nicolaus Copernicus University in Toru\'n, and by the Polish National Science Centre under Grant No. SONATINA 2022/44/C/ST9/00078. I am grateful to Pierre Mourier and \'Aron Szab\'o for constant discussions and commentaries on the manuscript. I thank L\'eo Brunswic and Fernando Pizaña for insightful discussions.

\begin{appendices}

\section{The NEN theory}
\label{sec::NEN}

In this section, we present the properties of the non-relativistic theory in non-Euclidean topologies, i.e. the NEN theory, developed in \citep{1976_Kunzle} and \citep[][section~5.6]{2022_Vigneron_b}. %As shown in Section~\ref{eq::R_Gal}, there is a little mistake

\subsection{Spacetime equations}
\label{app::def_NEN_4D}

The theory is defined on a 4-manifold $\CM = \mathbb{R}\times\Sigma$ where $\Sigma$ is closed. $\CM$ is equipped with a Galilean structure $(\hat h^{\alpha\beta}, \tau_\mu, \hat\nabla_\nu)$ as defined in Section~\ref{sec::Gal_struct}. The spacetime equations of the NEN theory are
\begin{align}
	&\hat h^{\mu\beta}{{\hat R}^\alpha}_{\gamma\mu\sigma} - \hat h^{\mu\alpha}{{\hat R}^\beta}_{\sigma\mu\gamma} = 0, \label{eq::_Bianchi_1}\\
	&\hat R_{\alpha\beta} - \frac{\hat\CR}{3}\bb{{\tilde G}}_{\alpha\beta} = \left(\frac{\kappa}{2} \rho - \Lambda\right)\tau_\alpha\tau_\beta, \label{eq::NEN_eq}\\
	&\hat\nabla_\mu T^{\mu\alpha} = 0, \label{eq::NEN_eq3}
\end{align}
where $\rho$ is the matter density; $T^{\alpha\beta}$ is the energy-momentum tensor; $\Lambda$ is the cosmological constant; $\bb{{\tilde G}}_{\alpha\beta}$ is the orthogonal projector (defined by relation~\eqref{eq::NC_def_bb}) to a vector $\tilde G^\alpha$ with ${\tilde G^\mu\tau_\mu \coloneqq 1}$ called \textit{the Galilean vector} (or Galilean observer) in $\CM$. It is defined by two conditions: (i) in coordinates adapted to the $\tau$-foliation and such that $\T\partial_t = \tilde{\T G}$, the time derivative of the spatial components of the space metric (which correspond to the expansion tensor of $\tilde{\T G}$) are
\begin{equation}
	\frac{1}{2}\partial_t \hat h^{ij} = -H(t)\hat h^{ij} - \Xi^{ij}, \label{eq::h_ij_Gal}
\end{equation}
and (ii) $\tilde{\T G}$ is vorticity free with respect to the Galilean structure, i.e.
\begin{equation}
	\hat h^{\mu[\alpha}\hat\nabla_\mu \tilde G^{\beta]} = 0. \label{eq::Omega_Gal}
\end{equation}
$H(t)$ is the expansion rate of the finite volume of the spatial sections defined by $\T \tau$, and $\Xi^{ij}$ is traceless-transverse and represents anisotropic expansion. Property~\eqref{eq::h_ij_Gal} ensures that, without anisotropic expansion, there exists an adapted coordinate system (the one related to~$\tilde{{\T G}}$) in the NEN theory such that the spatial metric can be separated in space and time dependence: $h_{ij}(t,x^k) = a(t)^2 \tilde h_{ij}(x^k)$. This coordinate system defines what we call an \textit{inertial frame} in classical Newton's theory. Therefore, property~\eqref{eq::h_ij_Gal} ensures its existence for the non-Euclidean version of the theory\footnote{In \citet{2022_Vigneron_b}, we give a fully covariant definition of the Galilean vector, which is equivalent to property~\eqref{eq::h_ij_Gal}.}.\saut

Equation~\eqref{eq::_Bianchi_1} is the Trautmann condition, also present in Newton's theory with the Newton-Cartan system. It ensures that no gravitomagnetism is present in the theory. In equation~\eqref{eq::NEN_eq}, $\frac{\hat\CR}{3}\bb{{\tilde G}}_{\alpha\beta}$ is an additional term with respect to the Newton-Cartan system. This term was added by \citep{1976_Kunzle, 2022_Vigneron_b} for the theory to be defined on spherical or hyperbolic topologies, as it implies the spatial Ricci tensor (from equation~\eqref{eq::NEN_eq}) to have the form $\hat\CR^{ij} = \pm 2/a(t)^2\,  \hat h^{ij}$, where $a$ is the scale factor of expansion.

\subsection{Spatial equations}
\label{sec::spatial_eq_NEN}

To allow for a more intuitive understanding of this theory, we present the spatial equations resulting from the above the spacetime equations. Assuming isotropic expansion ($\Xi_{ij} = 0$), and introducing the gravitational potential $\Phi$ defined via the 4-acceleration of the Galilean vector (i.e. $\hat D^\mu \Phi \coloneqq \tilde G^\nu\hat \nabla_\nu \tilde G^\mu$), we have (see \citep{2022_Vigneron_b} for a detailed derivation)
\begin{align}
	\dot v^a &= -\hat D^a \Phi - 2v^a H + (a_{\not= \rm grav})^a, \label{eq::NEN_eq1}\\
	\dot \rho/\rho &= -3H - \hat D_c v^c,\\
	\hat \Delta \Phi &= \frac{\kappa}{2} \left(\rho - \Saverage{\rho}\right), \label{eq::Poisson}\\
	\hat \CR_{ab} &= \frac{\pm 2}{a(t)^2}\hat h_{ab},
\end{align}
with the expansion law
\begin{align}
	3\ddot a/a &= -\frac{\kappa}{2} \Saverage{\rho} + \Lambda, \label{eq::NEN_eqn}
	%3H^2 &= \kappa\Saverage{\rho} + \Lambda, \label{eq::NEN_eqn}
\end{align}
where the dot derivative corresponds to $\partial_t + v^c\hat D_c$ with $\hat D_i$ the covariant derivative with respect to the spatial metric; $\T v$ is the spatial fluid velocity; $(a_{\not= \rm grav})^a$ is the non-gravitational spatial acceleration of the fluid; $\average{\psi}{\PerD}(t) \coloneqq \frac{1}{V_\PerD}\int_\PerD \psi \sqrt{\mathrm{det}(\hat h_{ij})}\dd^3 x$ is the average of a scalar field $\psi$ over the volume $V_\Sigma$ of $\Sigma$. This system is valid in an inertial frame, i.e. in coordinates where the spatial metric is separated is space and time dependence.\saut

We see that the spatial equations defining the NEN theory are algebraically equivalent to the cosmological Newton equations, but with the presence of a non-zero spatial Ricci tensor related to a spherical or hyperbolic topology. This shows that the NEN theory can be seen as an adaptation of Newton's theory (by adding a curvature) such that it is defined on non-Euclidean topologies.

\newpage
\section{Natural metrics on Thurston's topological class}
\label{app:Thurston_metric}

We give the natural metrics for each Thurston's topological class and some properties of their Ricci tensor (we denote $N_{\ker{}} \coloneqq \dim[\ker(\bar\CR_{ij})]$):
\begin{itemize}
	\item For Euclidean topologies (the covering space $\tilde\Sigma$ of $\Sigma$ is $\mE^3$):
	\begin{equation}
		\bar h_{ij} = \delta_{ij},
	\end{equation}
	with $\bar \CR_{ij} = 0$ and $N_{\ker{}} = 3$.
	\item For $\tilde\Sigma = \mS^3$ (spherical topologies):
	\begin{equation}
		\bar h_{ij}\dd x^i\dd x^j =\dd r^2 + \sin^2r \left(\dd\theta^2 + \sin^2\theta\dd\varphi^2\right),
	\end{equation}
		with $\bar\CR_{ij} = 2\bar h_{ij}$ and $N_{\ker{}} = 0$.
	\item For $\tilde\Sigma = \mH^3$ (hyperbolic topologies):
	\begin{equation}
		\bar h_{ij}\dd x^i\dd x^j =\dd r^2 + \sinh^2r \left(\dd\theta^2 + \sin^2\theta\dd\varphi^2\right),
	\end{equation}
		with $\bar\CR_{ij} = - 2\bar h_{ij}$ and $N_{\ker{}} = 0$.
	\item For $\tilde\Sigma = \mathbb{R}\times\mS^2$:
		\begin{equation}
			\bar h_{ij}\dd x^i\dd x^j = \dd r^2 + \sin^2 r \, \dd\varphi^2 + \dd z^2,
		\end{equation}
		with $\bar\CR = 2$, $\bar\CR_{\langle ij\rangle} \not=0$ and $N_{\ker{}} = 1$.
	\item For $\tilde\Sigma = \mathbb{R}\times\mH^2$:
		\begin{equation}
			\bar h_{ij}\dd x^i\dd x^j = \dd r^2 + \sinh^2 r \, \dd\varphi^2 + \dd z^2,
		\end{equation}
		with $\bar\CR = -2$, $\bar\CR_{\langle ij\rangle} \not=0$ and $N_{\ker{}} = 1$.
	\item For $\tilde\Sigma = \tilde{SL2\mathbb{R}}$:
		\begin{equation}
			\bar h_{ij}\dd x^i\dd x^j = \dd x^2 + \cosh^2 x \, \dd y^2 + \left(\dd z + \sinh^2x \, \dd y\right)^2,
		\end{equation}
		with $\bar\CR = -5/2$, $\bar\CR_{\langle ij\rangle} \not=0$ and $N_{\ker{}} = 0$.
	\item For \textit{Nil} topologies:
		\begin{equation}
			\bar h_{ij}\dd x^i\dd x^j = \dd x^2 + \dd y^2 + \left(\dd z - x \dd y\right)^2,
		\end{equation}
		with $\bar\CR = -1/2$, $\bar\CR_{\langle ij\rangle} \not=0$ and $N_{\ker{}} = 0$.
	\item For \textit{Sol} topologies:
		\begin{equation}
			\bar h_{ij}\dd x^i\dd x^j = e^{-2z}\dd x^2 + e^{2z}\dd y^2 + \dd z^2,
		\end{equation}
		with $\bar\CR = -2$, $\bar\CR_{\langle ij\rangle} \not=0$ and $N_{\ker{}} = 2$.
\end{itemize}

\section{Formulas for the NR limit}
\label{app::formulas}

In this section, the notion of spatial is related to the foliation defined by $\T \tau$ and to the metric $\hat{\T h}$ present in the Galilean structure.

\subsection{Leading order of the fluid orthogonal projector}
\label{app::Limit_Formulas}

The projector orthogonal to a g-timelike vector $\accentset{\lambda}{\T u}$ is defined as
\begin{equation}
	\tensor[^{u}]{\accentset{\lambda}{b}}{_\alpha_\beta} \coloneqq \accentset{\lambda}{g}_{\alpha\beta} + \lambda\accentset{\lambda}{u}_\alpha\accentset{\lambda}{u}_\beta.
\end{equation}
Its limit is
\begin{align}
	\tensor[^{u}]{\accentset{\lambda}{b}}{_\alpha_\beta}	&= \bb{B}_{\alpha\beta} + \tau_{\alpha}\tau_\beta \accentset{(0)}{u}^{\,\mu}\,\accentset{(0)}{u}^{\,\nu}\,\bb{B}_{\mu\nu} - 2\tau_{(\alpha}\accentset{(0)}{u}^{\,\mu}\,\bb{B}_{\beta)\mu} +  \mathcal{O}(\lambda), \label{eq::app_b_1} \\
	\tensor[^{u}]{\accentset{\lambda}{b}}{^\alpha_\beta}	&= \tensor{\delta}{^\alpha_\beta} - \accentset{(0)}{u}^{\,\alpha}\,\tau_\beta + \lambda\left[\accentset{(0)}{u}^{\,\alpha}\,\accentset{(0)}{u}_\beta - \accentset{(1)}{u}^{\,\alpha}\tau_\beta\right] + \mathcal{O}(\lambda^2), \\
	\tensor[^{u}]{\accentset{\lambda}{b}}{^\alpha^\beta}	&= \hat h^{\alpha\beta} + \lambda\left[\accentset{(0)}{u}^{\,\alpha}\,\accentset{(0)}{u}^{\,\beta} + \accentset{(1)}{g}^{\,\alpha\beta}\,\right] + \mathcal{O}(\lambda^2). \label{eq::app_b_3}
\end{align}
Equation~\eqref{eq::app_b_1} can also be written as
\begin{equation}
	\tensor[^{u}]{\accentset{\lambda}{b}}{_\alpha_\beta}	= \tensor[^{\overset{(0)}{u}}]{b}{_{\alpha\beta}} + \mathcal{O}(\lambda),
\end{equation}
where $\tensor[^{\overset{(0)}{u}}]{b}{_{\alpha\beta}}$ is the projector orthogonal to $\accentset{(0)}{u}^{\,\alpha}$, with respect to the Galilean structure (defined by relations~\eqref{eq::NC_def_bb}).

\subsection{Expansion and vorticity tensors}
\label{app::exp_vort}

The formalism of this section is presented in more details in \citep{2021_Vigneron}.\saut

For a Galilean structure, the expansion $\TT{\Theta}{U}^{\alpha\beta}$ and vorticity $\TT{\Omega}{U}^{\alpha\beta}$ tensors of a unit $\tau$-timelike vector $\T U$ are defined as
\begin{align}
	\TT{\Theta}{U}^{\alpha\beta}	&\coloneqq \hat h^{\mu(\alpha}\hat\nabla_\mu U^{\beta)}\\
							&= -\frac{1}{2}\Lie{\T U}\hat h^{\alpha\beta}, \label{eqapp::theta_lie}\\
	\TT{\Omega}{U}^{\alpha\beta}	&\coloneqq \hat h^{\mu[\alpha}\hat\nabla_\mu U^{\beta]}. \label{eqapp::omega_lie}
\end{align}
These tensors are spatial. For two unit $\tau$-timelike vectors $\T U$ and $\T T$, such that $\T U - \T T = \T w$, which is spatial, we have
\begin{align}
	\TT{\Theta}{U}^{\alpha\beta} - \TT{\Theta}{T}^{\alpha\beta}	&= \hat D^{(\alpha} w^{\beta)}, \label{eqapp::dif_theta}\\
	\TT{\Omega}{U}^{\alpha\beta} - \TT{\Omega}{T}^{\alpha\beta}	&= \hat D^{[\alpha} w^{\beta]}, \label{eqapp::dif_omega}
\end{align}
where $\hat D_\alpha$ is the spatial derivative induced by $\hat h^{\alpha\beta}$ on the $\tau$-foliation, with $\hat D^\alpha \coloneqq \hat h^{\alpha\mu} \hat D_\mu$ (see section II.C.4. in \citep{2021_Vigneron}). From the general definition of a Galilean vector $\tilde{\T G}$ given in Section~\ref{sec::Gal_existence}, and using \eqref{eqapp::dif_theta} and \eqref{eqapp::dif_omega}, for any unit $\tau$-timelike vector $\T U$ we have
\begin{align}
	\TT{\Theta}{U}^{\alpha\beta}	&= H(t)\hat h^{\alpha\beta} + \hat D^{(\alpha}\left(\TT{v}{U}^{\beta)}\right) + \Xi^{\alpha\beta}, \label{eqapp::theta_U_G} \\
	\TT{\Omega}{U}^{\alpha\beta}	&=  \hat D^{[\alpha}\left(\TT{v}{U}^{\beta]}\right), \label{eqapp::omega_U_G}
\end{align}
where $\TT{\T v}{U} \coloneqq \T U - \tilde{\T G}$. \saut%Therefore, from the existence of a Galilean observer, the vectors present in the expansion and the vorticity tensors are the same. A consequence is that, from

A symmetric tensor can be decomposed into a scalar part (i.e. proportional to the metric), a gradient part (i.e. the gradient of a vector) and a tensor part (a traceless-transverse tensor) \citep{1973_York}, called SVT decomposition. In general, the scalar part has spatial dependence. The existence of a Galilean vector implies that the expansion tensors (in the framework of Galilean structures) of any $\tau$-timelike vector have a spatially constant scalar mode as shown by relation~\eqref{eqapp::theta_U_G}. Another consequence is the form of the vorticity~\eqref{eqapp::omega_U_G} which is expressed as function of the vector present in the gradient part of the expansion tensor. These relations were already shown to hold in \citep{2021_Vigneron, 2022_Vigneron_b} for Newton-Cartan and the NEN theory.\saut

Finally, the definition~\eqref{eqapp::omega_lie} implies that the vector $\T B$ present in the Galilean connection in the NR limit [see equation~\eqref{eq::Limit_connection_coeff}] is irrotational ($\TT{\Omega}{B}^{\alpha\beta} = 0$). Therefore formulas~\eqref{eqapp::theta_U_G} and~\eqref{eqapp::omega_U_G} applied for $\T B$ give
\begin{align}
	\TT{\Theta}{B}^{\alpha\beta} = H(t)\hat h^{\alpha\beta} + \hat D^{\alpha}\hat D^\beta \sigma + \Xi^{\alpha\beta}, \label{eqapp::Theta_B}
\end{align}
where $\sigma$ is a scalar field and $B^\mu - \tilde G^\mu = \hat D^\mu \sigma$.

\subsection{First order of the Lorentzian connection}

From the relation of the difference between two connections, we have
\begin{align}
	\accl{\Gamma}^\gamma_{\alpha\beta} - {\hat \Gamma}^\gamma_{\alpha\beta} &= \accl{g}^{\gamma\mu}\left(\hat\nabla_{(\alpha}\accl{g}_{\beta)\mu} - \frac{1}{2}\hat\nabla_{\mu}\accl{g}_{\alpha\beta}\right).
\end{align}
This leads to
\begin{align}
	\accentset{(1)}{\Gamma}^{\gamma}_{\alpha\beta} &= \hat h^{\gamma\mu}\left(\hat\nabla_{(\alpha}\accentset{(1)}{g}_{\beta)\mu} - \frac{1}{2}\hat\nabla_{\mu}\accentset{(1)}{g}_{\alpha\beta}\right) + \accentset{(1)}{g}^{\,\gamma\mu}\left(\hat\nabla_{(\alpha}\accentset{(0)}{g}_{\beta)\mu} - \frac{1}{2}\hat\nabla_{\mu}\accentset{(0)}{g}_{\alpha\beta}\right). \label{eq::form_Gamma1} 
\end{align}
This formula is in agreement with the ``connection perturbation'' introduced by \citep{2011_Tichy_et_al} for the post-Newton-Cartan approximation (Table~II in \citep{2011_Tichy_et_al}).\saut

Using $k^{\alpha\beta} = -\hat h^{\alpha\mu}\hat h^{\beta\nu}\accentset{(1)}{g}_{\mu\nu}$, which follows from \eqref{eq::soupe}, where $k^{\alpha\beta}$ is defined with respect to the first order $\accentset{(1)}{g}^{\,\alpha\beta}$ in equation~\eqref{eq::g^1_limit}, we have
\begin{align}
	\accentset{(1)}{\Gamma}^\mu_{\mu\nu}\hat h^{\nu\alpha} &= \hat D^\alpha\left(\phi - \frac{1}{2} k^{\mu\nu}\bbB_{\mu\nu}\right), \label{eq::Gamma1_1} \\
	\accentset{(1)}{\Gamma}^\gamma_{\mu\nu}\hat h^{\mu\alpha}\hat h^{\nu\beta} &= \frac{1}{2} \hat D^\gamma k^{\alpha\beta} -  \hat D^{(\alpha}k^{\beta)\gamma} + B^\gamma \TT{\Theta}{B}^{\alpha\beta}. \label{eq::Gamma1_2}
\end{align}

\subsection{First order of the spacetime curvature tensor}

Using the formula for the difference between two Riemann tensors
\begin{equation}
	{\accl{R}^\alpha}_{\beta\gamma\sigma} - {\hat R^\alpha}_{\beta\gamma\sigma} = 2 \hat\nabla_{[\gamma}\accentset{\lambda}{\Delta}^\alpha_{\sigma]\beta} + 2\accentset{\lambda}{\Delta}^\alpha_{\mu[\gamma}\accentset{\lambda}{\Delta}^\mu_{\sigma]\beta},
\end{equation}
where $\accl{\Delta}^\alpha_{\beta\gamma} \coloneqq \accl{\Gamma}^\alpha_{\beta\gamma} - \hat \Gamma^\alpha_{\beta\gamma}$, we obtain:
\begin{align}
	{\accentset{(1)}{R}^\alpha}_{\beta\gamma\sigma} &= 2 \hat\nabla_{[\gamma}\accentset{(1)}{\Gamma}^\alpha_{\sigma]\beta}, \\
	\accentset{(1)}{R}_{\alpha\beta} &= 2 \hat\nabla_{[\mu}\accentset{(1)}{\Gamma}^\mu_{\alpha]\beta}.
\end{align}
Then,
\begin{align}
	\accentset{(1)}{R}_{\mu\nu}\hat h^{\mu\alpha}\hat h^{\nu\beta} = \hat\nabla_\gamma\left(\hat h^{\alpha\mu}\hat h^{\beta\nu}\accentset{(1)}{\Gamma}^\gamma_{\mu\nu}\right) - \hat h^{\alpha\mu}\hat\nabla_\nu\left(\accentset{(1)}{\Gamma}^\gamma_{\gamma\nu}\hat h^{\nu\beta}\right).
\end{align}
Using equations~\eqref{eq::Gamma1_1} and \eqref{eq::Gamma1_2} along with $\TT{\Theta}{B}^{\alpha\beta} = \hat h^{\mu\alpha}\hat\nabla_\mu B^{\beta}$ as $\TT{\Omega}{B}^{\alpha\beta} = 0$, we obtain
\begin{align}
	\accentset{(1)}{R}_{\mu\nu}\hat h^{\mu i}\hat h^{\nu j}
		&= \Lie{\T B}\TT{\Theta}{B}^{ij} +  \TT{\theta}{B} \TT{\Theta}{B}^{ij} + 2\TT{\Theta}{B}^{c(i}{\TT{\Theta}{B}^{j)}}_c\nonumber\\
		&\qquad +\frac{1}{2}\left( \hat D_c  \hat D^c k^{ij} +  \hat D^i \hat D^j {k^c}_c\right) - \hat D_c\hat D^{(i}k^{j)c} - \hat D^i \hat D^j \phi, \label{eqapp::Rij} \\
	\accentset{(1)}{R}_{\mu\nu}\hat h^{\mu\nu}
		&= \Lie{\T B} \TT{\theta}{B} + \TT{\theta}{B}^2 - \hat D_i\hat D^i\phi + \hat D_j \hat D^j {k^i}_i - \hat D_i \hat D_jk^{ij}, \label{eqapp::Rij_hij} 
\end{align}
where we only keep spatial components as $\accentset{(1)}{R}_{\mu\nu}\hat h^{\mu\alpha}\hat h^{\nu\beta}$ is spatial, and $\TT{\theta}{B} \coloneqq \TT{\Theta}{B}^{ij} \hat h_{ij}$.

\section{Spatial metric at the limit and existence of the Galilean vector}
\label{app::Gal_met}

\subsection{A solution to equation~\eqref{eq::R_ij=barR_ij}}
\label{app:R_ij=barR_ij_sol}

While we have not yet found the general solution for $\hat h_{ij}$ to the equation $\hat\CR_{ij} = \bar\CR_{ij}$, we give in this section a simple non-trivial solution. It can be found by writing the physical metric with a scalar-vector-tensor (SVT) decomposition with respect to the reference spatial metric $\bar h_{ij}$. In general, the contravariant and covariant components of the physical metric have a different decomposition:
\begin{align}
	\hat h_{ij} = A\bar h_{ij} + \bar D_{(i}A_{j)} + A_{ij}, \\
	\hat h^{ij} = B\bar h^{ij} + \bar D^{(i}B^{j)} + B^{ij},
\end{align}
where $A_{ij}$ and $B^{ij}$ are traceless-transverse tensors with respect to $\bar{h}_{ij}$, and $\bar D_i$ is the covariant derivative with respect to that same metric, with $\bar D^i \coloneqq \bar h^{ci}\bar D_c$. We recall that $\hat h_{ij}$ and $\hat h^{ij}$ are by definition inverse from each-other, and the same for $\bar h_{ij}$ and $\bar h^{ij}$.\saut

We can rewrite equation~\eqref{eq::R_ij=barR_ij} as
\begin{equation}
	2 \bar D_{[c}C^c_{j]i} + 2C^c_{d[c}C^d_{j]i} = 0,
\end{equation}
with $C^a_{ij} \coloneqq \hat h^{ak}\left(\bar D_{(i}\hat h_{j)k} - \frac{1}{2}\bar D_k\hat h_{ij}\right)$. Then, a sufficient condition on $\hat h_{ij}$ to be solution of the above equation is
\begin{equation}
	\bar D_{k}\hat h_{ij} = 0. \label{eq::cond_sol_R=R}
\end{equation}
This condition implies that $\hat h_{ij} - \hat h_{kl}\bar h^{kl}\,\bar h_{ij}/3$ is a traceless-transverse tensor with respect to $\bar{\T h}$. Then, from the uniqueness of the SVT decomposition, the gradient mode (i.e.~$\hat D_{(i} A_{j)}$) of $\hat h_{ij}$ is zero and we have
\begin{align}
	\hat h_{ij} = A&\bar h_{ij} + A_{ij}, \label{eq::gtf}\\
	\bar D_i A = 0 \quad &; \quad \bar D_k A_{ij} = 0. \label{eq::DA_DH}
\end{align}
The condition~\eqref{eq::cond_sol_R=R} implies $C^a_{ij} =0$, which is the difference between the physical $\hat D_i$ and the reference $\bar D_i$ connections. Therefore we have $\hat D_i = \bar D_i$, implying $\bar D_{k}\hat h^{ij} = 0$. Then, the contravariant gradient mode also vanishes and we have
\begin{align}
	\hat h^{ij} = B&\bar h^{ij} + B^{ij}, \\
	\bar D_i B = 0 \quad &; \quad \bar D_k B^{ij} = 0.
\end{align}

As for the homogeneity of $\hat\CR_{ij}$, we have
\begin{equation}
	\partial_i\hat\CR= \partial_i\left(B^{cd}\bar\CR_{cd}\right),
\end{equation}
using $\partial_i(\bar h^{cd}\bar\CR_{cd}) = 0$ and equation~\eqref{eq::R_ij=barR_ij}. Therefore, for spherical or hyperbolic topologies, where by definition $\bar\CR_{ij} = \pm2 \bar h_{ij}$, we obtain $\partial_i\hat\CR = 0$. In summary, from the guess~\eqref{eq::cond_sol_R=R}, for $\tilde{\Sigma} = \mS^3$ or $\mH^3$, the spatial scalar curvature of the Galilean structure is homogeneous. However, it is not necessarily isotropic as we have the following relation:
\begin{equation}
	\hat\CR_{ij} = \pm2/A \left( \hat h_{ij} - A_{ij}\right), \label{eq::Sol_R_ij}
\end{equation}
with $\hat h_{ij} - \frac{1}{3} h^{cd}A_{cd} \, \hat h_{ij} \not= 0$.
%\quentin{Does this mean the manifold at the limit is curvature homogeneous, but not necessarily homogeneous ? \citep{1989_Tricerri_et_al}}

\subsection{Expansion tensor of $G^\mu$}
\label{app::exp_G}

We consider a vector $\T G \in {\rm ker}(\bar R_{\mu\nu})$. Then, from relation~\eqref{eq::Lie_Rbar} and the definition of $\bar R_{\mu\nu}$, in a coordinate system $\{\T\partial_t, \T\partial x^i\}$ adapted to the $\tau$-foliation with $\T G = \T\partial_t$, we can choose~$\bar h_{ij}$ %"choose" is important here !!
such that
\begin{equation}
	\partial_{t|_G} \bar h_{ij} = 0. \label{eq::partial_th_ij}
\end{equation}
From that equation and relation~\eqref{eq::DA_DH}, the time derivative along $\T G$ of the physical metric $\hat h_{ij}$ is
\begin{equation}
	\partial_{t|_G} \hat h_{ij} = \frac{\partial_{t|_G} A}{A} \hat h_{ij} - \frac{\partial_{t|_G} A}{A} A_{ij} + \partial_{t|_G} A_{ij}. \label{eq::part_h_fgh}
\end{equation}
From the definition of the expansion tensor~\eqref{eqapp::theta_lie}, this leads to $\TT{\Theta}{G}_{ij} = \frac{1}{2}\partial_{t|_G} \hat h_{ij}$. Then from~\eqref{eq::part_h_fgh} and the fact that $\hat D_k A = 0$, $\hat D_k A_{ij} = 0$ and $\hat D_k \partial_{t|_G} A_{ij} = 0$ (using \eqref{eq::partial_th_ij} along with the property $\hat D_i = \bar D_i$ for the last relation), we have 
\begin{align}
	\hat D_k \TT{\Theta}{G}_{ij} = 0.
\end{align}
That property is equivalent as having (this follows from the unicity of the SVT decomposition)
\begin{align}
	\TT{\Theta}{G}_{ij} = H\hat h_{ij} + \Xi_{ij}, \label{eq::final_Theta_G}
\end{align}
with $\hat D_k H = 0$ and $D_k\Xi_{ij} = 0$. From $\TT{\Theta}{G}_{ij} = \frac{1}{2}\partial_{t|_G} \hat h_{ij}$ and the definition of the volume of $\Sigma$, i.e. $V_\Sigma \coloneqq \int_\Sigma\sqrt{{\rm det}\, \hat h_{ij}} \, \dd x^3$, we have $H(t) = \dot V_\Sigma/(3V_\Sigma)$. Result~\eqref{eq::final_Theta_G} holds once condition~\eqref{eq::cond_sol_R=R} is fulfilled, or if we directly impose $\hat\CR_{ij} = \hat\CR(t)\, \hat h_{ij}/3$ in the case $\tilde\Sigma = \mS^3$ or $\mH^3$. Equation~\ref{eq::final_Theta_G} is the first of the two conditions~\eqref{eq::def_G_theta} and \eqref{eq::def_G_omega} for $\T G$ to be a Galilean vector.\saut

From the general form of the physical spatial Ricci curvature~\eqref{eq::Sol_R_ij} (depending on the guess~\eqref{eq::cond_sol_R=R}) and with~\eqref{eq::gtf}, we see that imposing the physical spatial curvature to be only trace (i.e. $\hat\CR_{ij} \propto \hat h_{ij}$) leads to $A_{ij} = 0$, and $\partial_{t|_G} \hat h_{ij} \propto \bar h_{ij} \propto \hat h_{ij}$. Therefore, we have the following property:
\begin{align}
	\hat\CR_{ij} \propto \hat h_{ij} \Rightarrow \Xi_{ij} = 0.
\end{align}
That is: an isotropic spatial curvature implies an isotropic expansion. It is however not yet clear whether or not the reverse property is true.\saut

\begin{remark}{We expect that the general solution of equation~\eqref{eq::R_ij=barR_ij} would rather constrain $\Xi_{ij}$ to be only harmonic, i.e. $\hat\Delta \Xi_{ij} = 0$, instead of gradient free. Indeed, being traceless-transverse and harmonic is the expected condition for a tensor representing global anisotropic expansion.}
\end{remark}

\subsection{Vorticity of $G^\mu$}
\label{app::vort_G}

From the Galilean limit, the vector $\T B$ present in the first order of $\accl{g}^{\alpha\beta}$ is irrotational with respect to the Galilean structure. Using relation~\eqref{eqapp::dif_omega} for the difference between two vorticity tensors, this implies that $\TT{\Omega}{G}^{ij} = \hat D^{[i}w^{j]}$ where $\T w = \T G - \T B$ is a spatial vector. Then using $\hat R_{\mu\nu}\hat h^{\mu\alpha}G^\nu = 0$, which results from~\eqref{eq::Limit_Mod_E}, we obtain
\begin{equation}
	\hat D_c\hat D^{[i}w^{c]} - 2\hat D^i H = 0.
\end{equation}
From the unicity of the Hodge decomposition, we have $\hat D_c\hat D^{[i}w^{c]} = 0$, which implies $\hat D^{[i}w^{j]} =0$. We just showed that $\TT{\Omega}{G}^{ij} = 0$.\saut

In conclusion, the vector $\T G$ is a Galilean vector: $\T G = \tilde{\T G}$.

\subsection{Uniqueness of the Galilean vector?}
\label{app::Unique_G}

While $\T G$ is a Galilean vector, it is not uniquely defined if $\textrm{dim}[\ker(\bar R_{\alpha\beta})] > 1$ (see section~\ref{sec::prop_gbar}). For two independent Galilean vectors $\tilde{\T G}_{(1)}$ and $\tilde{\T G}_{(2)}$, we define the spatial vector ${\T w = \tilde{\T G}_{(2)} - \tilde{\T G}_{(1)}}$. Then from the difference between two expansion and vorticity tensors (equations~\eqref{eqapp::dif_theta} and~\eqref{eqapp::dif_omega}), the vector $\T v$ is constrained by:
\begin{align}
	\hat D_i w^j = 0 \quad ; \quad w^k\hat\CR_{ki} = 0.
\end{align}
The second property is a direct consequence of the first one when calculating $D_{[i}D_{j]} w^k$. In conclusion, if there is not a unique inertial frame at the NR limit, then this frame is define up to a global translation in the direction of zero spatial curvature. \saut

\begin{remark}{Even if there is not a unique inertial frame, the closedness of $\Sigma$ still implies the existence of a ``preferred'' one, due to topological properties \citep{2002_Uzan_et_al,2008_Roukema_et_al}. In other words, once we require the existence of a NR limit of either general relativity or the bi-connection theory, this implies the existence of a preferred observer in the Universe. \textit{\underline{Question:} could the primordial power spectrum of the CMB be explained by the fondamental presence of such a preferred frame in the theory, and therefore be of topological origin?}}
\end{remark}

\section{Derivation of the dictionary}
\label{app::dico}

In this section, we consider isotropic expansion ($\Xi_{ij} = 0$). Using the Taylor series~\eqref{eq::Limit_g^ab} and \eqref{eq::Limit_cov_g_lambda} of the spacetime metric, in a coordinate system $\{\T\partial_t, \T\partial x^i\}$ adapted to the $\tau$-foliation and where $\T\partial_t = \tilde{\T G}$, we have
\begin{align}
	\accentset{\lambda}{g}^{\alpha\beta} = 
	\left(\begin{array}{cc}
		-\lambda + \lambda^2 2\phi + \bigO{\lambda^3} & -\lambda \hat D^i \sigma + \bigO{\lambda^2} \\
		-\lambda \hat D^i \sigma + \bigO{\lambda^2} & \hat h^{ij} + \lambda (k^{ij} - \hat D^i \sigma \hat D^j \sigma) + \bigO{\lambda^2}
	\end{array}\right)
\end{align}
and 
\begin{align}
	\accentset{\lambda}{g}_{\alpha\beta} = 
	\left(\begin{array}{cc}
		-\frac{1}{\lambda} - 2(\phi -\frac{1}{2}\hat D^k \sigma \hat D_k \sigma)+ \bigO{\lambda} & - \hat D_i \sigma + \bigO{\lambda}\\
		- \hat D_i \sigma + \bigO{\lambda} & \hat h_{ij} - \lambda k_{ij} + \bigO{\lambda^2}
	\end{array}\right),
\end{align}
where, as shown in Appendix~\ref{app::exp_vort}, $B^\mu - \tilde G^\mu \eqqcolon \hat D^\mu \sigma$ with $\hat D_i \sigma \coloneqq \hat h_{ki}\hat D^k \sigma $. We also defined $k_{ij} \coloneqq k^{cd}\hat h_{ci} \hat h_{dj}$. We recall that $B^\mu$ and $\phi$ are defined in Section~\ref{sec::the_limit} where the NR limit is presented. For now, $\phi$ is just a scalar field related to the 4-acceleration $\tensor[^{B}]{a}{^\mu}$ of $B^\mu$, i.e. $\tensor[^{B}]{a}{^\mu} \coloneqq B^\nu \hat\nabla_\nu B^\mu = \hat D^\mu \phi$.\saut

The goal is to find if there is a relation between $\phi$, $\sigma$, $k^{ij}$ and the gravitational field~$\Phi$. A powerful property is the fact that in both Newton's theory and the NEN theory, $\Phi$ is given by the 4-acceleration of the Galilean vector \citep{2021_Vigneron, 2022_Vigneron_b}, i.e. the 4-acceleration of inertial frames:
\begin{align}
	\tensor[^{\tilde G}]{a}{^\mu} \coloneqq \tilde G^\nu\hat \nabla_\nu \tilde G^\mu = \hat D^\mu \Phi.
\end{align}
Using formula~(60) in \citep{2021_Vigneron}, we can express $\tensor[^{B}]{a}{^\mu}$ as function of $\tensor[^{\tilde G}]{a}{^\mu}$ and obtain (we keep only spatial indices as all the tensors involved are spatial):
\begin{align}
	\hat D^i \phi - \hat D^i \Phi = \partial_{t|_G} \hat D^i \sigma + 2\hat D_k \sigma\left(\tensor[^{B}]{\Theta}{^k^i} + \tensor[^{B}]{\Omega}{^k^i}\right) - \hat D^k \sigma\hat D_k \hat D^i \sigma.
\end{align}
Then using~\eqref{eqapp::Theta_B} and $\tensor[^{B}]{\Omega}{^k^i} =0$, we obtain $\hat D^i \left(\phi -\frac{1}{2} \hat D^k \sigma \hat D_k \sigma - \partial_{t|_G}\sigma\right) = \hat D^i\Phi$, which leads to
\begin{align}
	\phi -\frac{1}{2} \hat D^k \sigma \hat D_k \sigma - \partial_{t|_G}\sigma = \Phi, \label{eq::phi->Phi}
\end{align}
up to an integration constant. We now have a link between $\phi$, $\sigma$ and $\Phi$.\saut

Let us consider the SVT decomposition of $k^{ij} - \hat D^i \sigma \hat D^j \sigma$, as is usually done in the weak field limit:
\begin{align}
	k^{ij} - \hat D^i \sigma \hat D^j \sigma \eqqcolon 2\psi\hat h^{ij} - 2\hat D^i\hat D^j E - 2\hat D^{(i}F^{j)} - 2f^{ij}, \label{eq::SVT_k^ij}
\end{align}
with $\hat D_kF^k \coloneqq 0$ and $f^{ij}\hat h_{ij}  \coloneqq 0 \eqqcolon \hat D_k f^{ki}$. The first order of the (modified) Einstein equation~\eqref{eq::ModE2} is
\begin{equation}
	\accentset{(1)}{R}_{\mu\nu}\hat h^{\mu\alpha}\hat h^{\nu\beta} = \left(\frac{\kappa}{2}\rho + \Lambda\right)\hat h^{\alpha\beta}. \label{eq::first_Einstein_eq}
\end{equation}
This holds either in the Euclidean case (general relativity), or in the non-Euclidean case (bi-connection theory). Using formula~\eqref{eqapp::Rij} for $\accentset{(1)}{R}_{\mu\nu}$, with \eqref{eq::SVT_k^ij} and \eqref{eq::phi->Phi} we obtain
\begin{align}
\left(\frac{\kappa}{2}\rho + \Lambda\right) \hat h^{ij}	&= \hat h^{ij}(\dot H + 3H^2) + \hat h^{ij}\left[\hat\Delta\psi + \hat\Delta(H\sigma)\right] \nonumber\\
									&\qquad + \hat D^i\hat D^j\left[-\Phi + H\sigma + \psi + 4KE\right] \\
									&\qquad + 4K\hat D^{(i}F^{j)} \nonumber\\
									&\qquad + \left(6K - \hat\Delta\right)f^{ij},\nonumber
\end{align}
where $K$ is the spatial Gaussian curvature: $\hat \CR_{ij} = 2K/a^2\, \hat h_{ij}$. The unicity of the SVT decomposition applied on the above equation leads to
\begin{align}
&\frac{\kappa}{2}\rho + \Lambda = \dot H + 3H^2 +\hat\Delta\psi + \hat\Delta(H\sigma), \label{eq::SVT_1}\\
&-\Phi + H\sigma + \psi + 4KE = 0,  \label{eq::SVT_2}\\
&4K\hat D^{(i}F^{j)} = 0, \label{eq::SVT_3}\\
&\left(6K - \hat\Delta\right)f^{ij} = 0. \label{eq::SVT_4}
\end{align}
So for any $K$, we have $f^{ij} = 0$ (more precisely, $f^{ij}$ is harmonic), and using the Poisson equation~\eqref{eq::Poisson} for $\Phi$ with relation~\eqref{eq::SVT_1} we obtain
\begin{align}
	\psi + H\sigma = \Phi. \label{eq::psi->Phi}
\end{align}
In the Euclidean case (i.e. $K=0$), which corresponds to the NR limit of Einstein's equation and the usual Newton-to-GR dictionary, $E$ and $F^i$ are free. This is a usual gauge freedom in post-Newtonian theory. However, we see that in the non-Euclidean case, relations~\eqref{eq::SVT_2} and \eqref{eq::SVT_3} imply $E=0$ and $F^i=0$.\saut

Finally, the spacetime metric can be written as follows as function of the gravitational potential:
\begin{align}
	\accentset{\lambda}{g}^{\alpha\beta} = 
	\left(\begin{array}{cc}
		-\lambda + 2\lambda^2 \left(\Phi + \frac{1}{2}\hat D^k\sigma \hat D_k \sigma  + \partial_{t|_G}\sigma\right) + \bigO{\lambda^3} & -\lambda \hat D^i \sigma + \bigO{\lambda^2}\vspace{.2cm} \\
		-\lambda \hat D^i \sigma + \bigO{\lambda^2} & \begin{array}{c}
												\hat h^{ij} + \lambda \left[2(\Phi - H\sigma)\hat h^{ij} \right.\\
												\left.- 2\hat D^i\hat D^j E - 2\hat D^{(i}F^{j)}\right] + \bigO{\lambda^2}
											\end{array}
	\end{array}\right)
\end{align}
and 
\begin{align}
	\accentset{\lambda}{g}_{\alpha\beta} = 
	\left(\begin{array}{cc}
		-\frac{1}{\lambda} - 2 \left(\Phi  + \partial_{t|_G}\sigma\right) + \bigO{\lambda} & - \hat D_i \sigma + \bigO{\lambda} \vspace{.2cm} \\
		- \hat D_i \sigma + \bigO{\lambda} &  \begin{array}{c}
											\hat h_{ij} - \lambda \left[2(\Phi - H\sigma)\hat h_{ij} \right.\\
											\left.\quad - 2\hat D_i\hat D_j E - 2\hat D_{(i}F_{j)} + \hat D_i\sigma \hat D_j \sigma\right] + \bigO{\lambda^2}
										\end{array}
	\end{array}\right),
\end{align}
where $E\not=0$ and $F^i\not=0$ only if $K = 0$.\saut

It seems that in the non-Euclidean case, $E$ and $F^i$ are not anymore gauge variables and are necessarily zero, i.e. the metric needs to be written with the same form as in the standard post-Newtonian gauge with Newton's theory. In the next Appendix, we show that this peculiarity in the non-Euclidean case comes from a gauge fixing we implicitly made in Section~\ref{sec::Galilean_limit} by assuming that $\bar R_{\mu\nu}$ should not depend on $\lambda$.

\section{Gauge freedom in the NR limit}
\label{app::Gauge}

\subsection{Gauge freedom}

Let us consider $\mathbb{R}^4$ on which we define two Minkowski metrics $\eta_{\mu\nu}$ and $\tilde\eta_{\mu\nu}$. The set of foliations which are flat with respect to $\eta_{\mu\nu}$ are not necessarily flat with respect to $\tilde\eta_{\mu\nu}$. In other words, in a coordinate system where $\eta_{\mu\nu} = {\rm diag}(-1; 1; 1; 1)$, the relation $\tilde\eta_{\mu\nu} = {\rm diag}(-1; 1; 1; 1)$ does not necessarily hold. This is because there are four degrees of freedom in defining $\tilde\eta_{\mu\nu}$ with respect to $\eta_{\mu\nu}$, corresponding to a diffeomorphism $\varphi$ of $\mathbb{R}^4$ in itself. The freedom in changing $\tilde\eta_{\mu\nu}$ while preserving its properties of being a Minkowski metric is related to the differential $\dd\varphi : T\mathbb{R}^4 \rightarrow T\mathbb{R}^4$ of $\varphi$:
\begin{align}
	\tilde\eta_{\mu\nu} \overset{\rm freedom}{\longrightarrow} \dd\varphi\left(\tilde\eta_{\mu\nu}\right).
\end{align}
The same freedom applies for $R_{\mu\nu}$ and $\bar R_{\mu\nu}$ in the bi-connection theory. Therefore, the role of the bi-connection condition~\eqref{eq::metrics_cond1} is to constrain the four degrees of freedom in the definition of the physical connection with respect to the reference one (or inversely).\saut

This freedom arising when two tensors are defined on the same manifold is also present for the family of Lorentzian metrics $\{\accl{g}_{\mu\nu}\}_{\lambda>0}$ used in the NR limit: there is a diffeomorphism freedom in defining $\accentset{{\lambda}}{g}_{\mu\nu}$ with respect to $\accentset{{\lambda_0}}{g}_{\mu\nu}$, with $\lambda \not= \lambda_0$ for any $\lambda_0 > 0$. However, while in the example related to the Minkowski metrics, this diffeomorphism is general, in the case of the family $\{\accl{g}_{\mu\nu}\}_{\lambda>0}$, because we impose $\underset{\lambda\rightarrow\lambda_0}{\rm lim}\accentset{{\lambda}}{g}_{\mu\nu} = \accentset{{\lambda_0}}{g}_{\mu\nu}$, that diffeomorphism freedom is constrained to be the identity in the limit $\lambda\rightarrow\lambda_0$, and thus should also be parametrised by $\lambda$: $\underset{\lambda\rightarrow\lambda_0}{\rm lim} \accl{\phi} = {\rm Id}$.\saut

A consequence is that, at first order in $(\lambda - \lambda_0)$, the freedom in (re)defining $\accentset{{\lambda}}{g}_{\mu\nu}$ with respect to $\accentset{{\lambda_0}}{g}_{\mu\nu}$ is a Lie derivative (see \citep{2008_Malik_et_al} for a detailed explanation\footnote{Note that while \citep{2008_Malik_et_al} defines one manifold $\accl{\CM}$ for each value of $\lambda$ embedded in a 5-dimensional manifold $\mathcal{N}$ (with a diffeomorphism $\accl\phi$ between $\accentset{{\lambda_0}}{\CM}$ and $\accentset{{\lambda}}{\CM}$), this is equivalent as considering a single manifold $\CM$ with an auto-diffeomorphism, as is our case.}):
\begin{align}
	\accentset{{\lambda}}{g}_{\mu\nu} - \accentset{{\lambda_0}}{g}_{\mu\nu} \overset{\rm freedom}{\longrightarrow} \accentset{{\lambda}}{g}_{\mu\nu} - \accentset{{\lambda_0}}{g}_{\mu\nu} + (\lambda - \lambda_0)\Lie{\T \xi} \, \accentset{{\lambda_0}}{g}_{\mu\nu} + \bigO{(\lambda-\lambda_0)^2},
\end{align}
where $\xi^\mu$ is a vector field on $\CM$. In perturbation theory, that freedom is called \textit{a gauge freedom} as it does not change physics.

\subsection{Gauge invariant variables in the NR limit}
\label{app:gauge_NR}

Under the gauge transformation describe in the previous section, $\accentset{(1)}{g}^{\,\mu\nu}$ and $\accentset{(0)}{g}_{\mu\nu}$ transform as
\begin{align}
	\accentset{(1)}{g}^{\mu\nu} &\ \accentset{{\T \xi}}{\longrightarrow}\ \accentset{(1)}{g}^{\,\mu\nu} + \Lie{\T \xi}\hat h^{\mu\nu}, \label{eq::gauge_g^1} \\
	\accentset{(0)}{g}_{\mu\nu} &\ \accentset{{\T \xi}}{\longrightarrow}\ \accentset{(0)}{g}_{\mu\nu} + \Lie{\T \xi}(-\tau_\mu\tau_\nu). \label{eq::gauge_g_0}
\end{align}
The vector $\xi^\mu$ can be written as
\begin{align}
	\xi^\mu = N \tilde G^\mu + \hat D^\mu \CA + \CA^\mu,
\end{align}
where $\CA^\mu \tau_\mu \coloneqq 0 \eqqcolon \hat D_\mu \CA^\mu$, and $N$ and $\CA$ are scalar fields. We recall that $\tilde G^\mu$ is a Galilean vector. Then
\begin{align}
	\Lie{\T \xi}\hat h^{\mu\nu} &= -2NH\hat h^{\mu\nu} - 2N\Xi^{\mu\nu} - 2\hat h^{\alpha(\mu}\tilde G^{\nu)}\nabla_\alpha N - 2\hat D^\mu \hat D^\nu \CA - 2\hat D^{(\mu} \CA^{\nu)}, \label{eq::gauge_g^1_A}\\
	\Lie{\T \xi} (-\tau_\mu\tau_\nu) &= -2\tau_{(\mu}\nabla_{\nu)} N, \label{eq::gauge_g_0_A}
\end{align}
where we used $\hat h^{\alpha\mu}\hat\nabla_\alpha \tilde G^\nu = -2H\hat h^{\mu\nu} -2\Xi^{\mu\nu}$. Then using the SVT decomposition introduced in Appendix~\ref{app::dico} we have
\begin{align}
	\accentset{(1)}{g}^{\,\mu\nu} = - \tilde G^\mu \tilde G^\nu - 2\tilde G^{(\mu}\hat D^{\nu)}\sigma + 2\psi\hat h^{\mu\nu} - 2\hat D^\mu\hat D^\nu E - 2\hat D^{(\mu}F^{\nu)} - 2f^{\mu\nu}. \label{eq::vin}
\end{align}
Using the definition~\eqref{eq::g_0_limit} of $\accentset{(0)}{g}_{\mu\nu}$ along with formula~\eqref{eq::app_b_1} (in which we replace $\accentset{(0)}{u}^{\,\mu}$ by $B^\mu \eqqcolon \tilde G^\mu + \hat D^\mu\sigma$; and we replace $\bb{B}_{\mu\nu}$ by $\bb{{\tilde G}}_{\mu\nu}$), we get
\begin{align}
	\accentset{(0)}{g}_{\mu\nu} = \bb{{\tilde G}}_{\mu\nu} - 2\tau_{(\mu}\hat\nabla_{\nu)} \sigma + \tau_\mu\tau_\nu \left(\hat h^{\alpha\beta}\hat\nabla_\alpha\sigma \hat\nabla_\beta\sigma - 2\phi\right). \label{eq::biere}
\end{align}
Then, from~\eqref{eq::vin} and \eqref{eq::biere}, the gauge transformations \eqref{eq::gauge_g^1} and \eqref{eq::gauge_g_0} induce a transformation of the variables $\sigma$, $\psi$, $E$, $F^\mu$, $f^{\mu\nu}$ and $\phi$ as follows:
\begin{align}
	 \psi			&\ \accentset{{\T \xi}}{\longrightarrow}\  \psi - HN, \\
	\sigma		&\ \accentset{{\T \xi}}{\longrightarrow}\ \sigma + N, \label{eq:couille} \\
	 E			&\ \accentset{{\T \xi}}{\longrightarrow}\  E + \CA, \\
	 F^\mu		&\ \accentset{{\T \xi}}{\longrightarrow}\  F^\mu + \CA^\mu, \label{eq:quille}  \\
	 f^{\mu\nu}	&\ \accentset{{\T \xi}}{\longrightarrow}\  f^{\mu\nu} + N\Xi^{\mu\nu}, \\
	 \phi + \frac{1}{2}\hat h^{\alpha\beta}\hat\nabla_\alpha\sigma \hat\nabla_\beta\sigma
				&\ \accentset{{\T \xi}}{\longrightarrow}\  \phi + \frac{1}{2}\hat h^{\alpha\beta}\hat\nabla_\alpha\sigma \hat\nabla_\beta\sigma + G^\mu\hat\nabla_\mu N.
\end{align}
Therefore, the following variables are gauge invariant:
\begin{align}
	\tilde\Psi &\coloneqq \psi + H\sigma, \\
	\tilde\Phi &\coloneqq  \phi + \frac{1}{2}\hat h^{\alpha\beta}\hat\nabla_\alpha\sigma \hat\nabla_\beta\sigma - G^\mu\hat\nabla_\mu \sigma, \\
	\tilde f^{\mu\nu} &\coloneqq f^{\mu\nu} - \sigma\Xi^{\mu\nu}.
\end{align}

$\tilde\Psi$ and $\tilde\Phi$ are the equivalent of the Bardeen potentials in the cosmological perturbation theory (weak field limit around a homogeneous and isotropic solution). We showed in Appendix~\ref{app::dico} that $\tilde\Psi = \tilde\Phi$ either in the Euclidean case (i.e. with Newton's theory from Einstein's equation, as expected), or in the non-Euclidean case (from the bi-connection theory). Therefore, the gravitational slip (i.e. $\tilde\Psi - \tilde\Phi$) is still zero in the bi-connection theory. Furthermore, we have $\tilde\Psi = \tilde\Phi = \Phi$, the gravitational potential, showing that this variable is gauge-invariant, again either in the Euclidean or non-Euclidean case. Equation~\eqref{eq:couille}-\eqref{eq:quille} show that $\sigma$, $E$ and $F^\mu$ can be set to zero by a gauge transformation.

%It is also interesting to note that in the case anisotropic expansion is present, the traceless-transverse tensor $f^{ij}$ is not anymore gauge-invariant.

\subsection{Gauge freedom in the reference curvature}
\label{app:gauge_Rbar}

As mentioned above, there is a diffeomorphism freedom in defining $\bar R_{\mu\nu}$ with respect to $R_{\mu\nu}$. In the case of the NR limit, that freedom is present with respect to all the metrics $\accl{g}_{\mu\nu}$ and their curvatures. This implies that, once we consider a NR limit, the diffeomorphism freedom on defining $\bar R_{\mu\nu}$ with respect to the family of Lorentzian metrics is parametrised by $\lambda$, leading, at first order, to
\begin{align}
	\bar R_{\mu\nu} = \accentset{(0)}{\bar R}_{\mu\nu} + \lambda \Lie{X}\left(\accentset{(0)}{\bar R}_{\mu\nu}\right) + \bigO{\lambda^2},
\end{align}
for any vector field $X^\mu$. Under a gauge transformation we have $X^\mu \ \accentset{{\T\xi}}{\longrightarrow}\ X^\mu + \lambda \xi^\mu$. As the difference between $\bar R_{\mu\nu}$ and $\accentset{(0)}{\bar R}_{\mu\nu}$ comes solely from diffeomorphism freedom, these two tensors have the same property, i.e. they take the form~\eqref{eq::Riccbar_choice} in a certain (one for each) coordinate system\footnote{At first order, these coordinate systems differ by the transformation $x^\mu \ \accentset{{\T\xi}}{\longrightarrow}\ x^\mu + \lambda\xi^\mu$.}. In particular, as for $\bar R_{\mu\nu}$, there exists a vector $\accentset{(0)}{G}^\mu \in {\rm ker}(\accentset{(0)}{\bar R}_{\mu\nu})$ such that $\Lie{{\accentset{(0)}{\T G}}} \accentset{(0)}{\bar R}_{\mu\nu} = 0$.\saut

Because until now we supposed that $\bar R_{\mu\nu}$ was independent on $\lambda$, this means that the calculation of Section~\ref{sec::Galilean_limit} and Appendix~\ref{app::dico} was performed in a specific gauge where $X^\mu = 0$. This is why we found in Appendix~\ref{app::dico} that $E$ and $F^i$ needed to vanish. In what follows we explain why, without the gauge choice $X^\mu = 0$, $E$ and $F^i$ do not vanish anymore.\saut

In a general gauge, equation~\eqref{eq::hemoroides} takes the form
\begin{align}
	\hat\CR^{\mu\nu}\, \bb{{\accentset{(0)}{G}}}_{\alpha\mu}\bb{{\accentset{(0)}{G}}}_{\beta\nu} = \accentset{(0)}{\bar R}_{\mu\nu}, \label{eq::lacter}
\end{align}
with $\accentset{(0)}{G}^\mu \, \accentset{(0)}{\bar R}_{\mu\nu} \coloneqq 0$, and the first order equation~\eqref{eq::first_Einstein_eq} (used to derive the dictionary) becomes
\begin{align}
	\left(\accentset{(1)}{R}_{\mu\nu} - \Lie{X}\left(\accentset{(0)}{\bar R}_{\mu\nu}\right)\right)\hat h^{\mu\alpha}\hat h^{\nu\beta} = \left(\frac{\kappa}{2}\rho + \Lambda\right)\hat h^{\alpha\beta}. \label{eq::fghjf}
\end{align}
Because $\accentset{(0)}{\bar R}_{\mu\nu}$ has the same properties as ${\bar R}_{\mu\nu}$, the resolution of equation~\eqref{eq::lacter} is equivalent as solving equation~\eqref{eq::hemoroides} in Section~\ref{eq::R_Gal}. In particular, $\accentset{(0)}{G}^\mu$ corresponds to a Galilean vector, and in Euclidean, spherical or hyperbolic topologies without anisotropic expansion we have
\begin{align}
	\accentset{(0)}{\bar R}_{\mu\nu} = 2K\bb{{\tilde G}}_{\mu\nu}.
\end{align}
Then, writing $X^\mu \eqqcolon M\tilde G^\mu + \hat D^\mu \CX + \CX^\mu$ with $\CX^\mu \tau_\mu \coloneqq 0 \eqqcolon \hat D_\mu \CX^\mu$, the additional term in equation~\eqref{eq::fghjf}, compared to equation~\eqref{eq::first_Einstein_eq}, takes the form
\begin{align}
	\Lie{X}\left(\accentset{(0)}{\bar R}_{\mu\nu}\right)\hat h^{\mu\alpha}\hat h^{\nu\beta} = 4K \left(\hat D^\mu \hat D^\nu \CX + 2\hat D^{(\mu} \CX^{\nu)}\right).
\end{align}
A consequence is that reperforming the calculation of Section~\ref{app::dico}, we obtain
\begin{align}
\left(\frac{\kappa}{2}\rho + \Lambda\right) \hat h^{ij}	&= \hat h^{ij}(\dot H + 3H^2) + \hat h^{ij}\left[\hat\Delta\psi + \hat\Delta(H\sigma)\right] \nonumber\\
									&\qquad + \hat D^i\hat D^j\left[-\Phi + H\sigma + \psi + 4K(E + \CX)\right] \\
									&\qquad + 4K\hat D^{(i}\left(F^{j)} + \CX^{j)}\right) \nonumber\\
									&\qquad + \left(6K - \hat\Delta\right)f^{ij}. \nonumber
\end{align}
Therefore, $E$ and $F^i$ are not anymore imposed to be zero, instead the gauge-invariant variables $\CC \coloneqq E + \CX$ and $\CC^i \coloneqq F^i + \CX^i$ are zero.\saut

In conclusion, from either the Einstein equation (Euclidean case) or the bi-connection theory (non-Euclidean case), the spacetime metric $\accl{g}^{\mu\nu}$ (respectively $\accl{g}_{\mu\nu}$) takes the form~\eqref{eq::dico_g^munu} [respectively~\eqref{eq::dico_g_munu}] as function of the gravitational potential, with $\sigma$, $E$ and $F^i$ being free gauge fields in both cases.

\end{appendices}

%\section*{References}
% {When available this bibliography style features three different links associated with three different colors: links to the journal/editor website or to a numerical version of the paper are in \textcolor{LinkJournal}{red}, links to the ADS website are in \textcolor{LinkADS}{blue} and links to the arXiv website are in \textcolor{LinkArXiv}{green}.}\\

\IfFileExists{QV_mnras.bst}{\bibliographystyle{QV_mnras}}{\bibliographystyle{/Users/quentinvigneron/Documents/Travail/Research/tex_/QV_mnras}}
%\IfFileExists{QV_mnras_short_CNRS.bst}{\bibliographystyle{QV_mnras_short_CNRS}}{\bibliographystyle{/Users/quentinvigneron/Documents/Travail/Research/tex_/QV_mnras_short_CNRS}}

\IfFileExists{bib_General.bib}{\bibliography{bib_General}}{\bibliography{/Users/quentinvigneron/Documents/Travail/Research/tex_/bib_General}}

\begin{thebibliography}{}
\makeatletter
\relax
\def\mn@urlcharsother{\let\do\@makeother \do\$\do\&\do\#\do\^\do\_\do\%\do\~}
\def\mn@doi{\begingroup\mn@urlcharsother \@ifnextchar [ {\mn@doi@}
  {\mn@doi@[]}}
\def\mn@doi@[#1]#2{\def\@tempa{#1}\ifx\@tempa\@empty \href
  {http://dx.doi.org/#2} {doi:#2}\else \href {http://dx.doi.org/#2} {#1}\fi
  \endgroup}
\def\mn@eprint#1#2{\mn@eprint@#1:#2::\@nil}
\def\mn@eprint@arXiv#1{\href {http://arxiv.org/abs/#1} {{\tt arXiv:#1}}}
\def\mn@eprint@dblp#1{\href {http://dblp.uni-trier.de/rec/bibtex/#1.xml}
  {dblp:#1}}
\def\mn@eprint@#1:#2:#3:#4\@nil{\def\@tempa {#1}\def\@tempb {#2}\def\@tempc
  {#3}\ifx \@tempc \@empty \let \@tempc \@tempb \let \@tempb \@tempa \fi \ifx
  \@tempb \@empty \def\@tempb {arXiv}\fi \@ifundefined
  {mn@eprint@\@tempb}{\@tempb:\@tempc}{\expandafter \expandafter \csname
  mn@eprint@\@tempb\endcsname \expandafter{\@tempc}}}

\bibitem[\protect\citeauthoryear{{Bardeen}}{{Bardeen}}{1980}]{1980_Bardeen}
{Bardeen} J.~M.,  1980,  \textit{ {Gauge-invariant cosmological
  perturbations}}, \mn@doi [\textcolor{LinkJournal}{\prd}]
  {10.1103/PhysRevD.22.1882}, \href
  {https://ui.adsabs.harvard.edu/abs/1980PhRvD..22.1882B}
  {\textcolor{LinkADS}{22}}, 1882-1905

\bibitem[\protect\citeauthoryear{{Barrow}}{{Barrow}}{2020}]{2020_Barrow}
{Barrow} J.~D.,  2020,  \textit{ {Non-Euclidean Newtonian cosmology}}, \mn@doi
  [\textcolor{LinkJournal}{\cqg}] {10.1088/1361-6382/ab8437}, \href
  {https://ui.adsabs.harvard.edu/abs/2020CQGra..37l5007B}
  {\textcolor{LinkADS}{37}}, \href {http://arxiv.org/abs/2002.10155}
  {\textcolor{LinkArXiv}{125007}}

\bibitem[\protect\citeauthoryear{{Buchert} \& {Ehlers}}{{Buchert} \&
  {Ehlers}}{1997}]{1997_Buchert_et_al}
{Buchert} T.,  {Ehlers} J.,  1997,  \textit{ {Averaging inhomogeneous Newtonian
  cosmologies}},
  \href{http://aa.springer.de/bibs/7320001/2300001/small.htm}{\textcolor{LinkJournal}{\aap}},
  \href {https://ui.adsabs.harvard.edu/abs/1997A&A...320....1B}
  {\textcolor{LinkADS}{320}}, \href {http://arxiv.org/abs/astro-ph/9510056}
  {\textcolor{LinkArXiv}{1-7}}

\bibitem[\protect\citeauthoryear{{Dautcourt}}{{Dautcourt}}{1990}]{1990_Dautcourt_a}
{Dautcourt} G.,  1990,  \textit{ {On the Newtonian limit of general
  relativity}},
  \href{https://www.actaphys.uj.edu.pl/R/21/10/755}{\textcolor{LinkJournal}{Acta
  Phys. Pol. B}}, 21, 755-765

\bibitem[\protect\citeauthoryear{{Di Valentino}, {Melchiorri}  \& {Silk}}{{Di
  Valentino} et~al.}{2020}]{2020_Di-Valentino_et_al}
{Di Valentino} E.,  {Melchiorri} A.,   {Silk} J.,  2020,  \textit{ {Planck
  evidence for a closed Universe and a possible crisis for cosmology}}, \mn@doi
  [\textcolor{LinkJournal}{Nature Astronomy}] {10.1038/s41550-019-0906-9},
  \href {https://ui.adsabs.harvard.edu/abs/2020NatAs...4..196D}
  {\textcolor{LinkADS}{4}}, \href {http://arxiv.org/abs/1911.02087}
  {\textcolor{LinkArXiv}{196-203}}

\bibitem[\protect\citeauthoryear{{Dyer} \& {Hinterbichler}}{{Dyer} \&
  {Hinterbichler}}{2009}]{2009_Dyer_et_al}
{Dyer} E.,  {Hinterbichler} K.,  2009,  \textit{ {Boundary terms, variational
  principles, and higher derivative modified gravity}}, \mn@doi
  [\textcolor{LinkJournal}{\prd}] {10.1103/PhysRevD.79.024028}, \href
  {https://ui.adsabs.harvard.edu/abs/2009PhRvD..79b4028D}
  {\textcolor{LinkADS}{79}}, \href {http://arxiv.org/abs/0809.4033}
  {\textcolor{LinkArXiv}{024028}}

\bibitem[\protect\citeauthoryear{{Ehlers}}{{Ehlers}}{2019}]{2019_Ehlers}
{Ehlers} J.,  2019,  \textit{ {Republication of: On the Newtonian limit of
  Einstein's theory of gravitation}}, \mn@doi [\textcolor{LinkJournal}{\grg}]
  {10.1007/s10714-019-2624-0}, \href
  {https://ui.adsabs.harvard.edu/abs/2019GReGr..51..163E}
  {\textcolor{LinkADS}{51}}, 163

\bibitem[\protect\citeauthoryear{{Galloway}, {Khuri}  \& {Woolgar}}{{Galloway}
  et~al.}{2020}]{2020_Galloway_et_al}
{Galloway} G.~J.,  {Khuri} M.~A.,   {Woolgar} E.,  2020,  \textit{ {The
  Topology of General Cosmological Models}}arXiv e-prints, \href
  {https://ui.adsabs.harvard.edu/abs/2020arXiv201006739G}
  {\textcolor{LinkADS}{ADS link}}, \href {http://arxiv.org/abs/2010.06739}
  {\textcolor{LinkArXiv}{arXiv:2010.06739}}

\bibitem[\protect\citeauthoryear{{Handley}}{{Handley}}{2021}]{2021_Handley}
{Handley} W.,  2021,  \textit{ {Curvature tension: Evidence for a closed
  universe}}, \mn@doi [\textcolor{LinkJournal}{\prd}]
  {10.1103/PhysRevD.103.L041301}, \href
  {https://ui.adsabs.harvard.edu/abs/2021PhRvD.103d1301H}
  {\textcolor{LinkADS}{103}}, \href {http://arxiv.org/abs/1908.09139}
  {\textcolor{LinkArXiv}{L041301}}

\bibitem[\protect\citeauthoryear{{Hansen}, {Hartong}  \& {Obers}}{{Hansen}
  et~al.}{2020}]{2020_Hansen_et_al}
{Hansen} D.,  {Hartong} J.,   {Obers} N.~A.,  2020,  \textit{ {Non-relativistic
  gravity and its coupling to matter}}, \mn@doi
  [\textcolor{LinkJournal}{Journal of High Energy Physics}]
  {10.1007/JHEP06(2020)145}, \href
  {https://ui.adsabs.harvard.edu/abs/2020JHEP...06..145H}
  {\textcolor{LinkADS}{2020}}, \href {http://arxiv.org/abs/2001.10277}
  {\textcolor{LinkArXiv}{145}}

\bibitem[\protect\citeauthoryear{{Hartong}, {Obers}  \& {Oling}}{{Hartong}
  et~al.}{2022}]{2022_Hartong_et_al}
{Hartong} J.,  {Obers} N.~A.,   {Oling} G.,  2022,  \textit{ {Review on
  Non-Relativistic Gravity}}, \mn@doi [\textcolor{LinkJournal}{arXiv e-prints}]
  {10.48550/arXiv.2212.11309}, \href
  {https://ui.adsabs.harvard.edu/abs/2022arXiv221211309H}
  {\textcolor{LinkADS}{ADS link}}, \href {http://arxiv.org/abs/2212.11309}
  {\textcolor{LinkArXiv}{arXiv:2212.11309}}

\bibitem[\protect\citeauthoryear{{Heckmann} \& {Sch{\"u}cking}}{{Heckmann} \&
  {Sch{\"u}cking}}{1955}]{1955_Heckmann_et_al}
{Heckmann} O.,  {Sch{\"u}cking} E.,  1955,  \textit{ {Bemerkungen zur
  Newtonschen Kosmologie. I. Mit 3 Textabbildungen in 8
  Einzeldarstellungen}}\zap, \href
  {https://ui.adsabs.harvard.edu/abs/1955ZA.....38...95H}
  {\textcolor{LinkADS}{38}}, 95

\bibitem[\protect\citeauthoryear{Iwasaki}{Iwasaki}{1971}]{1971_Iwasaki}
Iwasaki Y.,  1971,  \textit{ {Quantum Theory of Gravitation vs. Classical
  Theory*): Fourth-Order Potential}}, \mn@doi [\textcolor{LinkJournal}{Progress
  of Theoretical Physics}] {10.1143/PTP.46.1587}, 46, \href
  {http://arxiv.org/abs/https://academic.oup.com/ptp/article-pdf/46/5/1587/5271183/46-5-1587.pdf}
  {\textcolor{LinkArXiv}{1587-1609}}

\bibitem[\protect\citeauthoryear{K{\"u}nzle}{K{\"u}nzle}{1972}]{1972_Kunzle}
K{\"u}nzle H.~P.,  1972,  \textit{ Galilei and Lorentz structures on space-time
  : comparison of the corresponding geometry and physics},
  \href{http://www.numdam.org/item/AIHPA\_1972\_\_17\_4\_337\_0}{\textcolor{LinkJournal}{Annales
  de l'I.H.P. Physique th{\'e}orique}}, 17, 337-362

\bibitem[\protect\citeauthoryear{{K{\"u}nzle}}{{K{\"u}nzle}}{1976}]{1976_Kunzle}
{K{\"u}nzle} H.~P.,  1976,  \textit{ {Covariant Newtonian limit of Lorentz
  space-times}}, \mn@doi [\textcolor{LinkJournal}{\grg}] {10.1007/BF00766139},
  \href {https://ui.adsabs.harvard.edu/abs/1976GReGr...7..445K}
  {\textcolor{LinkADS}{7}}, 445-457

\bibitem[\protect\citeauthoryear{{Lachieze-Rey} \& {Luminet}}{{Lachieze-Rey} \&
  {Luminet}}{1995}]{1995_La_Lu}
{Lachieze-Rey} M.,  {Luminet} J.,  1995,  \textit{ {Cosmic topology}}, \mn@doi
  [\textcolor{LinkJournal}{\physrep}] {10.1016/0370-1573(94)00085-H}, \href
  {https://ui.adsabs.harvard.edu/abs/1995PhR...254..135L}
  {\textcolor{LinkADS}{254}}, \href {http://arxiv.org/abs/gr-qc/9605010}
  {\textcolor{LinkArXiv}{135-214}}

\bibitem[\protect\citeauthoryear{{Lovelock}}{{Lovelock}}{1972}]{1972_Lovelock}
{Lovelock} D.,  1972,  \textit{ {The Four-Dimensionality of Space and the
  Einstein Tensor}}, \mn@doi [\textcolor{LinkJournal}{Journal of Mathematical
  Physics}] {10.1063/1.1666069}, \href
  {https://ui.adsabs.harvard.edu/abs/1972JMP....13..874L}
  {\textcolor{LinkADS}{13}}, 874-876

\bibitem[\protect\citeauthoryear{{Malament}}{{Malament}}{1986a}]{1986_Malament_a}
{Malament} D.~B.,  1986a,  \textit{ {Newtonian Gravity, Limits, and the
  Geometry of Space}}, in Colodny, R. (ed), From Quarks to Quasars. University
  of Pittsburgh Press

\bibitem[\protect\citeauthoryear{Malament}{Malament}{1986b}]{1986_Malament_b}
Malament D.,  1986b, in , Vol.~114, Studies in Logic and the Foundations of
  Mathematics.
Elsevier, pp 405--411

\bibitem[\protect\citeauthoryear{{Malik} \& {Matravers}}{{Malik} \&
  {Matravers}}{2008}]{2008_Malik_et_al}
{Malik} K.~A.,  {Matravers} D.~R.,  2008,  \textit{ {TOPICAL REVIEW: A concise
  introduction to perturbation theory in cosmology}}, \mn@doi
  [\textcolor{LinkJournal}{\cqg}] {10.1088/0264-9381/25/19/193001}, \href
  {https://ui.adsabs.harvard.edu/abs/2008CQGra..25s3001M}
  {\textcolor{LinkADS}{25}}, \href {http://arxiv.org/abs/0804.3276}
  {\textcolor{LinkArXiv}{193001}}

\bibitem[\protect\citeauthoryear{{Misner}, {Thorne}  \& {Wheeler}}{{Misner}
  et~al.}{1973}]{1973_MTW}
{Misner} C.~W.,  {Thorne} K.~S.,   {Wheeler} J.~A.,  1973,  \textit{
  Gravitation}, San Francisco: W.H. Freeman and Co.

\bibitem[\protect\citeauthoryear{{Morgan} \& {Tian}}{{Morgan} \&
  {Tian}}{2006}]{2006_Morgan_et_al}
{Morgan} J.~W.,  {Tian} G.,  2006,  \textit{ {Ricci Flow and the Poincare
  Conjecture}}, \mn@doi [\textcolor{LinkJournal}{arXiv Mathematics e-prints}]
  {10.48550/arXiv.math/0607607}, \href
  {https://ui.adsabs.harvard.edu/abs/2006math......7607M}
  {\textcolor{LinkADS}{ADS link}}, \href {http://arxiv.org/abs/math/0607607}
  {\textcolor{LinkArXiv}{math/0607607}}

\bibitem[\protect\citeauthoryear{{Peebles}}{{Peebles}}{1980}]{1980_Peebles}
{Peebles} P.~J.~E.,  1980,  \textit{ {The large-scale structure of the
  universe}}, Princeton University Press

\bibitem[\protect\citeauthoryear{{Rendall}}{{Rendall}}{1992}]{1992_Rendall}
{Rendall} A.~D.,  1992,  \textit{ {On the definition of post-Newtonian
  approximations}}, \mn@doi [\textcolor{LinkJournal}{Proceedings of the Royal
  Society of London Series A}] {10.1098/rspa.1992.0111}, \href
  {https://ui.adsabs.harvard.edu/abs/1992RSPSA.438..341R}
  {\textcolor{LinkADS}{438}}, 341-360

\bibitem[\protect\citeauthoryear{{Rosen}}{{Rosen}}{1980}]{1980_Rosen}
{Rosen} N.,  1980,  \textit{ {General relativity with a background metric}},
  \mn@doi [\textcolor{LinkJournal}{Foundations of Physics}]
  {10.1007/BF00708416}, \href
  {https://ui.adsabs.harvard.edu/abs/1980FoPh...10..673R}
  {\textcolor{LinkADS}{10}}, 673-704

\bibitem[\protect\citeauthoryear{{Rosen}}{{Rosen}}{1985}]{1985_Rosen_a}
{Rosen} N.,  1985,  \textit{ {Localization of gravitational energy}}, \mn@doi
  [\textcolor{LinkJournal}{Foundations of Physics}] {10.1007/BF00732842}, \href
  {https://ui.adsabs.harvard.edu/abs/1985FoPh...15..997R}
  {\textcolor{LinkADS}{15}}, 997-1008

\bibitem[\protect\citeauthoryear{{Roukema} \& {Bajtlik}}{{Roukema} \&
  {Bajtlik}}{2008}]{2008_Roukema_et_al}
{Roukema} B.~F.,  {Bajtlik} S.,  2008,  \textit{ {Homotopy symmetry in the
  multiply connected twin paradox of special relativity}}, \mn@doi
  [\textcolor{LinkJournal}{\mnras}] {10.1111/j.1365-2966.2008.13734.x}, \href
  {https://ui.adsabs.harvard.edu/abs/2008MNRAS.390..655R}
  {\textcolor{LinkADS}{390}}, \href {http://arxiv.org/abs/astro-ph/0606559}
  {\textcolor{LinkArXiv}{655-664}}

\bibitem[\protect\citeauthoryear{{Roukema} \& {R{\'o}{\.z}a{\'n}ski}}{{Roukema}
  \& {R{\'o}{\.z}a{\'n}ski}}{2009}]{2009_Roukema_et_al}
{Roukema} B.~F.,  {R{\'o}{\.z}a{\'n}ski} P.~T.,  2009,  \textit{ {The residual
  gravity acceleration effect in the Poincar{\'e} dodecahedral space}}, \mn@doi
  [\textcolor{LinkJournal}{\aap}] {10.1051/0004-6361/200911881}, \href
  {https://ui.adsabs.harvard.edu/abs/2009A&A...502...27R}
  {\textcolor{LinkADS}{502}}, \href {http://arxiv.org/abs/0902.3402}
  {\textcolor{LinkArXiv}{27-35}}

\bibitem[\protect\citeauthoryear{{Stichel}}{{Stichel}}{2016}]{2016_Stichel}
{Stichel} P.~C.,  2016,  \textit{ {Cosmological model with dynamical
  curvature}}arXiv e-prints, \href
  {https://ui.adsabs.harvard.edu/abs/2016arXiv160107030S}
  {\textcolor{LinkADS}{ADS link}}, \href {http://arxiv.org/abs/1601.07030}
  {\textcolor{LinkArXiv}{arXiv:1601.07030}}

\bibitem[\protect\citeauthoryear{{Thurston}}{{Thurston}}{1982}]{1982_Thurston}
{Thurston} W.~P.,  1982,  \textit{ {Three dimensional manifolds, Kleinian
  groups and hyperbolic geometry}},
  \href{https://projecteuclid.org/journals/bulletin-of-the-american-mathematical-society-new-series/volume-6/issue-3/Three-dimensional-manifolds-Kleinian-groups-and-hyperbolic-geometry/bams/1183548782.full}{\textcolor{LinkJournal}{\bams}},
  6, 357-381

\bibitem[\protect\citeauthoryear{{Tichy} \& {Flanagan}}{{Tichy} \&
  {Flanagan}}{2011}]{2011_Tichy_et_al}
{Tichy} W.,  {Flanagan} {\'E}.~{\'E}.,  2011,  \textit{ {Covariant formulation
  of the post-1-Newtonian approximation to general relativity}}, \mn@doi
  [\textcolor{LinkJournal}{\prd}] {10.1103/PhysRevD.84.044038}, \href
  {https://ui.adsabs.harvard.edu/abs/2011PhRvD..84d4038T}
  {\textcolor{LinkADS}{84}}, \href {http://arxiv.org/abs/1101.0588}
  {\textcolor{LinkArXiv}{044038}}

\bibitem[\protect\citeauthoryear{{Uzan}, {Luminet}, {Lehoucq}  \&
  {Peter}}{{Uzan} et~al.}{2002}]{2002_Uzan_et_al}
{Uzan} J.-P.,  {Luminet} J.-P.,  {Lehoucq} R.,   {Peter} P.,  2002,  \textit{
  {The twin paradox and space topology}}, \mn@doi
  [\textcolor{LinkJournal}{European Journal of Physics}]
  {10.1088/0143-0807/23/3/306}, \href
  {https://ui.adsabs.harvard.edu/abs/2002EJPh...23..277U}
  {\textcolor{LinkADS}{23}}, \href {http://arxiv.org/abs/physics/0006039}
  {\textcolor{LinkArXiv}{277-284}}

\bibitem[\protect\citeauthoryear{{Vigneron}}{{Vigneron}}{2021a}]{2021_Vigneron_PhD}
{Vigneron} Q.,  2021a, PhD thesis, Universit\'e Claude Bernard Lyon 1, \url
  {https://www.theses.fr/en/2021LYSE1117}

\bibitem[\protect\citeauthoryear{{Vigneron}}{{Vigneron}}{2021b}]{2021_Vigneron}
{Vigneron} Q.,  2021b,  \textit{ 1+3 -Newton-Cartan system and Newton-Cartan
  cosmology}, \mn@doi [\textcolor{LinkJournal}{\prd}]
  {10.1103/PhysRevD.103.064064}, \href
  {https://ui.adsabs.harvard.edu/abs/2021PhRvD.103f4064V}
  {\textcolor{LinkADS}{103}}, \href {http://arxiv.org/abs/2012.10213}
  {\textcolor{LinkArXiv}{064064}}

\bibitem[\protect\citeauthoryear{{Vigneron}}{{Vigneron}}{2022a}]{2022_Vigneron}
{Vigneron} Q.,  2022a,  \textit{ {{\mockalphA{1}}Is backreaction in cosmology a
  relativistic effect? On the need for an extension of Newton's theory to
  non-Euclidean topologies}}, \mn@doi [\textcolor{LinkJournal}{\prd}]
  {10.1103/PhysRevD.105.043524}, \href
  {https://ui.adsabs.harvard.edu/abs/2022PhRvD.105d3524V}
  {\textcolor{LinkADS}{105}}, \href {http://arxiv.org/abs/2109.10336}
  {\textcolor{LinkArXiv}{043524}}

\bibitem[\protect\citeauthoryear{{Vigneron}}{{Vigneron}}{2022b}]{2022_Vigneron_b}
{Vigneron} Q.,  2022b,  \textit{ {\mockalphA{2}}On non-Euclidean Newtonian
  theories and their cosmological backreaction}, \mn@doi
  [\textcolor{LinkJournal}{\cqg}] {10.1088/1361-6382/ac7a87}, \href
  {https://ui.adsabs.harvard.edu/abs/2022CQGra..39o5006V}
  {\textcolor{LinkADS}{39{\mockalphA{222}}}}, \href
  {http://arxiv.org/abs/2201.02112} {\textcolor{LinkArXiv}{155006}}

\bibitem[\protect\citeauthoryear{{Vigneron} \& {Poulin}}{{Vigneron} \&
  {Poulin}}{2023}]{2023_Vigneron_et_al_b}
{Vigneron} Q.,  {Poulin} V.,  2023,  \textit{ {Is expansion blind to the
  spatial curvature?}}, \mn@doi [\textcolor{LinkJournal}{\prd}]
  {10.1103/PhysRevD.108.103518}, \href
  {https://ui.adsabs.harvard.edu/abs/2022arXiv221200675V}
  {\textcolor{LinkADS}{108}}, \href {http://arxiv.org/abs/2212.00675}
  {\textcolor{LinkArXiv}{103518}}

\bibitem[\protect\citeauthoryear{{Vigneron} \& {Roukema}}{{Vigneron} \&
  {Roukema}}{2023}]{2023_Vigneron_et_al_a}
{Vigneron} Q.,  {Roukema} B.~F.,  2023,  \textit{ {Gravitational potential in
  spherical topologies}}, \mn@doi [\textcolor{LinkJournal}{\prd}]
  {10.1103/PhysRevD.107.063545}, \href
  {https://ui.adsabs.harvard.edu/abs/2022arXiv220109102V}
  {\textcolor{LinkADS}{107}}, \href {http://arxiv.org/abs/2201.09102}
  {\textcolor{LinkArXiv}{063545}}

\bibitem[\protect\citeauthoryear{{York}}{{York}}{1973}]{1973_York}
{York} James~W. J.,  1973,  \textit{ {Conformally invariant orthogonal
  decomposition of symmetric tensors on Riemannian manifolds and the
  initial-value problem of general relativity}}, \mn@doi
  [\textcolor{LinkJournal}{Journal of Mathematical Physics}]
  {10.1063/1.1666338}, \href
  {https://ui.adsabs.harvard.edu/abs/1973JMP....14..456Y}
  {\textcolor{LinkADS}{14}}, 456-464

\makeatother
\end{thebibliography}

%\bibliographystyle{QV_mnras}
%\bibliography{bib_General}


\end{document}